\documentclass[useAMS,usenatbib]{mn2e}

\usepackage[flushleft]{threeparttable}
\usepackage{graphicx}
\usepackage{xtab,afterpage}
\usepackage{amsmath}
\usepackage{epstopdf}
\usepackage{url}
\usepackage[flushleft]{threeparttable}
\usepackage{booktabs,fixltx2e}
\usepackage{amssymb}
\usepackage{textcomp}
\usepackage{hyperref}

\newcommand{\icarus}{Icarus}

\newcommand{\pasp}{{\it PASP~\/}}

\newcommand{\apj}{ApJ}
\newcommand{\apjl}{ApJ}

\newcommand{\aap}{A \& A}

\newcommand{\aj}{AJ}
\newcommand{\mnras}{MNRAS}

\def\jgr{\rmfamily{J.~Geophys.~Res.~}}

\renewcommand{\v}{\ensuremath{\mathbf{v}}}

\makeatletter
\newcommand*{\rom}[1]{\expandafter\@slowromancap\romannumeral #1@}
\makeatother

\title[Detection of gas around HD 129590]{Survey of planetesimal belts with ALMA: gas detected around the Sun-like star HD 129590}
\author[Q. Kral et al.]{Quentin Kral,$^{1}$\thanks{E-mail: quentin.kral@obspm.fr} Luca Matr{\`a},$^{2,3}$ Grant M. Kennedy,$^{4,5}$ Sebastian Marino$^{6}$ and
\newauthor{Mark C. Wyatt$^{7}$}\\
$^{1}$LESIA, Observatoire de Paris, Universit{\'e} PSL, CNRS, Sorbonne Universit{\'e}, Univ. Paris Diderot,\\ Sorbonne Paris Cit{\'e}, 5 place Jules Janssen, 92195 Meudon, France\\
$^{2}$School of Physics, National University of Ireland Galway, University Road, Galway, Ireland\\
$^{3}$Harvard-Smithsonian Center for Astrophysics, 60 Garden Street, Cambridge, MA 02138, USA\\
$^{4}$Department of Physics, University of Warwick, Gibbet Hill Road, Coventry CV4 7AL, UK\\
$^{5}$Centre for Exoplanets and Habitability, University of Warwick, Gibbet Hill Road, Coventry CV4 7AL, UK\\
$^{6}$Max Planck Institute for Astronomy, K{\"o}nigstuhl 17, D-69117 Heidelberg, Germany\\
$^{7}$Institute of Astronomy, University of Cambridge, Madingley Road, Cambridge CB3 0HA, UK\\
}

\begin{document}

\date{Accepted 1928 December 15. Received 1928 December 14; in original form 1928 October 11}

\pagerange{\pageref{firstpage}--\pageref{lastpage}} \pubyear{2002}

\maketitle

\label{firstpage}

\begin{abstract}
Gas detection around main sequence stars is becoming more common with around 20 systems showing the presence of CO. However, more detections are needed, especially around later spectral type stars to better understand the origin of this gas and refine our models. To do so, we carried out a survey of 10 stars with predicted high likelihoods of secondary CO detection using ALMA in band 6. We looked for continuum emission of mm-dust as well as gas emission (CO and CN transitions). The continuum emission was detected in 9/10 systems for which we derived the discs' dust masses and geometrical properties, providing the first mm-wave detection of the disc around HD 106906, the first mm-wave radius for HD 114082, 117214, HD 15745, HD 191089 and the first radius at all for HD 121191. A crucial finding of our paper is that we detect CO for the first time around the young 10-16 Myr old G1V star HD 129590, similar to our early Sun. The gas seems colocated with its planetesimal belt and its total mass is likely between $2-10 \times 10^{-5}$ M$_\oplus$. This first gas detection around a G-type main-sequence star raises questions as to whether gas may have been released in the Solar System as well in its youth, which could potentially have affected planet formation. We also detected CO gas around HD 121191 at a higher S/N than previously and find that the CO lies much closer-in than the planetesimals in the system, which could be evidence for the previously suspected CO viscous spreading owing to shielding preventing its photodissociation. Finally, we make estimates for the CO content in planetesimals and the HCN/CO outgassing rate (from CN upper limits), which we find are below the level seen in Solar System comets in some systems.
\end{abstract}


\section{Introduction}
There are currently about 20 main sequence stars known with ages $>10$ Myr that are surrounded by circumstellar gas \citep[e.g.][]{1995Natur.373..494Z,2014Sci...343.1490D,2014A&A...563A..66C,2016MNRAS.460.2933M,2017ApJ...842....9M,2018arXiv181108439K}. Thanks to many recent surveys of this gas with ALMA \citep{2011ApJ...740L...7M,2016ApJ...828...25L,2017ApJ...849..123M,2019arXiv190809685M}, this number keeps increasing at a high rate.

Most of this gas is found to orbit A-stars, for which CO is detected in the sub-mm \citep[e.g.][]{2013ApJ...776...77K,2015ApJ...814...42M,2016MNRAS.461.3910G}. Indeed, \citet{2017ApJ...849..123M} find a higher detection rate of CO gas ($69^{+9}_{-13}\%$) around main-sequence A-stars with high fractional luminosity planetesimal belts compared to later-type stars ($7^{+13}_{-2}\%$), but this is expected from models and observational biases \citep{2017MNRAS.469..521K,2019AJ....157..117M,Marino}. Recently, neutral carbon has been shown by models to potentially be a more sensitive tracer of this gas in the sub-mm \citep{2017MNRAS.469..521K}, and it is now targeted with ALMA, leading to the first detections \citep{2017ApJ...839L..14H,2018arXiv181108439K,2019arXiv190807032H,Catal,2019arXiv190407215C}. This carbon component may be more extended inwards \citep[according to models, e.g.][]{2016MNRAS.461..845K} and outwards \citep[according to models and hinted at by recent high resolution observations,][]{2019arXiv190807032H} compared to CO. We also note that thanks to Herschel a few systems have ionised carbon and neutral oxygen detected \citep{2012A&A...546L...8R,2014A&A...565A..68R,2014A&A...563A..66C,2016MNRAS.461..845K, 2016A&A...591A..27B} as well as heavier metals (e.g. Ca, Na, Fe, ...) detected in the UV and optical but only in a very few systems \citep[mostly $\beta$ Pic, e.g.][]{2004A&A...413..681B,2013ApJ...771...69R, 2012A&A...544A.134N,2019A&A...621A.121W}. 

CO gas masses derived from system to system can vary by orders of magnitude, going from low gas mass discs \citep[$\sim10^{-7}$ M$_\oplus$,][]{2017ApJ...842....9M} to high gas mass discs \citep[$> 10^{-2}$ M$_\oplus$,][]{2017ApJ...849..123M,2019arXiv190809685M} with CO masses comparable to protoplanetary disc levels. In low mass systems, CO is consistent with being colocated with the systems' planetesimal belts \citep[e.g.][]{2016MNRAS.460.2933M} and the observed CO is always optically thin to photodissociating UV radiation \citep[even considering unreasonable amounts of H$_2$ present in the systems,][]{2017ApJ...842....9M} and therefore photodissociates on timescales of order 100 years \citep{2009A&A...503..323V}. Hence the observed CO must be recently produced, implying a recent release event \citep[e.g.][]{2019arXiv190407215C} or a continuous process to release gas, which is thought to be released from planetesimals \citep[e.g.][]{2012ApJ...758...77Z,2016MNRAS.461..845K}.

The current framework that best explains all the observables is one where the released CO is secondary \citep[as released from planetesimals rather than being a remnant of the primordial phase,][]{2012ApJ...758...77Z} and then photodissociates into carbon and oxygen, hence creating an atomic gas disc that can viscously spread \citep{2016MNRAS.461..845K}. This model also works for the most massive CO discs observed, where it is thought that a lot of carbon had time to accumulate, which then shielded CO from photodissociating, which in turn can also accumulate \citep{2018arXiv181108439K}. These massive discs are called shielded secondary discs and in those, the longer CO lifetime allows for CO to viscously spread as well (similar to atomic species such as carbon or oxygen) and therefore CO gas discs could be more extended than their planetesimal belts \citep{2018arXiv181108439K,Marino} as may be observed in HD 21997 \citep{2013ApJ...776...77K}. This simple model explains most of the current detections and non-detections \citep{2017MNRAS.469..521K,2018arXiv181108439K,Marino}.

In this picture, if most of the released gas is CO, this secondary gas is expected to be H$_2$-poor, in contrast with gas in protoplanetary discs and in the ISM. There is mounting evidence for this, including small scale heights measured in these discs (indicative of large mean molecular weights), CO line ratios implying low excitation temperatures and sub-Solar H content in $\beta$ Pic \citep{2017ApJ...839...86H,2017MNRAS.464.1415M,2017A&A...599A..75W}. One important motivation of this model is that one can start extracting the volatile composition of planetesimals in these belts around main sequence stars. Indeed, comparing the CO and dust production rates derived from observations, we find that the CO(+CO$_2$) ice mass fractions measured in these belts are similar to Solar System comets \citep[within one order of magnitude,][]{2017ApJ...842....9M}, therefore making it possible to detect gas from Solar System-like exocomets (i.e. icy planetesimals) orbiting at tens of au.

Depending on the composition of planetesimals, and as may be expected from Solar System comets, other species than CO may also be released. Many gas release mechanisms can co-exist and they all predict efficient outgassing: UV photodesorption \citep{2007A&A...475..755G}, high velocity collisions \citep{2007ApJ...660.1541C}, planetesimal breakup
\citep{2012ApJ...758...77Z}, sublimation \citep[e.g.][]{1990A&A...236..202B}, and giant impacts \citep[e.g.][]{2014MNRAS.440.3757J}. Therefore, other molecular species may be expected to be released together with CO in these discs. Their shorter photodissociation timescales makes it difficult to detect them with current facilities, except perhaps for HCN or CN, which may potentially be detectable with ALMA in the near future \citep{2018ApJ...853..147M}. However, products of their photodissociation, such as atomic N, could be detected. There are indications of this with the presence of atomic N gas in $\beta$ Pic \citep{2019A&A...621A.121W}. This is also suggested by the low C/O ratio found in the gas phase around $\beta$ Pic, which seems to show that oxygen may also be released from, e.g., H$_2$O or CO$_2$ \citep{2016A&A...591A..27B,2016MNRAS.461..845K}, but is at a level that is undetectable with current facilities \citep[e.g.][]{2019A&A...628A.127C}. 

This gas may also play a key role in the late stages of planet formation. Recently, \citet{Kral19} showed that gas accretion in this late phase is very efficient onto terrestrial planets embedded in these discs and can create massive CO atmospheres with masses going from an Earth-atmosphere worth of mass to an atmosphere with mini-Neptune-like pressures, hence totally resetting the initial primordial atmospheres of these planets. Atmosphere formation can then last for tens of millions of years so this could be the way atmospheres form around terrestrial planets after photoevaporation or giant impacts desiccated their atmospheres. Owing to this very efficient accretion, any planet in these gas discs could create cavities in carbon and oxygen (as well as CO in shielded discs) that could be used as an alternative planet detection method to infer the location of low to high mass planets at a few to tens of au from their host stars using the high resolution and sensitivity of ALMA. Finally, the detection of this gas is also a new opportunity to study mechanisms that can transport angular momentum in discs and may be beneficial for our understanding of the magnetorotational instability in low density environments (with non ideal effects such as ambipolar diffusion), which may be potentially active in these discs \citep{2016MNRAS.461.1614K}. 

In this paper, we present a new deep CO and CN survey of 10 targets with ALMA. We start by presenting the new dust and gas observations in Section~\ref{obs}. We then carry on by modelling the data and present the results in Section~\ref{model}. We then discuss our findings in Section~\ref{discu} before concluding in Section~\ref{ccl}.

\section{ALMA observations}\label{obs}

\subsection{Observed sample}

We looked for dust and gas around the 10 following stars: HD 106906, HD 114082, HD 117214, HD 121191, HD 129590, HD 143675, HD 15745, HD 191089, HD 69830, and HR 4796. These systems were chosen from a large set of nearby stars, from which we prioritised systems with late spectral types (which have less detections so far) and bright debris discs, which were predicted to have the highest likelihood of secondary CO detection \citep[][using the limited knowledge of planetesimal belt radii at the time]{2017MNRAS.469..521K} and could be detected with ALMA assuming a CO mass fraction of 6\% in the planetesimals. The systems' characteristics (fractional luminosity, distance, blow-out size and stellar parameters) are listed in Table~\ref{tabsys}. 

Our sample is composed of 3 A stars, 5 F stars and 2 G stars. The discs in our sample have fractional luminosities (L$_{\rm IR}/$L$_\star$) between $0.12-5.8 \times 10^{-3}$ and are within 140 pc from Earth. We note that HD 69830 is much older \citep[$10.6\pm4$ Gyr][]{2015ApJ...800..115T} than the other 9 targets.

\begin{table*}
  \centering
  \caption{Disc and stellar parameters of our sample of 10 stars observed with ALMA. From left to right, we list the name of the system, its fractional luminosity, its distance (GAIA DR2), the stellar type, luminosity, mass, temperature of the star, and the blowout size of the grains \citep[computed from][using L$_\star$ and M$_\star$ listed, assuming $Q_{\rm PR}=1$]{1979Icar...40....1B}.}

  \begin{threeparttable}
  \label{tabsys}
  \begin{tabular}{|l|c|c|c|c|c|c|c|c|}
   \toprule
   Systems & L$_{\rm IR}/$L$_\star$  & $d$ & Spectral type & L$_\star$ & M$_\star$ & T$_{\rm eff}$ & $s_{\rm blow}$ \\ 
    & & (pc) & & (L$_\odot$) & (M$_\odot$) & ($K$) & ($\mu$m)  \\ 

    \midrule
    \midrule
    HD 106906 & $1.2 \times 10^{-3}$ & 103.3 & F5V & 6.6 & 1.6 & 6490 & 1.8   \\
    HD 114082 & $3.7 \times 10^{-3}$ & 95.7 & F3V & 3.8 & 1.4 & 6590 & 1.2   \\
    HD 117214 & $2.4 \times 10^{-3}$ & 107.6 & F6V & 5.7 & 1.5 & 6340 & 1.6    \\
    HD 121191 & $2.4 \times 10^{-3}$& 132.1 & A5V & 7.2 & 1.6 & 7690 & 1.9    \\
    HD 129590 & $5.8 \times 10^{-3}$& 136.0 & G1V & 3.0 & 1.3 & 5810 & 0.98   \\
    HD 143675 & $5 \times 10^{-4}$ & 139.2 & A5V & 8.9 & 1.7 & 7890 & 2.2   \\
    HD 15745 & $8 \times 10^{-4}$ & 72.0 & F0V & 4.2 & 1.4 & 6830 & 1.3    \\
    HD 191089  & $1.5 \times 10^{-3}$ & 50.1 & F5V & 2.7 & 1.3 & 6460 & 0.91    \\
    HD 69830 & $1.2 \times 10^{-4}$ & 12.6 & G8V & 0.6 & 0.9 & 5410 & 0.29    \\
    HR 4796A & $3.7 \times 10^{-3}$ & 71.9 & A0V & 25.7 & 2.3 & 9810 & 4.8    \\

   \bottomrule
  \end{tabular}
\begin{tablenotes}
      \small
      \item References: From \citet[][and references therein]{2017MNRAS.469..521K} updated with GAIA DR2 distances \citep{2016A&A...595A...1G,2018A&A...616A...1G}
    \end{tablenotes}
  \end{threeparttable}
\end{table*}

\subsection{ALMA observing configuration}

Our sample of 10 stars were targeted by ALMA in band 6 between $7^{\rm th}$ April and $12^{\rm th}$ August 2018 as part of the cycle 5 project 2017.1.00704.S. The observations were carried out using 43 antennas with baselines ranging from 15 to 500 m.

The ALMA correlator is divided in four spectral windows, two of which focused on observing the dust continuum with 128 channels centered at 243.1 and 245.1 GHz (bandwidth of 2 GHz each). 
The third and the fourth spectral windows targeted the CO J=2-1 line at 230.538 GHz and the hyperfine transitions of the CN N=2-1 line at $\sim$227 GHz, respectively (i.e., rest wavelengths of 1.30040 mm for CO, and $\sim$1.32 mm for the strongest CN transition).
The latter two spectral windows have a channel width of 488.281 kHz (0.635 or 0.645 km/s at the rest frequency of the CO and dominant CN line) over 3840 channels (i.e., a bandwidth of 1.875 GHz each) centered at 230.1 GHz for CO and 227.2 GHz for CN.

The pipeline provided by ALMA was used to apply calibrations and extract the calibrated visibilities.

\subsection{Dust continuum observations}\label{dustobs}

As part of the procedure to extract the continuum data, we removed the channels close to the CO J=2-1 line and summed the four spectral windows to obtain a continuum image at $\sim$ 1.27mm ($\sim$ 236 GHz). We then used the CLEAN algorithm to image the data using natural weightings. This yielded the synthesised beam sizes and position angles listed in Table~\ref{tabobscon}. For each image, we extracted the rms noise (far away from the source and across many beam widths) as listed in Table~\ref{tabobscon}.


\begin{table*}
  \centering
  \caption{Continuum observations of our sample of 10 stars. From left to right, the columns correspond to the system's name, the beam size and its position angle for a given observation, the rms that is reached and the total integrated flux on the source (for the non-detection around HD 143675, we give the 3$\sigma$ value in one beam). }

  \label{tabobscon}
  \begin{tabular}{|l|c|c|c|c|}
   \toprule
   Systems & Beam size  & Beam PA  & rms & F$_\nu$ at 1.27mm\\ 
    & (\arcsec) & ($^\circ$)& ($\mu$Jy/beam) & (mJy) \\ 

    \midrule
    \midrule
    HD 106906 & $1.62 \times 1.29$ & 77.2 & 18 & $0.35 \pm 0.04$     \\
    HD 114082 & $1.49 \times 1.42$ & 48.6 & 24 & $0.71 \pm 0.07$     \\
    HD 117214 & $1.66 \times 1.27$ & 62.9 & 26 & $0.78 \pm 0.08$     \\
    HD 121191 & $0.92 \times 0.86$& -52.2 & 16 & $0.45 \pm 0.05$     \\
    HD 129590 & $1.57 \times 1.14$& 83.0 & 22 & $1.25 \pm 0.1$     \\
    HD 143675 & $0.91 \times 0.8$ & 78.5 & 13 & $<0.04$     \\
    HD 15745 & $1.68 \times 1.19$ & 24.2 & 42 & $1.2 \pm 0.1$     \\
    HD 191089 & $1.57 \times 1.08$ & 74.9 & 16 & $1.83 \pm 0.2$     \\
    HD 69830 & $0.95 \times 0.76$ & 64.1 & 14 & $0.05 \pm 0.01$     \\
    HR 4796A & $1.73 \times 1.26$ & -87.9 & 25 & $5.7 \pm 0.06$     \\

   \bottomrule
  \end{tabular}

\end{table*}

\begin{figure*}
   \centering
   \includegraphics[width=11cm]{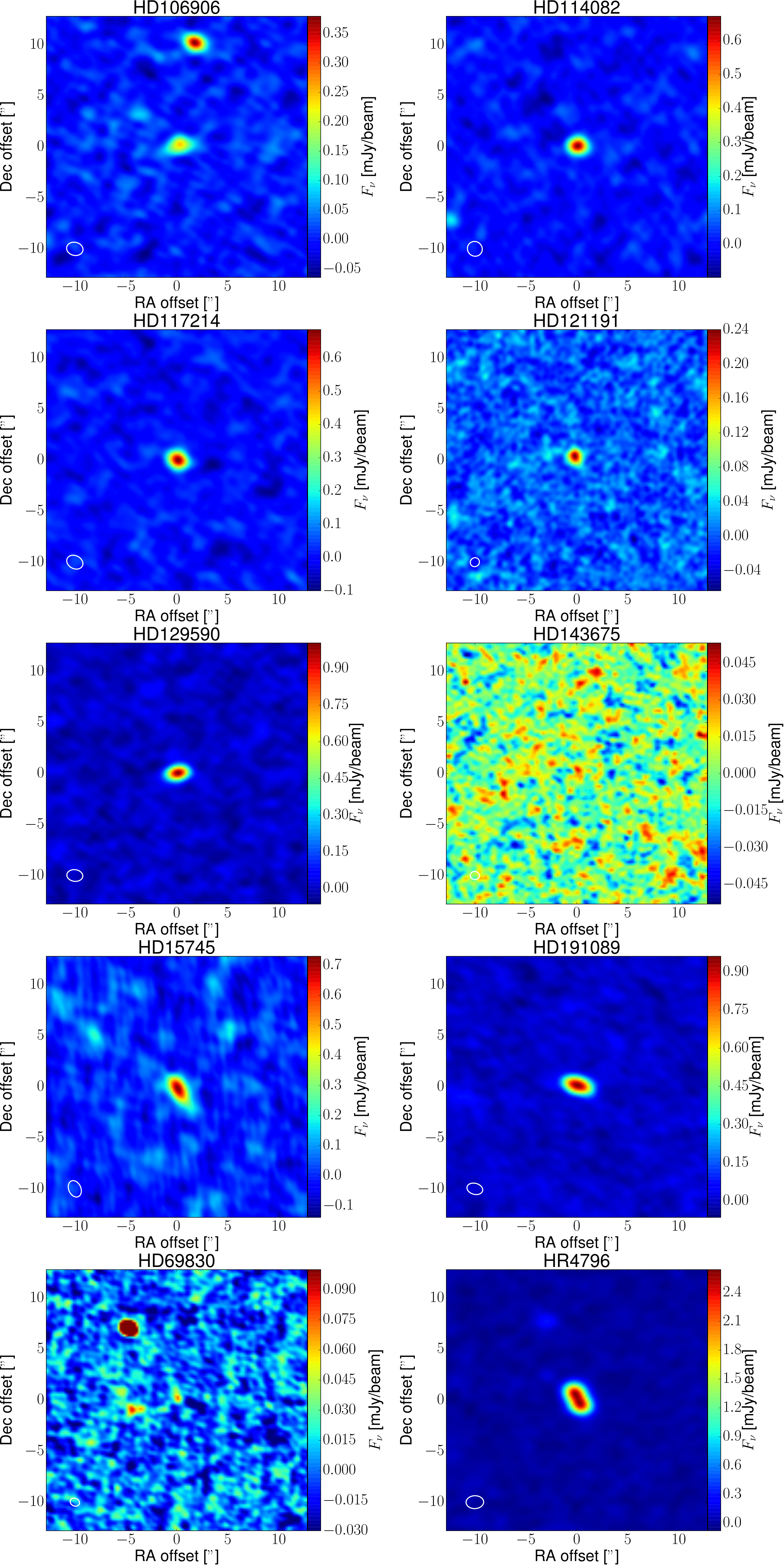}
   \caption{\label{figcont} Continuum images obtained for our survey of 10 stars. These are naturally-weighted cleaned images (see Sec.~\ref{dustobs} for the reduction method used). The physical parameters derived from the modelling part (see Sec.~\ref{dustfitt}) for each observation are listed in Table~\ref{tabbestfitcont}. We have rescaled the colourbar for the HD 69830 image as the flux was otherwise dominated by the bright (now saturated) galaxy observed at the North-East of the image.}
\end{figure*}

In Fig.~\ref{figcont}, we show the results of our continuum survey. We see by eye that the continuum is detected in 9/10 systems. Only the observation of HD 143675 yielded no clear detection but rather a marginal 2-$\sigma$ level peak\footnote{However, this non detection puts a constraint on the slope of the modified black body (and hence size distribution of the grains) that fits the SED as explained further in Sec.~\ref{new}.}. The detection around HD 69830 is $>3\sigma$, but it is actually consistent with detection of the stellar photosphere (as it is only 12.6 pc away from Earth) rather than a dust component as explained in further detail in Sec.\ref{star}\footnote{HD 69830 is surrounded by warm dust very close to its host star, which yields almost no emission at 1.27 mm.}. For the 9 systems with detections, we calculated their total flux in the natural map within an ellipse slightly larger than where the emission comes from (i.e., where S/N $\gtrsim$ 2) as listed in the last column of Table~\ref{tabobscon} where the errors were computed from the image noise and flux calibration uncertainties (10\%) added in quadrature (where the latter typically dominates).

\subsection{Gas observations}\label{gasobs}

\begin{figure*}
   \centering
   \includegraphics[width=11cm]{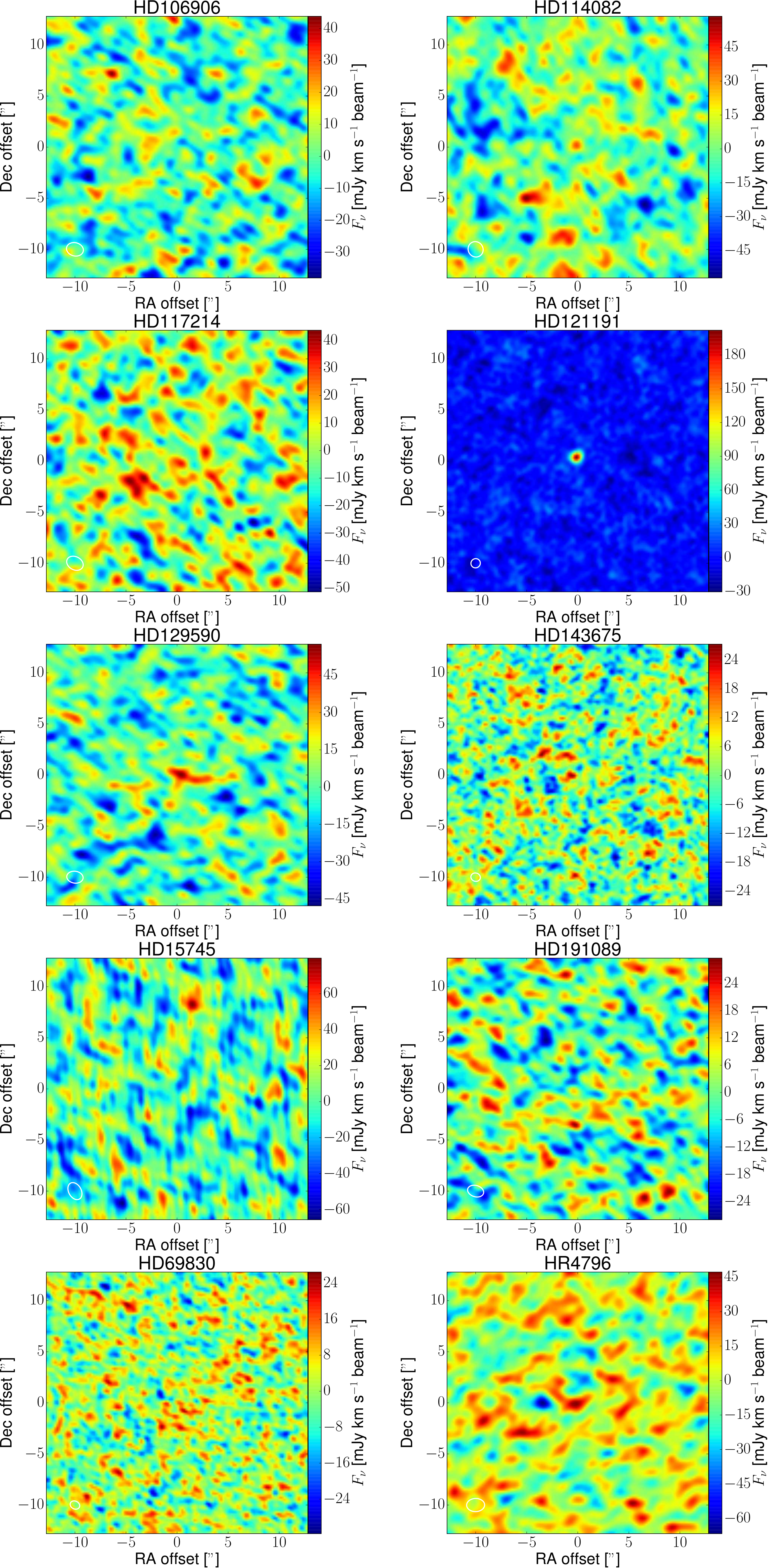}
   \caption{\label{figcodetec} CO J=2-1 moment-0 images integrated between $\pm 20$km/s for our survey of 10 stars. These are naturally-weighted images (see Sec.~\ref{gasobs} for the reduction method used). The parameters derived from our model (see Sec.~\ref{gasfit}) associated with each observation are listed in Table~\ref{tabbestfitgas}.}
\end{figure*}

From the calibrated visibilities and using the two spectral windows with a higher spectral resolution (channel width of 488.281 kHz), we also extracted data cubes around the CO J=2-1 line (at a rest frequency of 230.538 GHz) and close to the strongest CN transition (where the Einstein coefficient is the highest around 226.875 GHz, i.e. it has the highest transition probability). To study gas emission, we fitted and subtracted continuum emission from the cube directly from visibilities (using the task \texttt{uvcontsub} in CASA), avoiding channels around the line frequencies. This yielded naturally weighted image cubes with the synthesised beam sizes and position angles listed in Table~\ref{tabobsgas}. For each cube, we extracted the rms noise per channel (far away from the source and across many beam widths) as listed in Table~\ref{tabobsgas}. Fig.~\ref{figcodetec} shows the moment-0 (i.e. spectrally integrated) images of the CO J=2-1 transition for all systems where we decided to be agnostic about the Keplerian velocity of a potential CO gas disc (that may or may not be rotating at the same velocity as the planetesimal belt) and integrated the cubes in velocity between $\pm20$km/s. This was refined on a case-by-case basis, using multiple velocity integration ranges for each source as well as different disc extents, and by plotting the resulting spectra to confirm the cases with non detections.

\begin{figure}
   \centering
      \includegraphics[width=8cm]{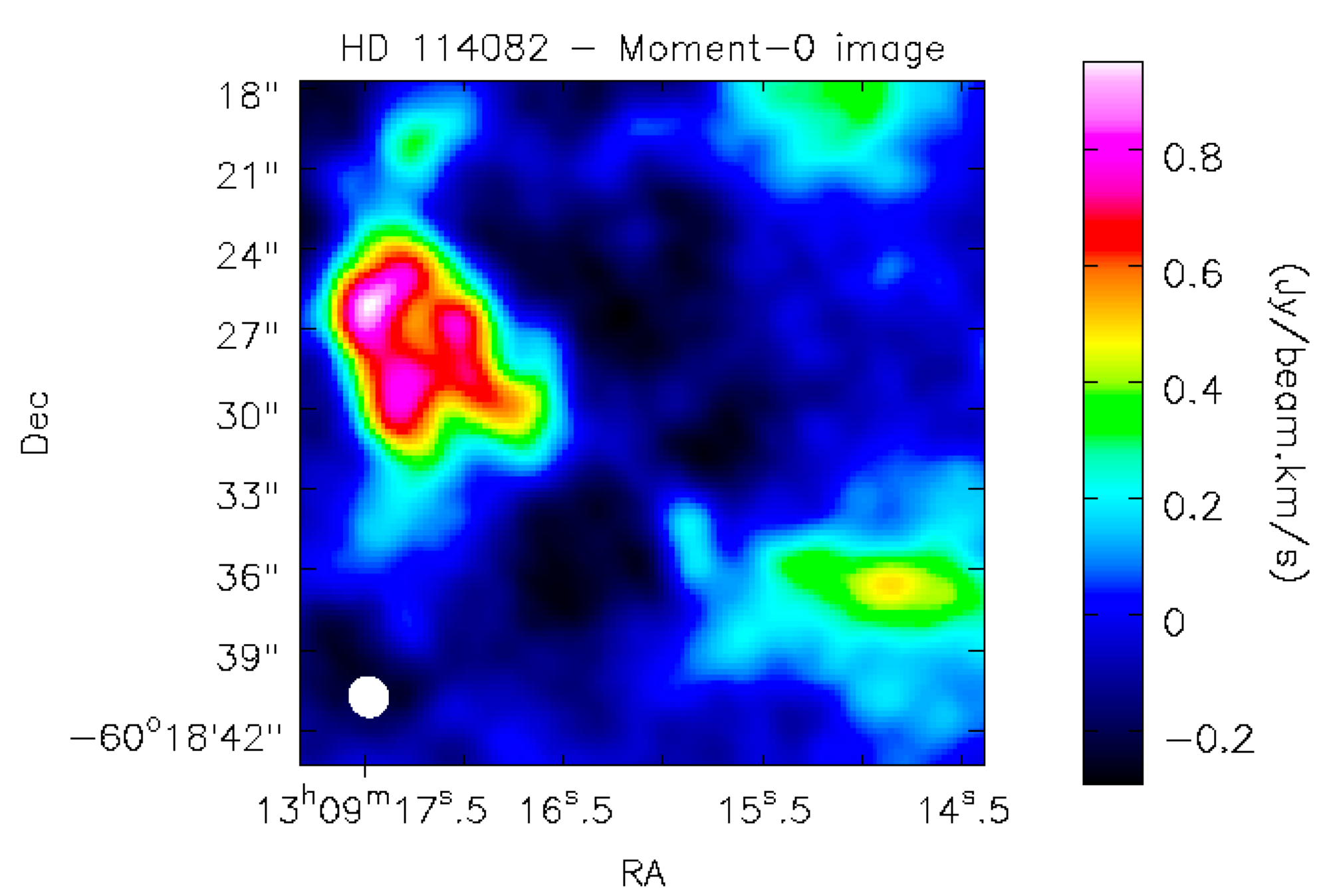}
   \caption{\label{figcocloud} CO emission around HD 114082 that might be due to a cloud along the line of sight. Moment-0 image integrated between -61 and -45 km/s.}
\end{figure}

\begin{figure}
   \centering
      \includegraphics[width=8cm]{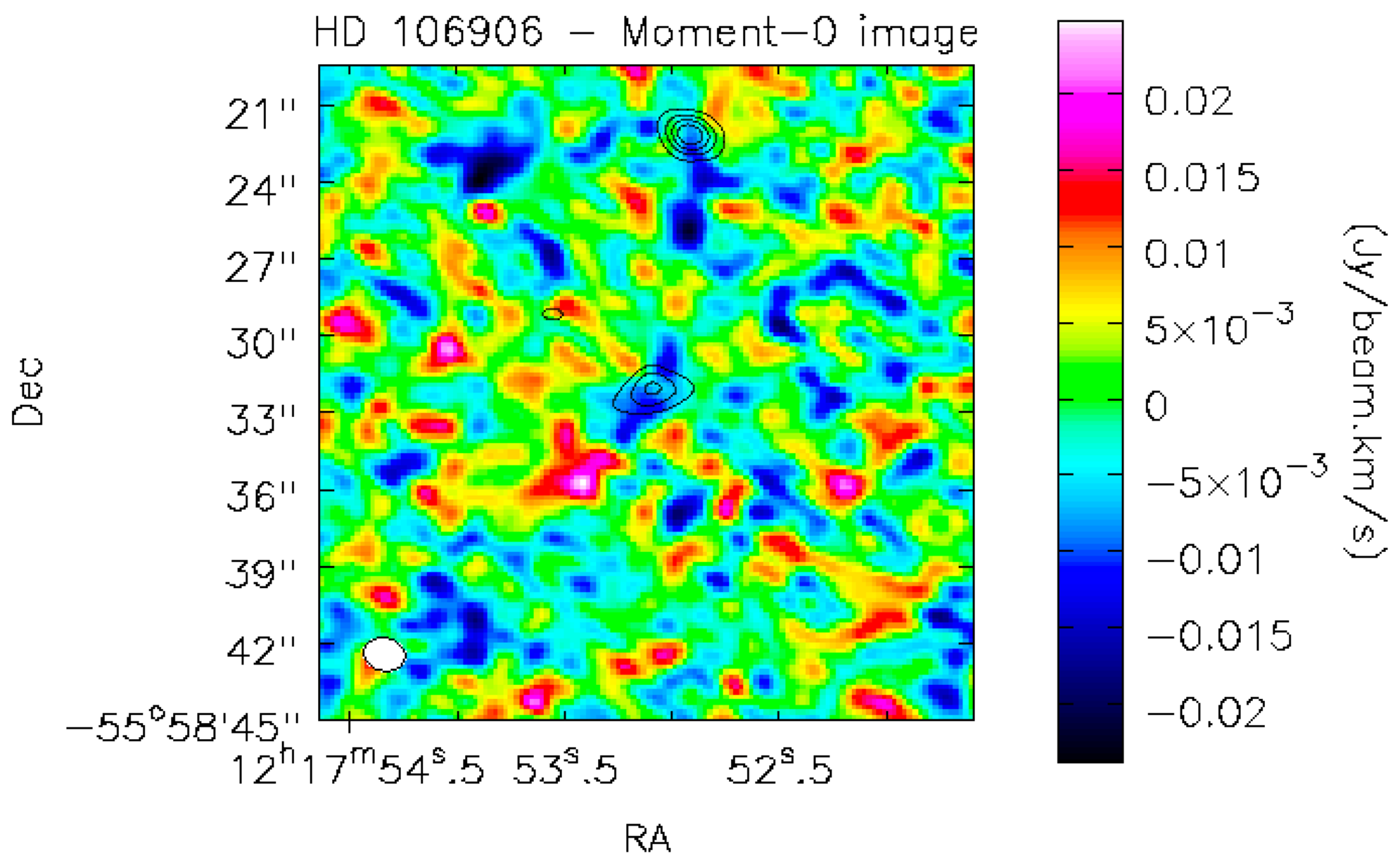}
   \caption{\label{fig106} Potential extra CO emission in HD 106906, not aligned with the dust disc and at larger distances (at the South-East) but with a spectrum showing a double-peaked profile.}
\end{figure}

In total, 4/10 systems yielded CO detections (2 of which are indeed circumstellar gas, see Sect.~\ref{gasfit}) and no CN was detected. To look for gas (both CO and CN), we analysed each cube and plotted their spectra assuming different spatial extents (starting with the continuum extent and exploring around that value) and averaging spatially. For HD 129590 and HD 121191, the CO detections are detected at $>4\sigma$ and colocated with the star (see Fig.~\ref{figcodetec}). For HD 114082, the CO gas is not colocated with the star but rather very extended and has a very different radial velocity (shifted by -60 km/s compared to the central star) and might be due to a cloud on the line of sight (see Fig.~\ref{figcocloud}). For HD 106906, the CO putative emission is at the South East of the dust disc a few arcsec away but it does not seem to belong to the circumstellar disc even though it is present over several channels and shows a double-peaked profile (see the moment-0 image in Fig.~\ref{fig106} and the discussion in Sec.~\ref{1069}). The CO would move radially with a velocity around 1 km/s, slightly different than the radial velocity of the system of $10\pm2$ km/s \citep{2006AstL...32..759G}. As for CN and non-detections in CO, we only provide 3$\sigma$ upper limits as listed in Table~\ref{tabobsgas}, where we assumed the same spatial extent as the continuum images \citep[except for HD 143675, which is not detected in the continuum and we assume it is extended over only one beam based on an SED fit whose temperature measurement enables determination of a true radius of about  50 au, i.e. well within one beam,][]{2017MNRAS.469..521K}.

\begin{table*}
  \centering
  \caption{Gas observations of our sample of 10 stars. From left to right, the columns correspond to the system's name, the beam size and its position angle for a given observation, the rms that is reached per channel and the total integrated flux in a 10 km/s channel (for CO detections we take the actual linewidth instead) on the source in the CO/CN moment-0 images. For non-detections, we give the 3$\sigma$ value and assume the same spatial extent as for the continuum image, i.e. it is more than one beam when resolved.}

  \label{tabobsgas}
  \begin{tabular}{|l|c|c|c|c|c|c|}
   \toprule
   Systems & Beam size  & Beam PA  & rms CO & S$_{12_{\rm CO}}$ & rms CN & S$_{{\rm CN}}$ \\ 
    & (\arcsec) & ($^\circ$)& (mJy/beam/channel) & (Jy km/s) & (mJy/beam/channel) & (Jy km/s)  \\ 

    \midrule
    \midrule
    HD 106906 & $1.67 \times 1.30$ & 75.9 & 1.18 & $<0.04$ & 1.1 & $<0.04$     \\
    HD 114082 & $1.53 \times 1.45$ & 34.1 & 1.55 & $<0.02$ & 1.4 & $<0.02$    \\
    HD 117214 & $1.70 \times 1.29$ & 62.6 & 1.60 & $<0.02$ & 1.5 & $<0.02$    \\
    HD 121191 & $0.93 \times 0.90$& -56.8 & 1.02 & $0.21 \pm 0.02$  & 0.9 & $<0.03$   \\
    HD 129590 & $1.61 \times 1.16$& 81.9 & 1.52 & $0.056 \pm 0.014$   & 1.3 & $<0.05$  \\ 
    HD 143675 & $0.93 \times 0.81$ & 73.6 & 0.88 & $<0.01$  & 0.8 & $<0.01$   \\   
    HD 15745 & $1.78 \times 1.22$ & 29.2 & 2.35 & $<0.1$   & 2.3 & $<0.1$  \\
    HD 191089 & $1.61 \times 1.11$ & 73.7 & 0.98 & $<0.05$  & 0.9 & $<0.05$   \\
    HD 69830 & $0.97 \times 0.77$ & 63.9 & 0.93 & $<0.01$ & 0.85 & $<0.01$    \\
    HR 4796A & $1.76 \times 1.29$ & -88.6 & 1.6 & $<0.1$ & 1.5 & $<0.1$    \\

   \bottomrule
  \end{tabular}

\end{table*}

The 1$\sigma$ noise level near the disc in the moment-0 images are 6 and 7 mJy$\,$km/s/beam for HD 129590 and HD 121191, respectively, giving peak S/N in the images of 9 and 30. The total integrated fluxes are respectively 0.056$\pm$0.014 and 0.21$\pm$0.022 Jy$\,$km/s, which were measured by integrating over a region that encompasses all disc emission. The quoted errors take into account the noise in the images and flux calibration uncertainties that were added in quadrature. 

\section{Modelling Results}\label{model}

\begin{table*}
  \centering
  \caption{Results of a 2-D Gaussian fit to continuum images. From left to right, the columns correspond to the system's name, the fitted size (major axis $\times$ minor axis of the 2-D Gaussian, which is deconvolved from the beam when resolved and convolved otherwise), disc (or beam if unresolved) position angle and the last column gives information on whether the disc was resolved in the observations. }

  \label{tabresgausscon}
  \begin{tabular}{|l|c|c|c|c|}
   \toprule
   Systems & Fitted size  & Disc/Beam PA  & Resolved?\\ 
    & (\arcsec) & ($^\circ$) & (yes/no/marginally) \\ 

    \midrule
    \midrule
    HD 106906 & $2.41\pm0.23 \times 1.30\pm0.07$ & $102.8 \pm 3.3$ &  yes     \\
    HD 114082 & $1.62\pm0.06 \times 1.38\pm0.05$ & $93 \pm 8$ &  marginally    \\
    HD 117214 & $1.69\pm0.07 \times 1.43\pm0.05$ & $161 \pm 58$ &  marginally    \\
    HD 121191 & $1.30\pm0.1 \times 1.13\pm0.07$ & $30 \pm 19$  &  yes   \\
    HD 129590 & $1.87\pm0.05 \times 1.22\pm0.02$ & $101.2 \pm 1.5$  &  yes    \\
    HD 143675$^{*}$ & $1.4\pm0.6 \times 0.9\pm0.3$ & $168 \pm 25$ &  no  \\
    HD 15745 & $2.69\pm0.21 \times 1.43\pm0.06$ & $38.9 \pm 5.6$  &  yes  \\
    HD 191089 & $2.36\pm0.05 \times 1.36\pm0.02$ & $72.8 \pm 1.4$  &  yes    \\
    HD 69830 & $1.29\pm0.33 \times 0.72\pm0.1$ & $15 \pm 9$  &  no   \\
    HR 4796A & $2.59\pm0.03 \times 1.69\pm0.01$ & $25.5 \pm 0.8$ & yes  \\

    \bottomrule
   \multicolumn{5}{l}{$^{*}$\footnotesize{We note that this disc in not considered as detected as the S/N is only of about 2-$\sigma$.}} \\
  \end{tabular}
   \footnotetext{Footnote}

\end{table*}

\subsection{Fit of the continuum data}\label{dustfitt}
We first fit our data with a simple 2-D Gaussian model to check whether the discs are resolved and what their PAs are. We then use a radiative transfer code and fit a Gaussian ring model to the data to find the physical parameters that best fit the data.

\subsubsection{Fit with a simple 2-D Gaussian model}
We use the CASA function \texttt{imfit} to find the best fit parameters of a 2-D Gaussian to our continuum images. The function returns the best-fit beam major and minor axes as well as the best disc PA (when resolved or beam PA otherwise) and the integrated flux. We list the results in Table~\ref{tabresgausscon} (and verified that all integrated fluxes are consistent with values listed in Table~\ref{tabobscon}). We see that all our targets are resolved in the continuum except for HD 69830 and HD 143675 (but the latter is only a $2 \sigma$ detection). However, we find that HD 114082 and HD 117214 are only marginally resolved (the derived FWHM of the Gaussians are consistent with the beam size).

\subsubsection{Fit with an MCMC radiative transfer model}

We now use an axisymmetric dust model to fit the continuum observations to place constraints on the geometry of these discs as well as their total dust masses. We parameterise a disc as being a ring centered at a radius $R_0$ with a Gaussian radial profile of full width at half maximum (FWHM) $W_{\rm ring}$. The vertical profile (in the $Z$ direction) of the dust disc is also assumed to be Gaussian with an aspect ratio $h=H/R$ ($H$ being the scaleheight) such that the dust density distribution is given by

\begin{equation}\label{dustmodel}
 \rho_d(R,Z)=\rho_0 \exp\left(-\frac{(R-R_0)^2}{2\sigma_d^2}\right) \exp\left(-\frac{Z^2}{2H^2} \right),
\end{equation}

\noindent where $\rho_0$ is the density at $R_0$ in the midplane and $\sigma_d=W_{\rm ring}/(2 \sqrt{2 \log 2})$.

The RADMC-3D code \citep{2012ascl.soft02015D} is then used to compute images for a given dust model at 1.27 mm to be compared to observations \citep[following the same procedure as in][]{2018arXiv181108439K}. To make the images, we assume that the grains are produced from a collisional cascade, which implies a size distribution with a power law index close to -3.5 as predicted from theory and models \citep{1969JGR....74.2531D,2006A&A...455..509K,2007A&A...472..169T}. We assume a maximum grain size of 1 cm (as larger grains will not significantly contribute to the flux observed in band 6). As for the minimum size, we use the blowout size \citep[derived from][]{1979Icar...40....1B}, which is the size below which grains become unbound and leave the system on a dynamical timescale\footnote{We note that there could be a copious amount of unbound grains in the most luminous belts in our sample \citep[with L$_{\rm IR}/$L$_\star$ close to $5 \times 10^{-3}$, see][]{2019TBO} but the contribution of these small submicron-sized grains is negligible in the mm.}.
For the composition of the grains, we assume astrosilicates with a density of 2.7 g$\,$cm$^{-3}$ \citep{2003ApJ...598.1017D} and use a mass-weighted mean opacity of $\kappa \sim 1.3$ cm$^2$g$^{-1}$ at 1.27 mm that is computed using the Mie theory code from \citet{1983asls.book.....B}. The composition of the grains we chose is arbitrary and the mass is given mainly for comparative purposes but it should be understood that it has large systematic uncertainties.

We use a Bayesian Markov Chain Monte Carlo (MCMC) approach to find the best fits to our free parameters that are: $R_0, W_{\rm ring}, h, {\rm inclination}\,(i)$, PA and $M_{\rm dust}$, where the latter is the total dust mass up to 1 cm bodies. To account for astrometric uncertainties in the ALMA data on the star position, we allow an offset in RA (offset $x$) and Dec (offset $y$). We fit the dirty images (i.e. not cleaned) directly, which is closest to fitting the visibilities and allow for a more accurate derivation of the error bars.
For the MCMC method to converge fast and accurately, we use the emcee module \citep[see][for the details of the method]{2010CAMCS...5...65G,2013PASP..125..306F}. We assume uniform priors and the posterior distributions of our parameters are given by the product of the prior distribution function and the likelihood function assumed to be $\propto \exp(-\chi^2/2)$, with $\chi^2= \sum_{\rm pixels} (F_{\rm obs}-F_{\rm mod})^2/\sigma_F^2$, where $\sigma_F$ is the variance of the data taking into account the number of independent beams in the image. $F_{\rm obs}$ and $F_{\rm mod}$ are the observed and model (which we convolve with the dirty beam) fluxes in a given pixel and the $\chi^2$ is computed over an area much larger than where the disc produces emission, including side-lobes from the dirty beam down to the noise level. 
The MCMC simulations were run with 100 walkers and for 1000 steps after the burn-in period.

\begin{figure*}
   \centering
   \includegraphics[width=13cm]{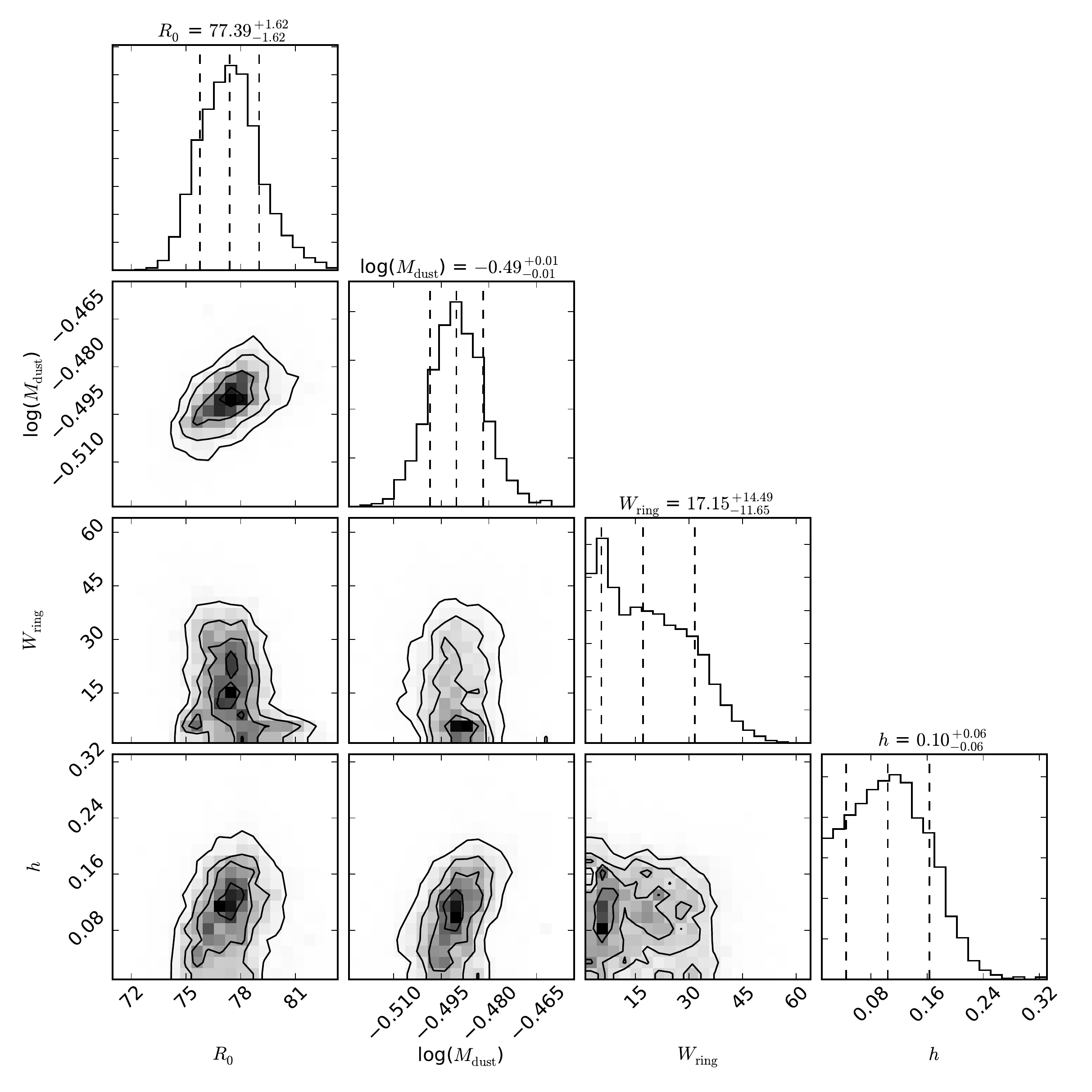}
   \caption{\label{figcorner} Corner plot from our MCMC fitting procedure of the ALMA continuum image observed for HR 4796A. $R_0$ is the centre of a Gaussian ring of FWHM equal to $W_{\rm ring}$ with an aspect ratio equal to h. $M_{\rm dust}$ corresponds to the total dust mass up to 1 cm bodies. The three vertical dashed lines are the median and the 16$^{\rm th}$ and 84$^{\rm th}$ percentiles of the marginalised distributions.}
\end{figure*}

\begin{table*}
  \centering
  \caption{Table describing the best-fit parameters for the continuum images of our sample of 10 stars using an MCMC method (see Sec.~\ref{dustfitt}). We list the median $\pm$ uncertainties, which are based on the 16$^{\rm th}$ and 84$^{\rm th}$ percentiles of the marginalised distributions.}
  \begin{threeparttable}

  \label{tabbestfitcont}
  \begin{tabular}{|l|c|c|c|c|c|c|c|c|}
   \toprule
   Systems & $R$  & W$_{\rm ring}$  & M$_{\rm dust}$ & $h$ & $i$ & PA & RA offset & Dec offset\\ 
    & (au) & (au) & (M$_\oplus$) & & ($^\circ$) &  ($^\circ$) & (\arcsec) & (\arcsec) \\ 

    \midrule
    \midrule
    HD 106906 & $85.4^{+12.4}_{-13.1}$ & $<101$ & $0.054^{+0.007}_{-0.006}$ & $<0.5$ & $>59$ & $112.1^{+7.1}_{-6.9}$ & $-0.04^{+0.07}_{-0.07}$ & $-0.12^{+0.05}_{-0.05}$    \\    
    HD 114082 & $24.1^{+11.0}_{-8.8}$ & $<41$ & $0.07^{+0.01}_{-0.02}$ & - & - & - & $0.1^{+0.04}_{-0.04}$ & $-0.16^{+0.03}_{-0.03}$    \\    
    HD 117214 & $32.0^{+8.9}_{-9.1}$ & $<41$ & $0.1^{+0.02}_{-0.02}$ & - & - & - & $0.1^{+0.04}_{-0.03}$ & $-0.19^{+0.03}_{-0.03}$    \\
    HD 121191$^b$ & $52.1^{+10.2}_{-10.5}$ & $<61$ & $0.1^{+0.01}_{-0.01}$ & - & - & - & $0.38^{+0.03}_{-0.03}$ & $0.14^{+0.03}_{-0.03}$    \\
    HD 129590 & $74.2^{+6.1}_{-5.9}$ & $<75$ & $0.39^{+0.02}_{-0.02}$ & $<0.33$ & $>65$ & $117.0^{+4.5}_{-4.5}$ & $0.11^{+0.02}_{-0.02}$ & $-0.18^{+0.02}_{-0.01}$    \\
    HD 143675 & - & - & $<0.08$ & - & - & - & - & -    \\    
    HD 15745$^b$ & $72.0^{+6.6}_{-6.4}$ & $<82$ & $0.10^{+0.01}_{-0.01}$ & $<0.55$ & $>52$ & $38.8^{+5.4}_{-5.2}$ & $-0.02^{+0.05}_{-0.05}$ & $-0.59^{+0.06}_{-0.05}$    \\
    HD 191089 & $43.4^{+2.8}_{-2.9}$ & $<45$ & $0.06^{+0.004}_{-0.004}$ & $<0.52$ & $>52$ & $72.7^{+4.1}_{-3.9}$ & $0.11^{+0.04}_{-0.04}$ & $-0.07^{+0.03}_{-0.02}$    \\
    HD 69830$^a$ & $<12$ & $<12$ & $<0.00011$ & - & - & - & $0.28^{+0.18}_{-0.34}$ & $0.14^{+0.36}_{-0.28}$    \\
    HR 4796A & $77.4^{+1.6}_{-1.6}$ & $<50$ & $0.32^{+0.01}_{-0.01}$ & $<0.26$ & $>72$ & $25.9^{+1.2}_{-1.2}$ & $0.12^{+0.01}_{-0.01}$ & $-0.09^{+0.02}_{-0.01}$    \\

   \bottomrule
  \end{tabular}
\begin{tablenotes}
      \small 
      \item $^a$ The fit was carried out without subtracting the stellar contribution so that it is a conservative upper limit.
      \item $^b$ The astrometric errors on the star position for these two targets are slightly larger than expected from ALMA astrometric accuracy. Besides phase stability effects, these errors could be due to several reasons according to the ALMA helpdesk. For instance, HD 15745 was at 29$^\circ$ elevation on the sky due to its very Northern declination and moreover the phase calibrator QSO J0237+2848 was 8$^\circ$ away from the source, which may have impacted the phase transfer. We also note that part of the asymmetry may be real and could be due to these dust discs being asymmetric.
    \end{tablenotes}
  \end{threeparttable}
\end{table*}

In Fig.~\ref{figcorner}, we show the result of the MCMC simulation for the system HR 4796A for the 4 most important parameters and the derived best-fit parameters are listed in Table~\ref{tabbestfitcont} for all systems.
We analyse the results for HR 4796A and compare to the study by \citet{2018MNRAS.475.4924K} to familiarise the reader with what we extract from the ALMA data for all systems. Our MCMC simulation converges toward a Gaussian profile centered at $R_0=77.4\pm1.6$ au with $W_{\rm ring}<50$ au and an aspect ratio $h<0.26$. There is no lower limit on the width of the belt due to the limited resolution of our observations ($\sim 1.5$ arcsec). The upper limits in this paper are given at the 99.7\% level.
This is consistent with the results by \citet{2018MNRAS.475.4924K}, who did a similar analysis with ALMA data at higher resolution in band 7. Indeed, they find $R_0=78.6\pm 0.2$ au with a FWHM $10 \pm 1$ au and an aspect ratio of $0.04 \pm 0.01$. We also find good agreement for the inclination $>72^\circ$ and position angle $25.9^\circ \pm 1.2$ (to be compared to $76.6^\circ \pm0.2$ and $26.7^\circ \pm 0.1$, respectively). Our results for the total mass of bodies up to 1 cm are also quite similar as we find $0.32 \pm 0.01$ M$_\oplus$ to be compared to $0.35 \pm 0.04$ M$_\oplus$. However, we note that the final mass depends on the assumptions concerning the stellar luminosity (which affects the dust temperature) and also on the composition of the bodies, which were slightly different between the two studies. Finally, the offsets in RA and Dec are also consistent with the ALMA astrometric uncertainties (i.e. a maximum accuracy of one tenth of the beam size, see ALMA technical handbook\footnote{\url{https://almascience.eso.org/documents-and-tools/cycle6/alma-technical-handbook}}).

Not all observations provide as many constraints on the free parameters of our model. We have 3 different cases. Either the continuum is not detected (HD 143675), or it is detected but not resolved (HD 69830), or it is detected and resolved (e.g. HR 4796A), which is the case for the 8 other sources (though HD 114082 and HD 117214 are only marginally resolved). For the non-detection around HD 143675, we only derive an upper limit on the mass of the system. When it is unresolved, we cannot constrain the inclination and position angle of the system. Some systems are marginally resolved, which leads to a variety of other cases, with all of the results summarised in Table~\ref{tabbestfitcont} and where unconstrained parameters are indicated with ``-''.

\subsection{Fit of the gas data}\label{gasfit}

\subsubsection{Fit with a simple 2-D Gaussian model}
We have started by fitting the moment-0 images for the two detections (HD 129590 and HD 121191) with a 2-D Gaussian following the same procedure as for the dust. We find that the disc around HD 121191 is not spatially resolved and that around HD 129590 is at best marginally resolved. Conversely, HD121191's mm-dust disc was resolved, implying a smaller radial extent in CO compared to the dust (which we will now investigate further).
In the following subsection we model these images using radiative transfer models to extract spatial information for data at a range of velocities and also because it includes the other less than $2\sigma$ pixels, which implicitly place upper limits on the emission further out.

\subsubsection{Fit with an MCMC radiative transfer model}

We now fit the gas cubes we have for our ten targets. The method we use to fit the gas data is similar to what is described in the previous section~\ref{dustfitt} for the continuum images and was already used in \citet{2018arXiv181108439K} to fit the [CI] fine structure line at 609.14$\mu$m in HD 131835. What differs from the previous section is that we now fit a whole cube using the above method on each frequency channel (which traces the different radial velocities of the gas) of the data cube. As described in Sec.~\ref{gasobs}, we subtract the continuum from the cube. We then produce dirty cubes (i.e. without cleaning the images) around the CO line and use RADMC-3D to produce CO model cubes to compare to our observations in an MCMC fashion. For the $\chi^2$ fit, we account for the correlation between adjacent channels (using 2.667 as the number of channels per effective spectral resolution element\footnote{as described in the ALMA spectral response document that follows: \url{https://safe.nrao.edu/wiki/pub/Main/ALMAWindowFunctions/Note_on_Spectral_Response.pdf}}) and the number of independent beams in the image. For the detections, we run the simulations with 100 walkers and for $\sim$10,000 steps (which takes about 4 weeks per target running on 40 cpus).

To interpret our results, we first assume that CO is in local thermal equilibrium (LTE), which is likely a good approximation for CO in most systems that can be detected with ALMA in band 6 at $>50$ pc as explained in detail in \citet{2017MNRAS.469..521K}. For completeness, we will also provide simplified non-LTE results at the end of this section to give a plausible range of masses corresponding to our observations. For massive gas discs, CO lines can be optically thick, which RADMC-3D takes into account. We note however that our non-LTE results assume optically thin emission.

As the two circumstellar gas detections (see Fig.~\ref{figcodetec}) do not show any obvious asymmetries, we restrict ourselves to an axisymmetric gas model. In a similar way as the dust continuum fitting procedure, we assume a ring with a Gaussian radial profile and the gas surface density follows

\begin{equation}
   \Sigma(R) = \Sigma_0  \exp\left(-\frac{(R-R_0)^2}{2\sigma_g^2}\right) ,     
\end{equation}

\noindent where $\Sigma_0$ is the surface density at $R_0$ and with $\sigma_g=\Delta R/(2 \sqrt{2 \log 2})$, where $\Delta R$ is the width (FWHM) of the gas disc. We then assume that the temperature follows a double power law profile \citep[motivated from][]{2016MNRAS.461..845K} defined by

\begin{equation}\label{Teq}
   T(R) =
    \begin{cases}
      T_0 \left(\frac{R}{R_0}\right)^{-\beta_1^t} & \text{for $R<R_0$}\\
      T_0 \left(\frac{R}{R_0}\right)^{-\beta_2^t} & \text{for $R>R_0$}\\
    \end{cases}    
    .   
\end{equation}

We then set the scaleheight of the disc from $T(R)$ such that $H=c_s/\Omega$, where $\Omega$ is the orbital frequency and $c_s=\sqrt{kT/(\mu m_H)}$ is the sound speed. As explained in detail in \citet{2017MNRAS.469..521K, 2018arXiv181108439K}, the mean molecular mass $\mu$ in secondary discs can be much higher than in protoplanetary discs as most of the gas is in the form of carbon, oxygen or CO, rather than H$_2$. To be consistent with previous studies, we assume $\mu=14$. From preliminary test runs, we found that $\beta_1^t$ and $\beta_2^t$ were not constrained and we fixed them to 0 \citep[motivated from the thermodynamical model in][]{2016MNRAS.461..845K} for exploring the parameter space of interest faster.

\begin{figure*}
   \centering
   \includegraphics[width=18cm]{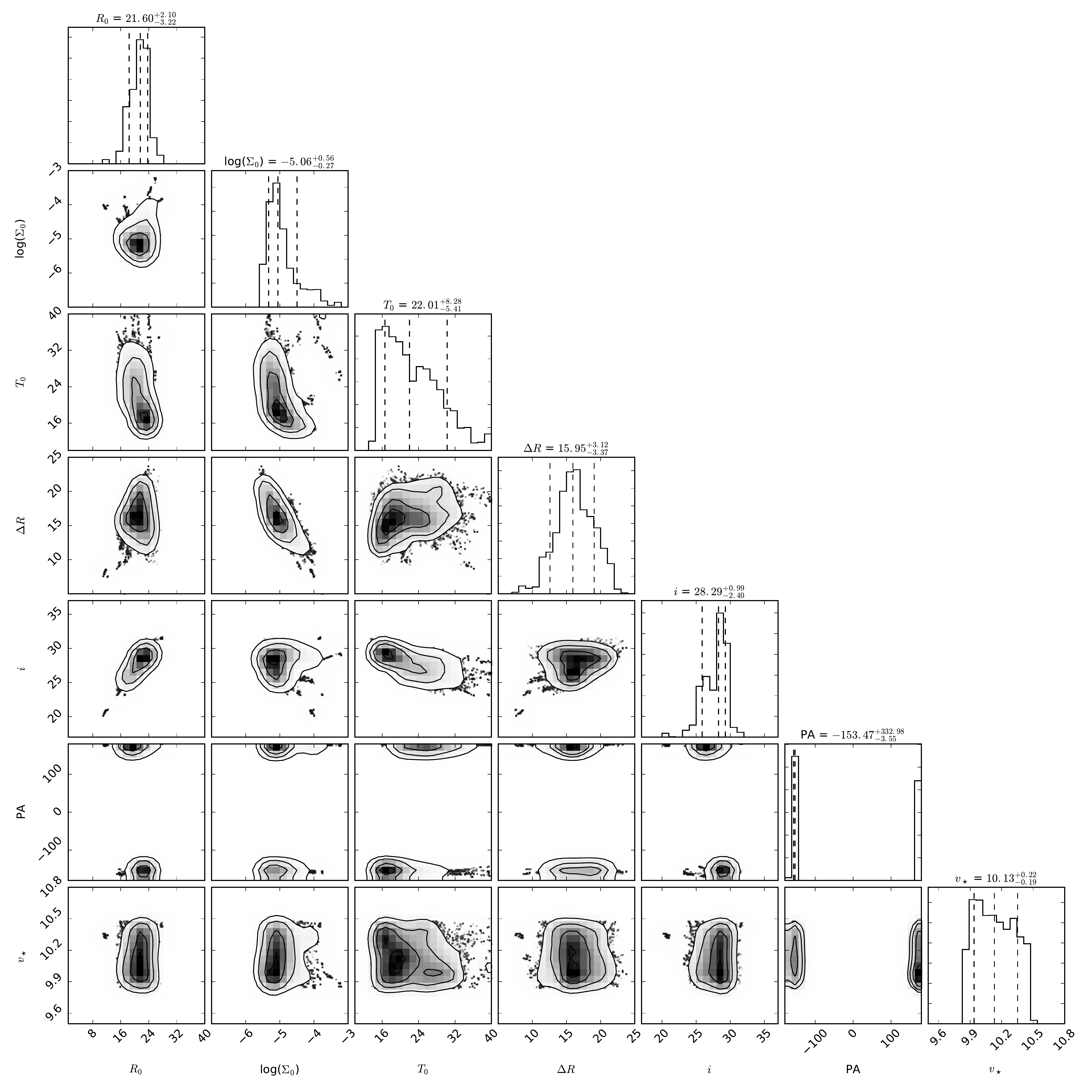}
   \caption{\label{figcornergas} Corner plot from our MCMC fitting procedure with a Gaussian ring model of the ALMA image cube observed for HD 121191. $R_0$ is the centre of a Gaussian ring of FWHM equal to $\Delta R$ whose surface density at $R_0$ equals $\Sigma_0$. The temperature is defined as being constant and equal to $T_0$. The disc's inclination, position angle and stellar velocity are $i$, PA, and $v_\star$ respectively. The three vertical dashed lines are the median and the 16$^{\rm th}$ and 84$^{\rm th}$ percentiles of the marginalised distributions.}
\end{figure*}


\begin{figure*}
   \centering
   \includegraphics[width=8cm]{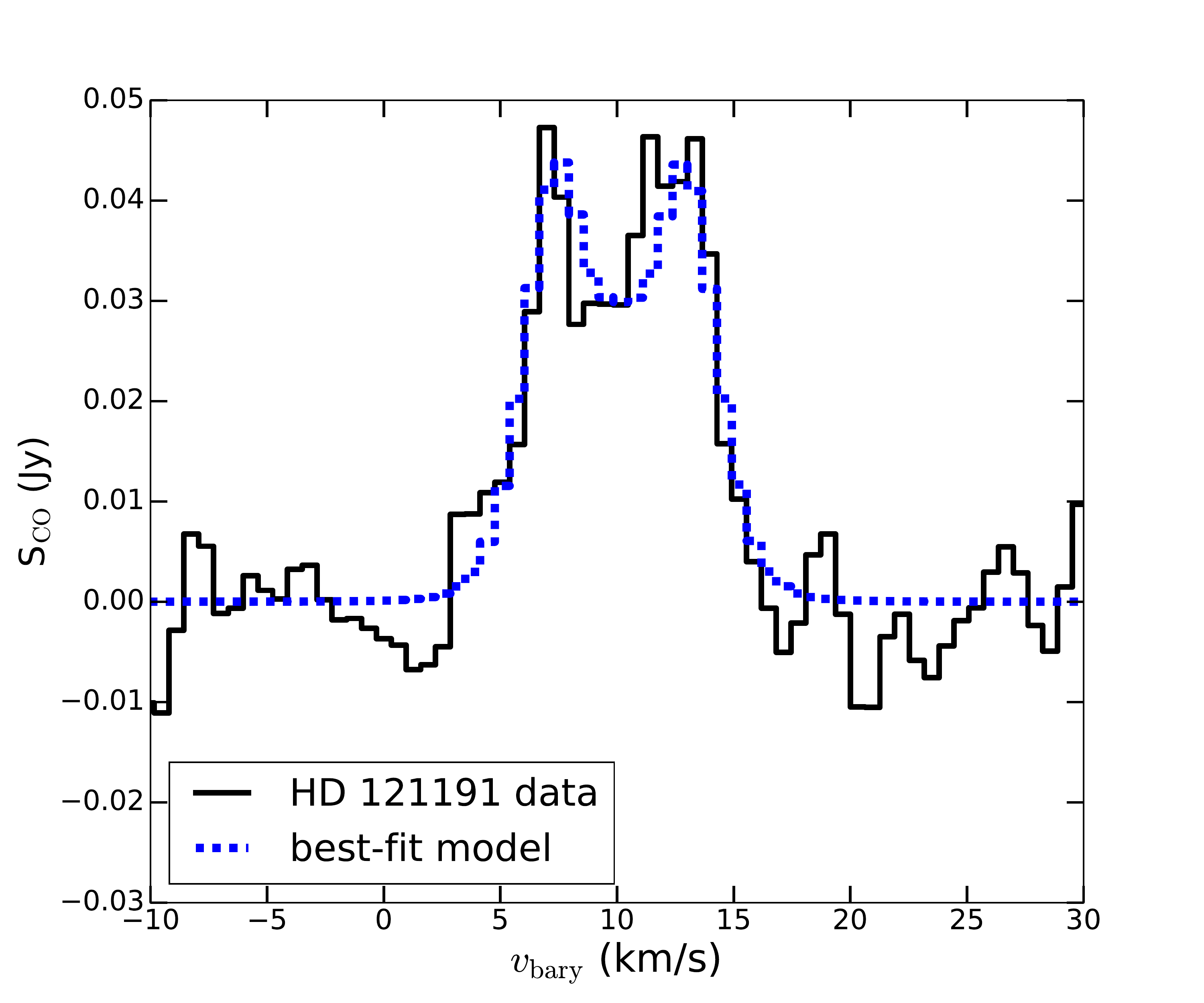}
      \includegraphics[width=8cm]{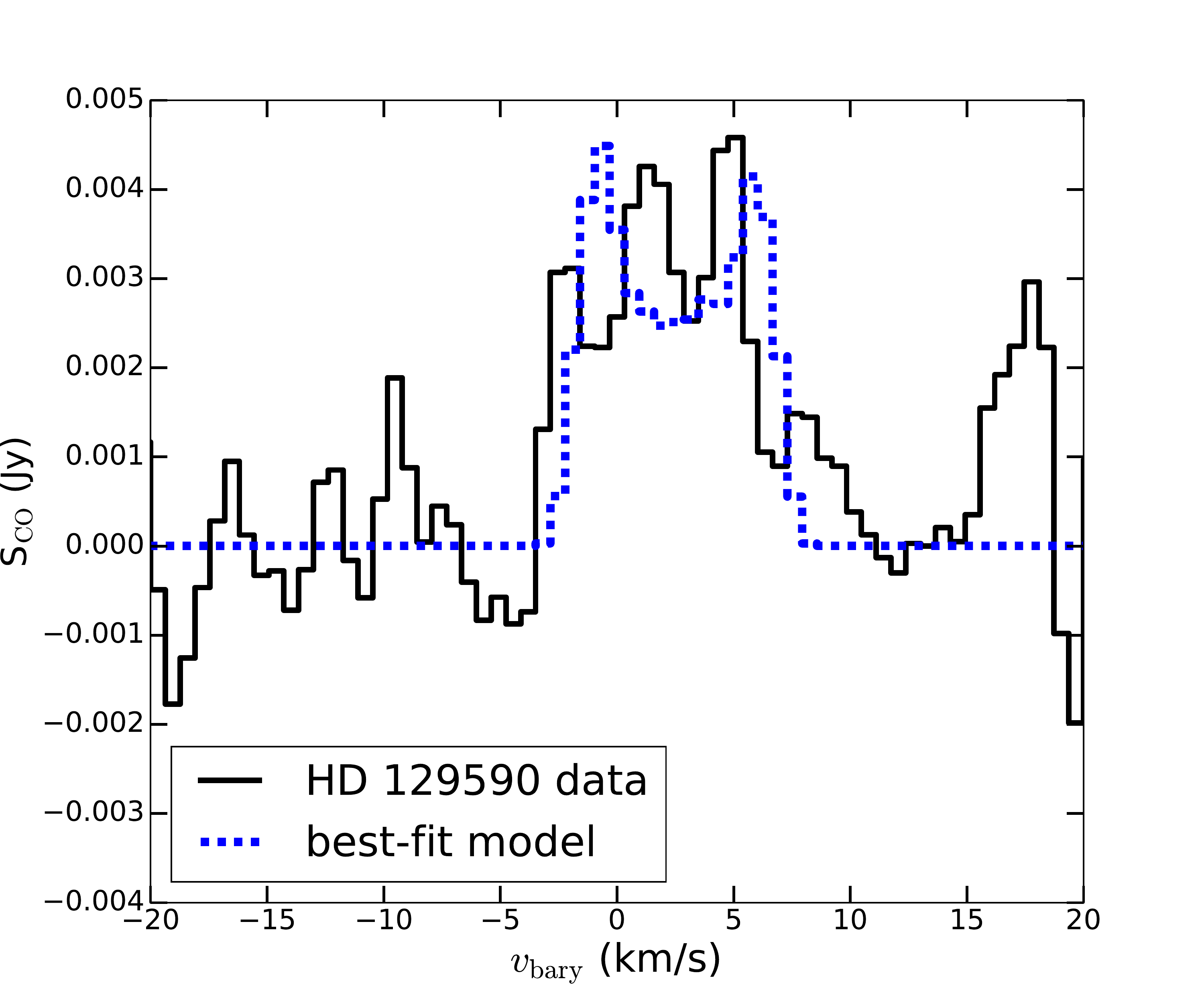}
   \caption{\label{figspectrum} Observed spectra of HD 121191 (left) and HD 129590 (right) in black along with our best-fit model for the gas disc around HD 121191 in dashed blue and a fit for HD 129590 to estimate its linewidth.}
\end{figure*}


We also fit for the radial velocity of the star $v_\star$ to verify that indeed the gas is co-moving with the star when detected. The free parameters of this gas model are thus $R_0$, $\Delta R$, $\Sigma_0$, $T_0$, $v_\star$, and (similar to the dust model) $i$, PA and the offsets in RA and Dec. We show an example of the results of our fitting procedure for the case of HD 121191 in Fig.~\ref{figcornergas}. We list the results for the 10 systems, i.e., the best-fit parameters for each system, in Table~\ref{tabbestfitgas}.

We compare our results on the CO detection in HD 121191 to \citet{2017ApJ...849..123M}. They find a total CO flux of $0.23\pm0.04$ Jy km/s, where we have $0.21\pm0.02$ Jy km/s. They find a radial velocity for the star equal to $9\pm1$ km/s (in barycentric coordinates), where we find $10.1\pm0.25$ km/s, consistent with their value. The total CO mass they find (from fitting the $^{13}$CO line) is equal to $2.7\pm0.9 \times 10^{-3}$ M$_\oplus$, which is in agreement with the value derived from our MCMC simulations equal to $2.3^{+2.3}_{-0.9} \times 10^{-3}$ M$_\oplus$. We also derive the geometrical and thermal properties of this disc for the first time. We find that the gas disc can be modelled with a Gaussian of centre $R_0 = 22^{+2}_{-3}$ au, with width $\Delta R = 16\pm4$ au, with a surface density equal to $\Sigma_0=0.87^{+2.3}_{-0.6} \times 10^{-5}$ kg/m$^2$ at $R_0$. The inclination and position angle found from our MCMC simulations are $i=28^{+1}_{-3}$ deg and $160<$PA$<240$ deg, respectively. Finally, we constrain the temperature to be $T_0=22^{+9}_{-6}$ K. 

In Fig.~\ref{figspectrum} (left), we show the 1-D spectrum for HD 121191 along with the best-fit we find. The large width of the line (extending from 6 to 14 km/s, i.e. a disc with a radial velocity of $\sim$ 4 km/s) is indeed best-fit with a gas belt closer-in (at $\sim$ 20 au) than the dust belt (at $\sim$ 52 au) owing to the Keplerian velocity at 20 au which is equal to 8.4 km/s (i.e. a radial velocity of 4 km/s for a 28 deg inclination). We checked that indeed a belt further out would not provide such a wide line and would not as well fit the 1-D spectrum.

For HD 129590, the signal-to-noise is equal to 4 in the moment-0 image and our MCMC fit using individual channels (with less than $3\sigma$ signals) gives no constraints on the physical parameters of the disc. However, we find a total flux of $0.056\pm0.014$ Jy km/s, i.e. roughly a factor 4 smaller than for HD 121191. The radial velocity of the star (from fitting the spectrum in Fig.~\ref{figspectrum}) is $2.3^{+1.3}_{-0.6}$ km/s, which is consistent with that found from previous studies \citep[$2.3\pm1.3$ km/s, see][]{2007AN....328..889K}. From the spectrum we also evaluate the width of the line to be 12 km/s across, which places the position of the inner part of the disc at $\sim$30 au \citep[assuming an inclination of 75 deg as observed in scattered light,][]{2017ApJ...843L..12M}. From our 2D-Gaussian fitting of the moment-0 image, we also find that the disc should be smaller than 82 au (if not it would be resolved).



\begin{table*}
  \centering
  \caption{Best-fit parameters for the gas images of our sample of 10 stars using an MCMC method (see Sec.~\ref{gasfit}). We list the median $\pm$ uncertainties, which are based on the 16$^{\rm th}$ and 84$^{\rm th}$ percentiles of the marginalised distributions. For HD 129590, we list results from non-LTE calculations (see text).}

  \label{tabbestfitgas}
  \begin{tabular}{|l|c|c|c|c|c|c|c|c|c|c|c|c|}
   \toprule
   Systems & $R$  & $\Delta R$  & $\Sigma_0 $ & $T_0$ & $v_\star$ & $i$ & PA  \\ 
    & (au) & (au) & (kg/m$^2$) & (K) & (km/s) & ($^\circ$) &  ($^\circ$) \\ 

    \midrule
    \midrule
    HD 106906 & - & - & $<4\times10^{-8}$ & - & - & - & -  \\    
    HD 114082 & - & - & $<2\times10^{-7}$ & - & - & - & -   \\    
    HD 117214 & - & - & $<1.6\times10^{-7}$ & - & - & - & -   \\       
    HD 121191 & $22^{+2}_{-3}$ & $16\pm4$ & $0.87^{+2.3}_{-0.6} \times 10^{-5}$ & $22^{+9}_{-6}$  & $10.1\pm0.25$ & $28^{+1}_{-3}$ & $-155\pm5$   \\
    HD 129590 & $30-82$ & $<52$ & $9\times10^{-7} - 6\times10^{-6}$ & - & $2.3^{+1.3}_{-0.6}$ & - & -   \\
    HD 143675 & - & - & $<1.5\times10^{-7}$ & - & - & - & -  \\        
    HD 15745 & - & - & $<8\times10^{-8}$ & - & - & - & -   \\    
    HD 191089 & - & - & $<6\times10^{-8}$ & - & - & - & -   \\    
    HD 69830 & - & - & $<1\times10^{-6}$ & - & - & - & -   \\    
    HR 4796A & - & - & $<1.5\times10^{-7}$ &- & - & - & -  \\    

   \bottomrule
  \end{tabular}

\end{table*}

\begin{figure*}
   \centering
\begin{tabular}{cc}
\includegraphics[width=7cm]{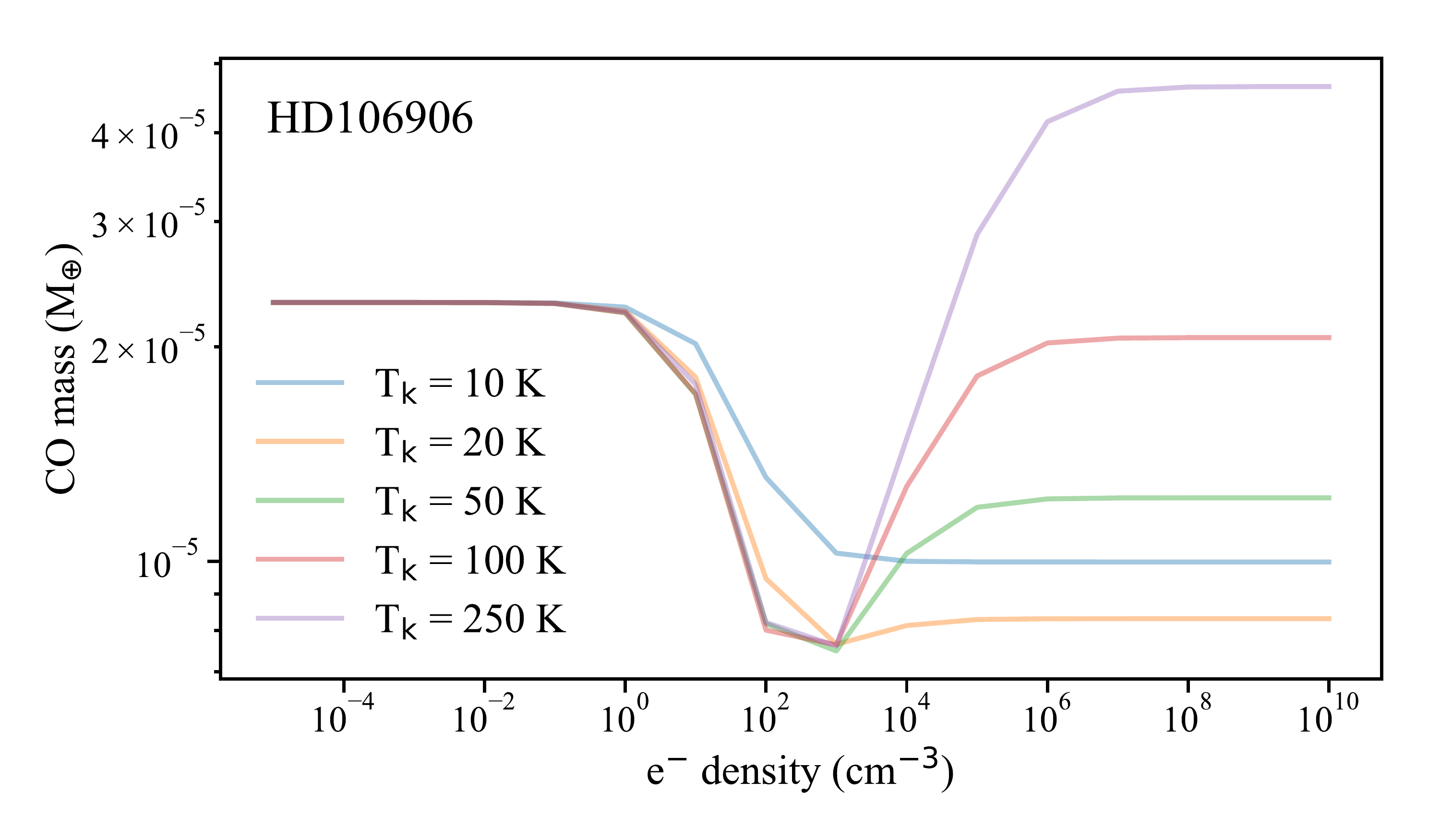}&
\includegraphics[width=7cm]{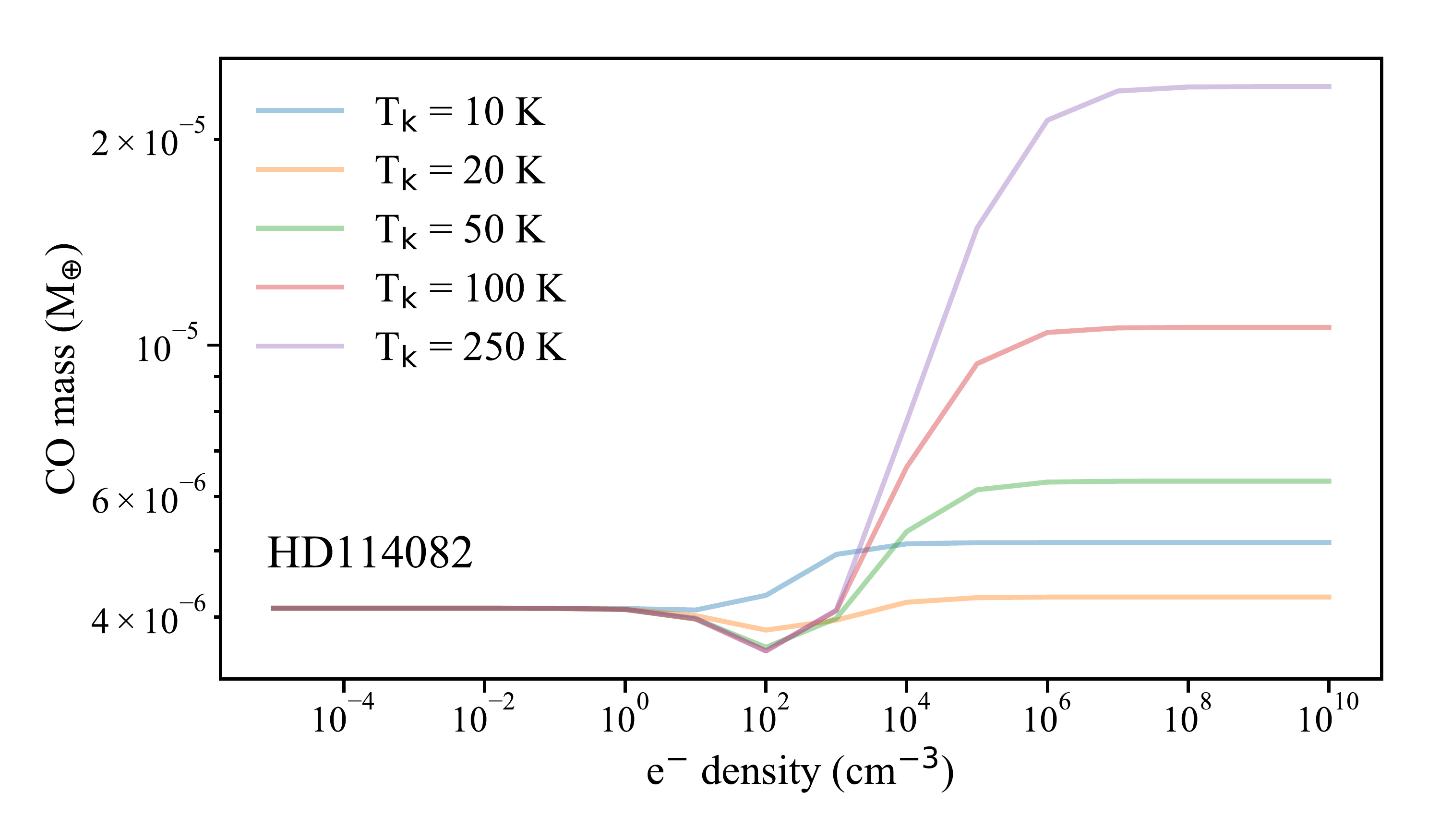}\\

\includegraphics[width=7cm]{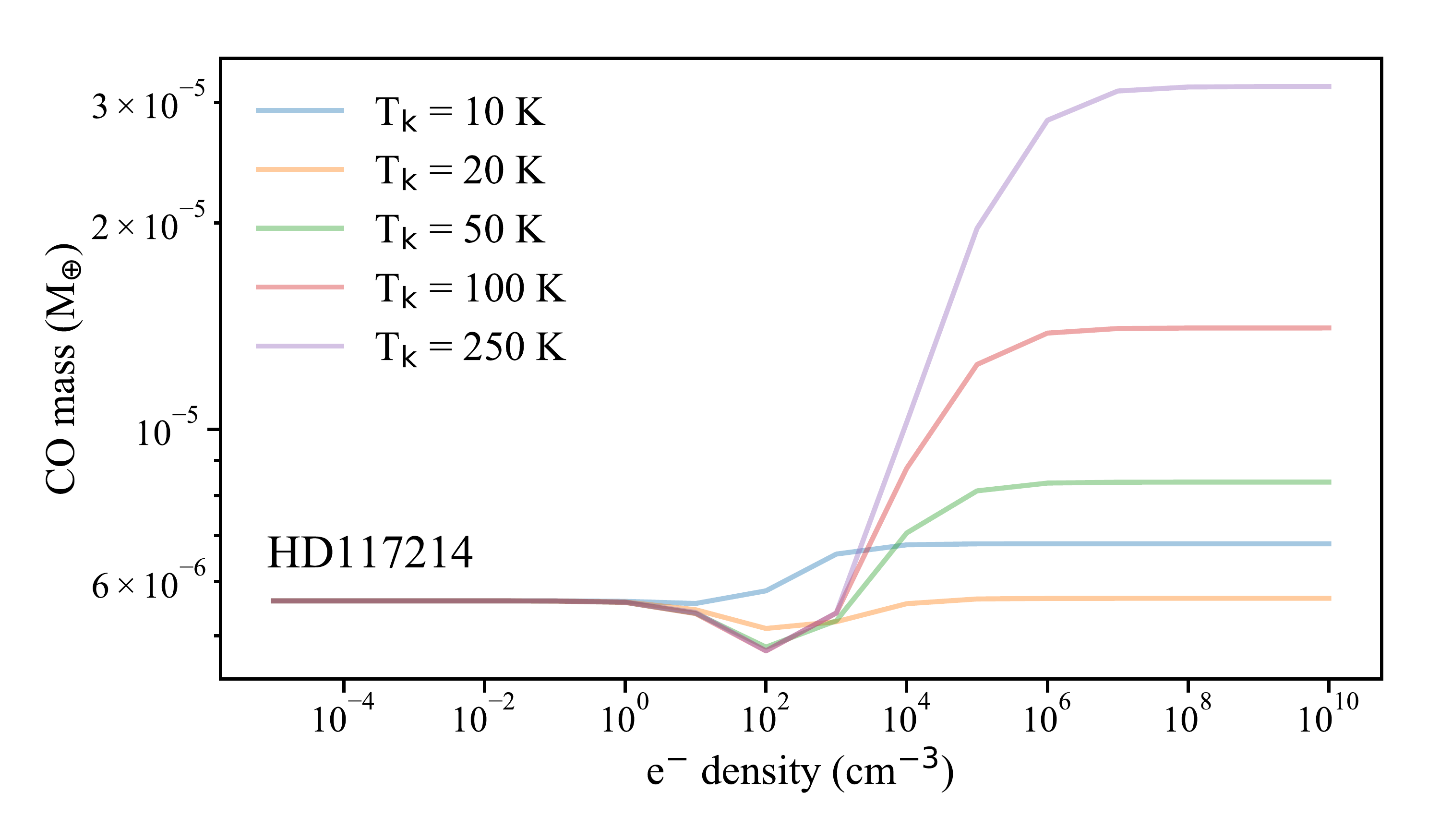}&
\includegraphics[width=7cm]{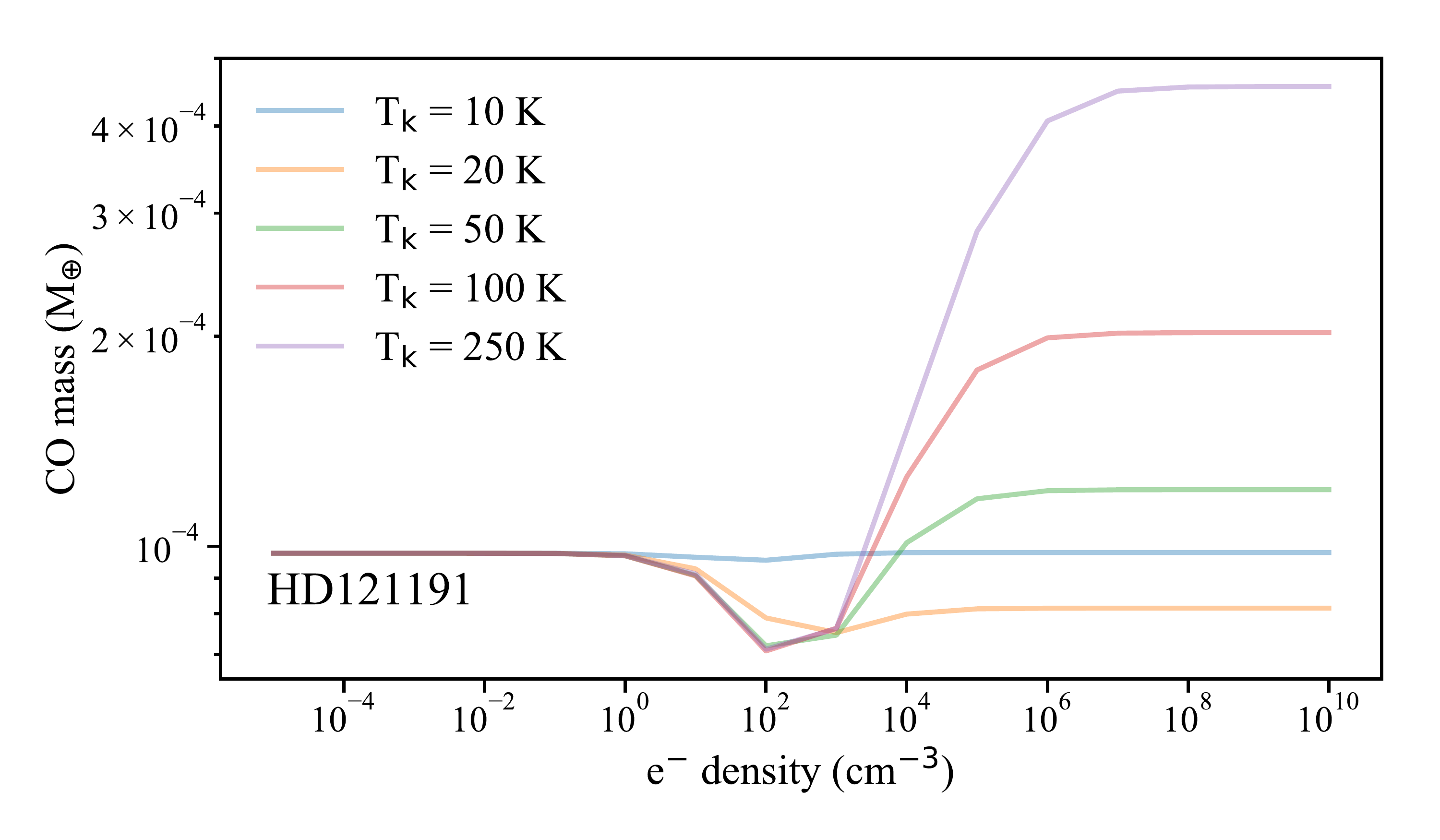}\\

\includegraphics[width=7cm]{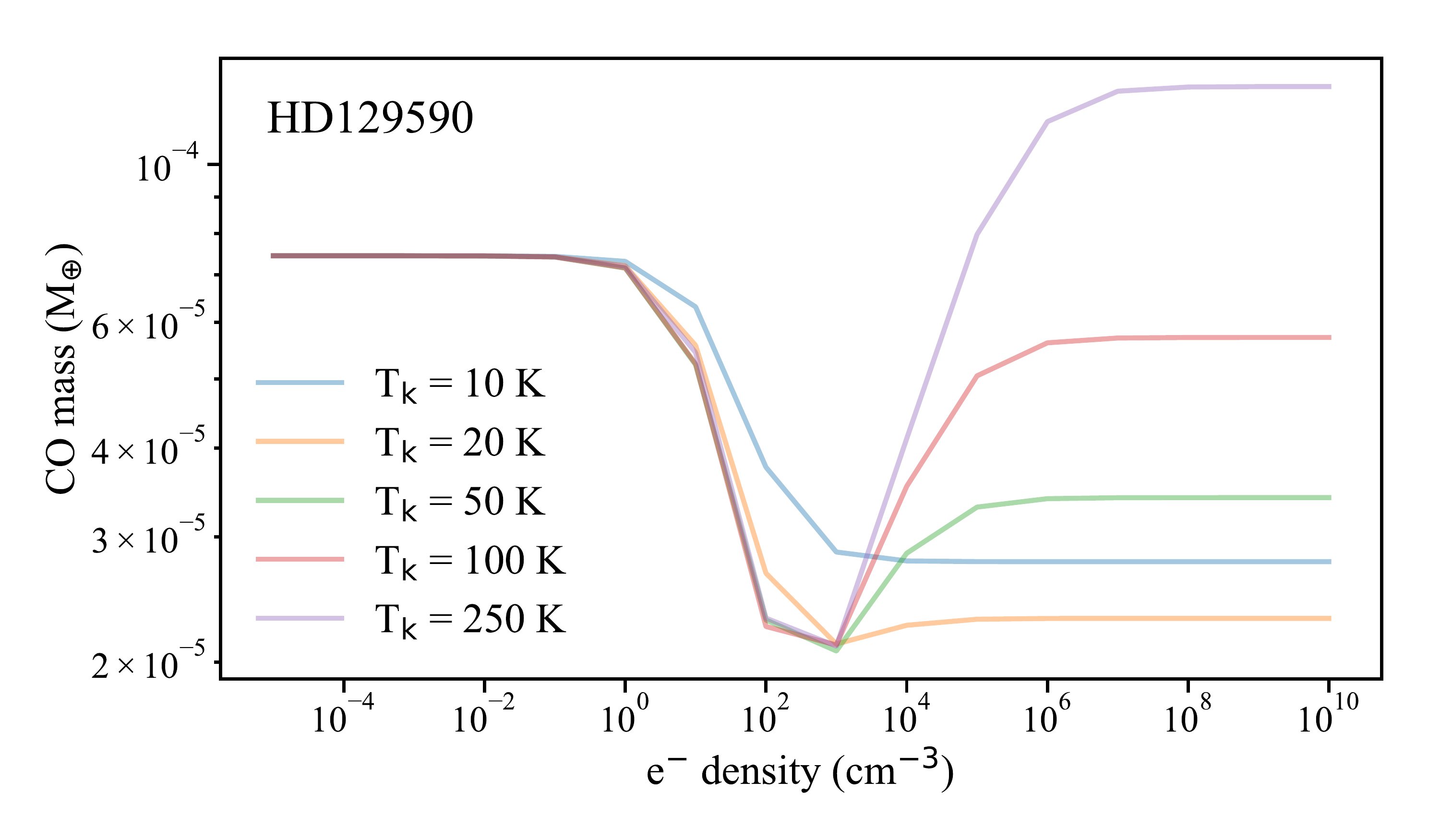}&
\includegraphics[width=7cm]{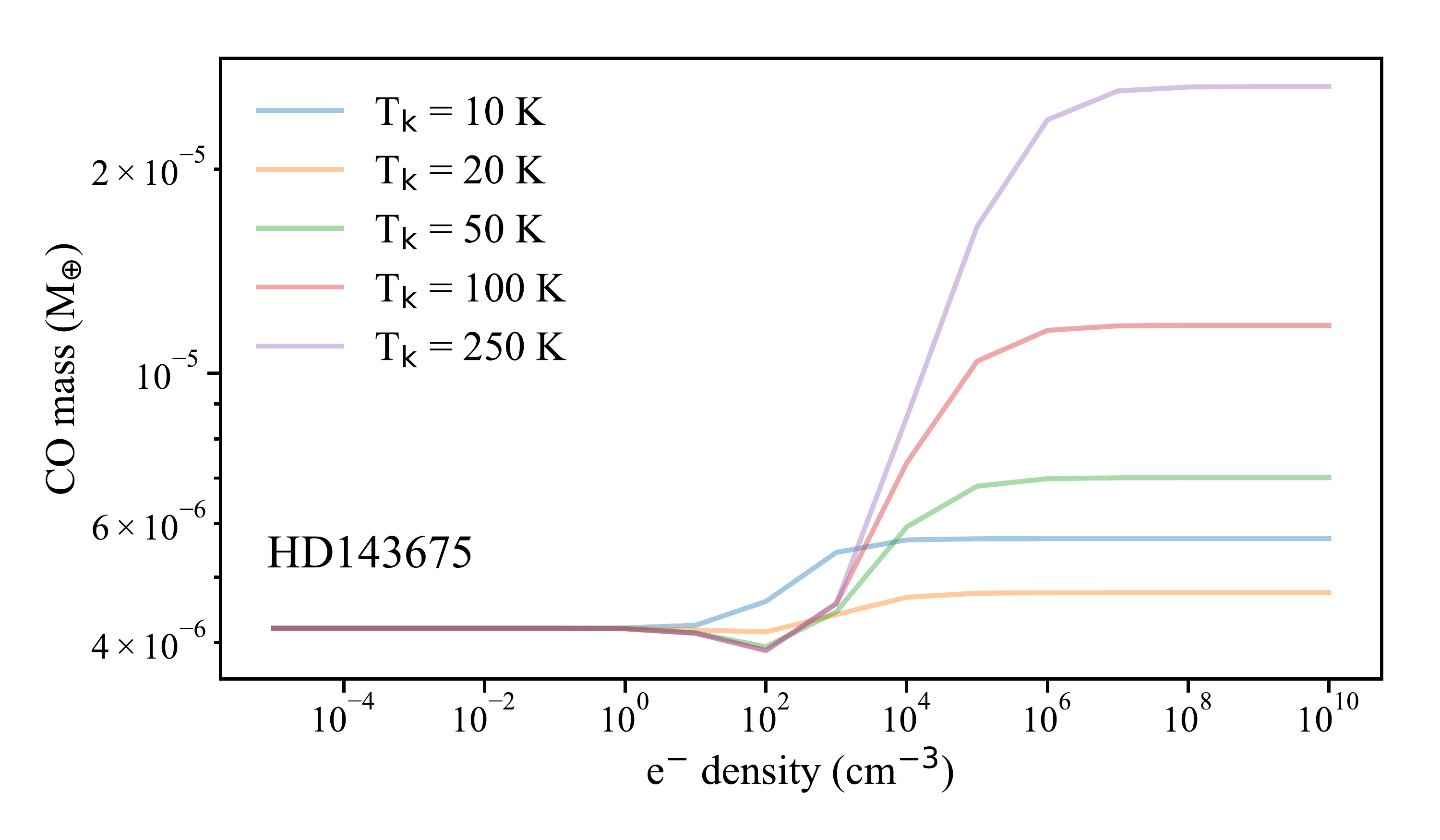}\\

\includegraphics[width=7cm]{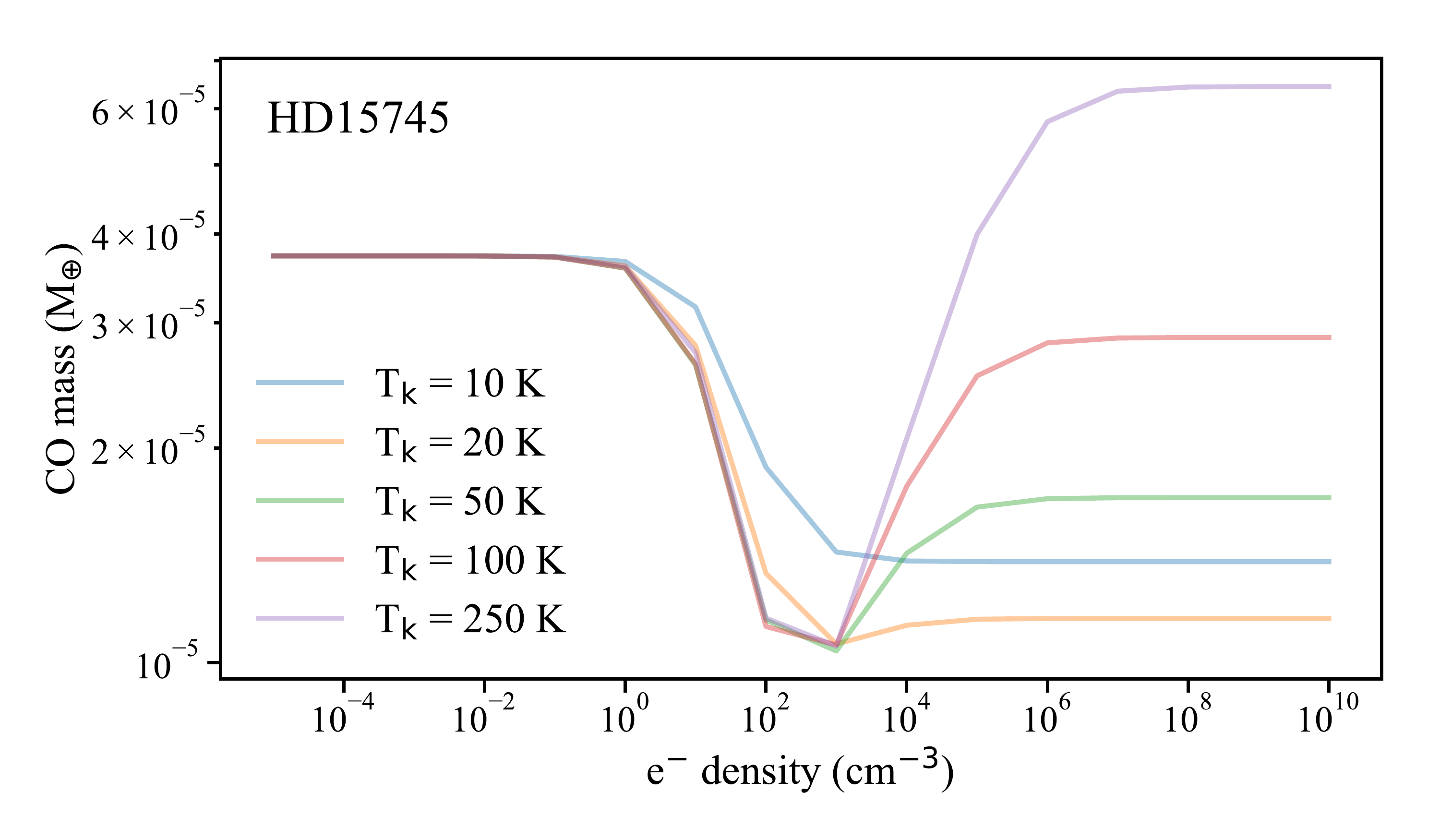}&
\includegraphics[width=7cm]{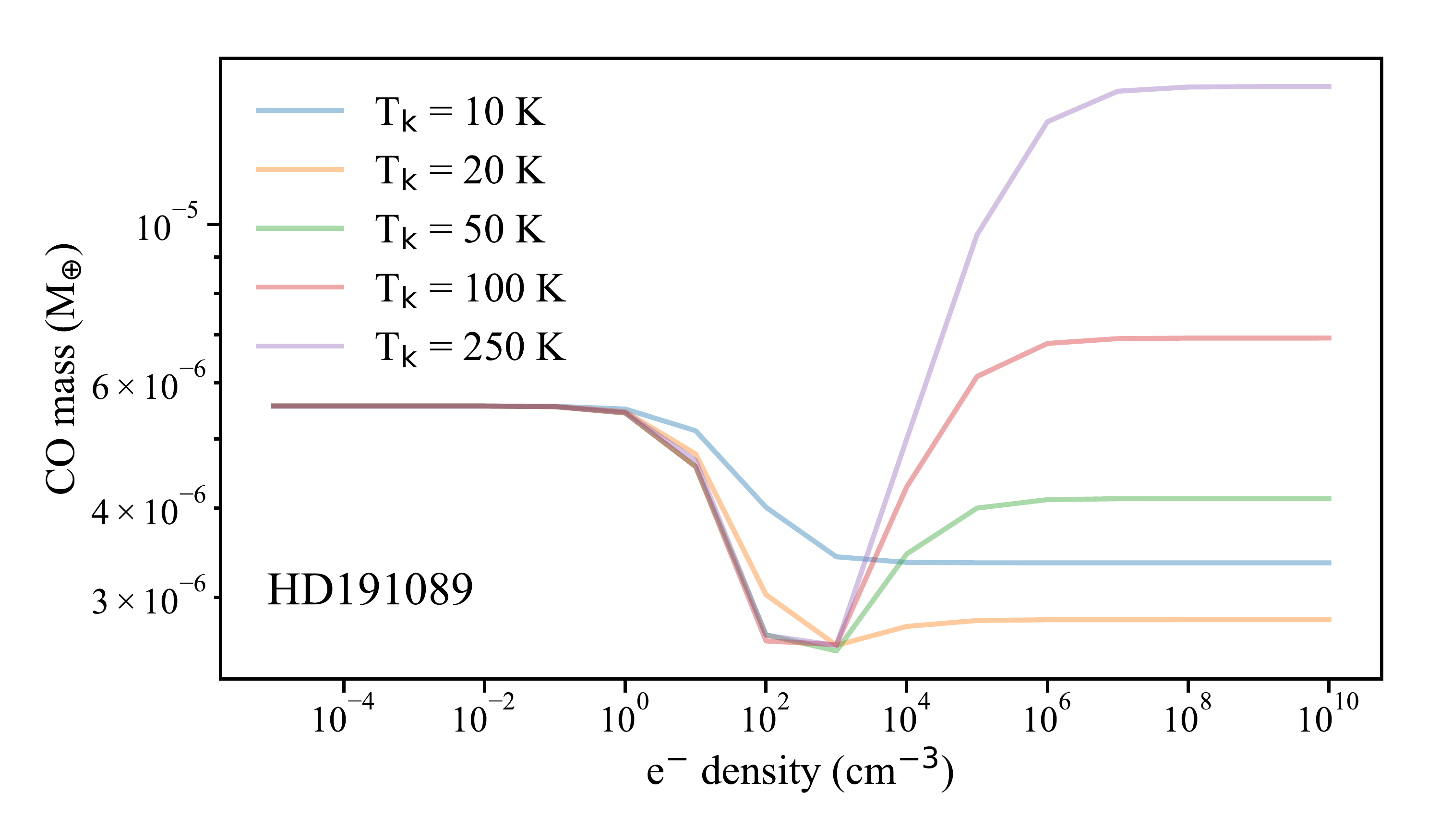}\\

\includegraphics[width=7cm]{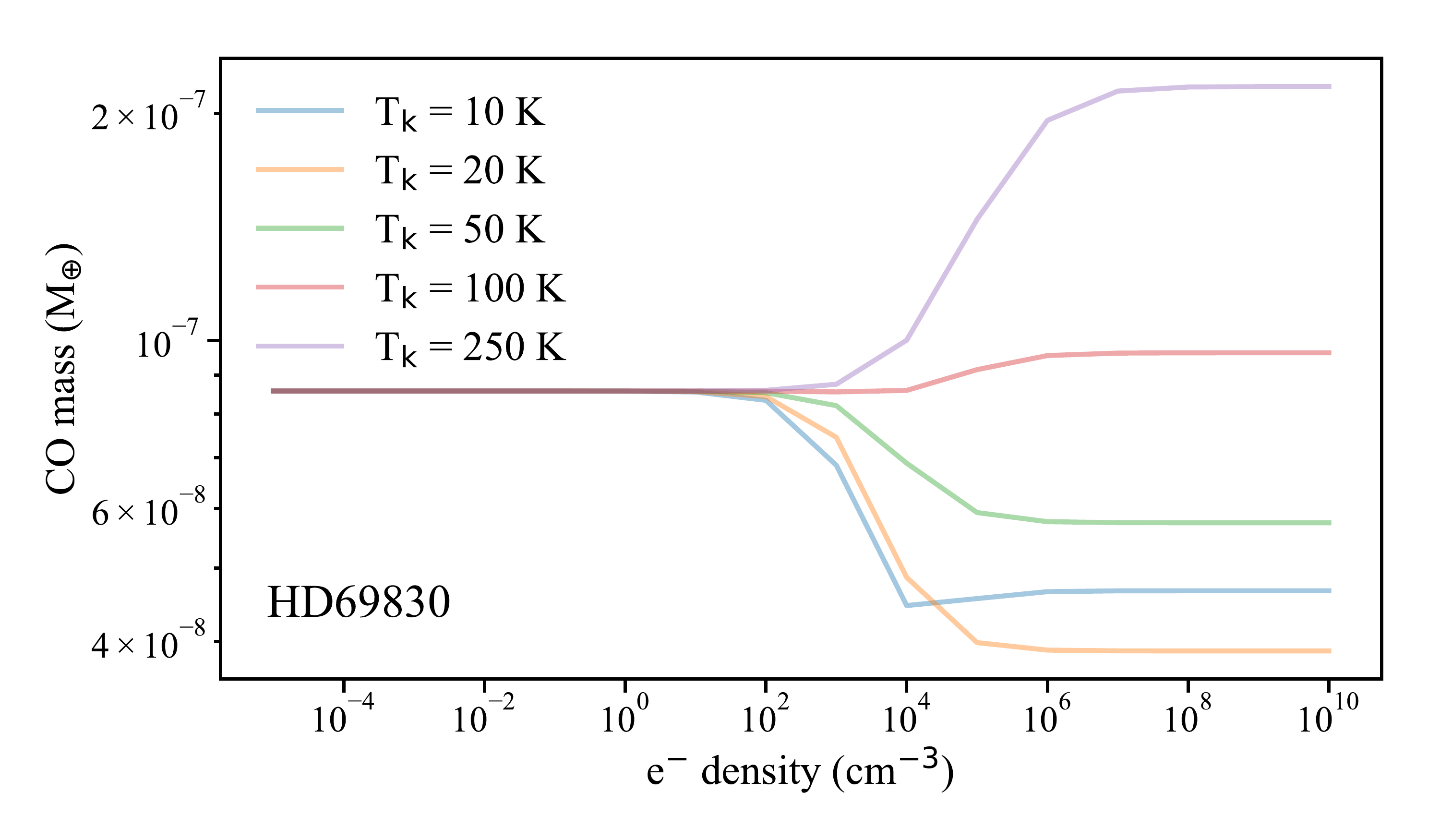}&
\includegraphics[width=7cm]{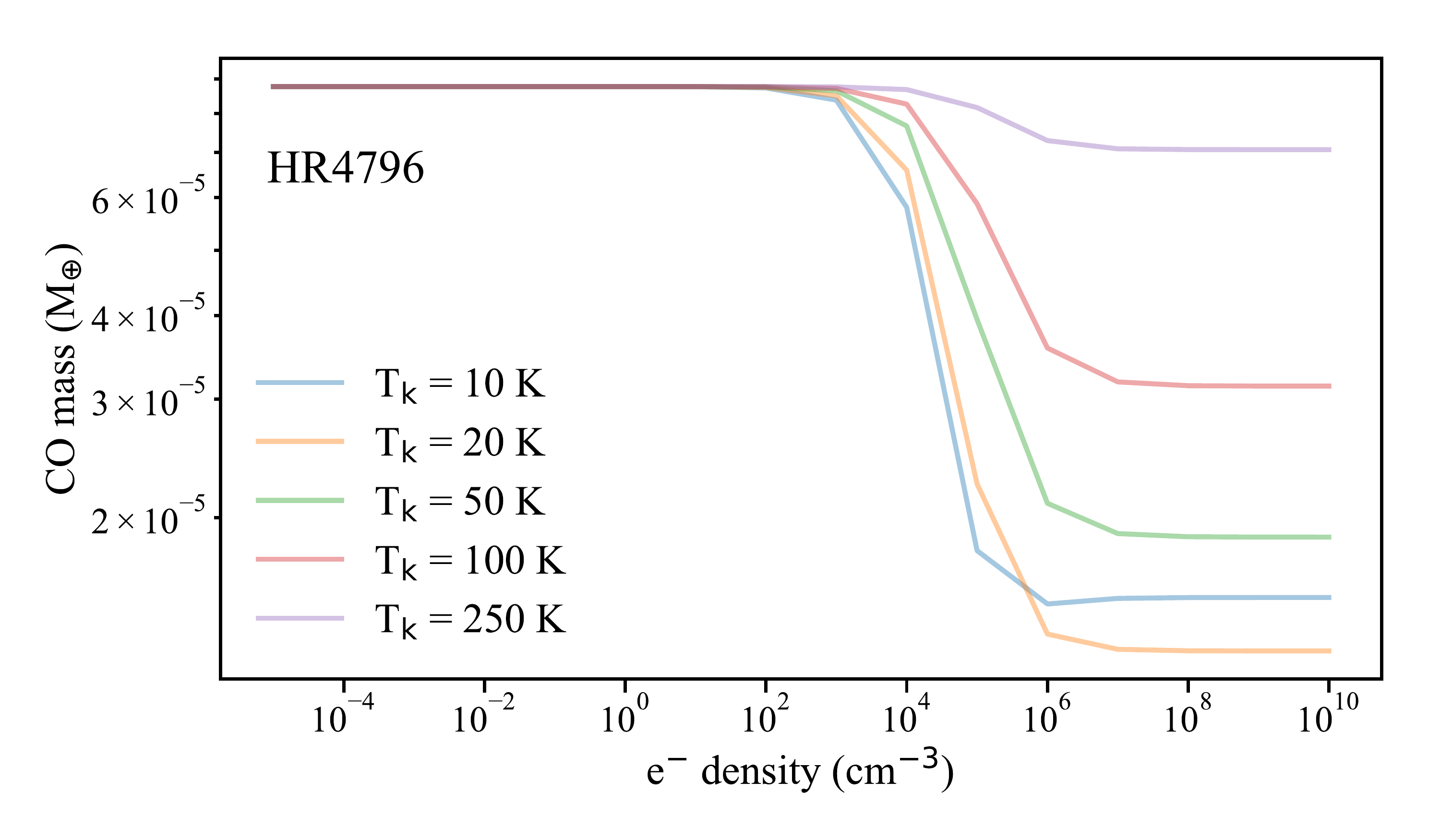}

\end{tabular}

    \caption{CO mass Vs. electron density for a range of kinetic temperatures (from 10 to 250 K). The masses are estimated from fluxes shown in Table~\ref{tabobsgas} using a non-LTE code \citep[including fluorescent excitation,][]{2015MNRAS.447.3936M,2018ApJ...853..147M}.}
    \label{fignlte} 
\end{figure*}

We also derive mass estimates from non-LTE calculations assuming the CO line is optically thin \citep[and including fluorescent excitation,][]{2018ApJ...853..147M} from the upper limits as well as for the detections. The CO/electron collision rates where taken from \citet{1975JPhB....8.2846D} and for CN/electron, we use \citet{1971JChPh..55.4342A} and the LAMDA database \citep{2005A&A...432..369S}. We show the results from the non-LTE calculations in Fig.~\ref{fignlte}, where it can be seen that the derived masses depend on the level of excitation of the gas and its temperature \citep[except in the radiation regime when the collider density is small,][]{2015MNRAS.447.3936M}. In these gas discs the main colliders are electrons \citep{2016MNRAS.461..845K} and it can be seen on the right hand side of plots in Fig.~\ref{fignlte} that when the electron density goes beyond a certain level (critical collider density), LTE is reached and the obtained masses only depend on the kinetic temperature (which we vary from 10 to 250 K). To account for our lack of knowledge of the temperature and electron density in each system, we assume wide ranges for both and then give a mass range estimate computed as being the minimum and maximum values reached on these plots and we list the derived non-LTE mass ranges (or upper limits for non-detections) for each target in Table~\ref{tabnlte}.

For HD 129590, we find that (the optically thin) mass range is $2.1\times10^{-5}$ - $1.3\times10^{-4}$ M$_\oplus$. This translates to a surface density range of $3\times10^{-7} - 2\times10^{-6}$ kg/m$^2$ (assuming a disc centered on the planetesimal belt at 74 au, with a 40 au total width, i.e. colocated with its planetesimal belt).

\begin{table*}
  \centering
  \caption{CO/CN masses computed from our models. The first column is the system's name, the second is yes if circumstellar CO is detected and no otherwise, and the last three columns are the LTE, non-LTE CO masses and CN masses (or upper limits) for CO detections (and non-detections) derived from our study.}

  \label{tabnlte}
  \begin{tabular}{|l|c|c|c|c|}
   \toprule
   Systems & CO detection  & M$^{\rm LTE}_{\rm CO}$ & M$^{\rm NLTE}_{\rm CO}$ & M$^{\rm NLTE}_{\rm CN}$ \\ 
    & (yes/no) & (M$_\oplus$) & (M$_\oplus$) & (M$_\oplus$) \\ 

    \midrule
    \midrule
    HD 106906 & n & $<8.5\times10^{-6}$ &  $<7.5\times10^{-6}$ - $4.6\times10^{-5}$ & - \\  
    HD 114082 & n & $<5\times10^{-6}$ &  $<4.6\times10^{-6}$ - $2.4\times10^{-5}$ & - \\  
    HD 117214 & n & $<5.2\times10^{-6}$ &  $<4.8\times10^{-6}$ - $3.2\times10^{-5}$  & -\\    
    HD 121191 & y & $2.3^{+2.3}_{-0.9} \times 10^{-3}$ &  $7.1\times10^{-5}$ - $4.6\times10^{-4}$ & $<1.2\times10^{-7}$ - $6.5\times10^{-7}$  \\  
    HD 129590 & y & $3\times10^{-5}$ - $1.6\times10^{-4}$ &  $2.1\times10^{-5}$ - $1.3\times10^{-4}$ & $<2.0\times10^{-7}$ - $1.2\times10^{-6}$  \\  
    HD 143675 & n & $<5.3\times10^{-6}$ &  $<3.9\times10^{-6}$ - $2.6\times10^{-5}$ & - \\      
    HD 15745 & n & $<1.2\times10^{-5}$ &  $<1.0\times10^{-5}$ - $6.4\times10^{-5}$ & - \\      
    HD 191089 & n & $<2.9\times10^{-6}$ &  $<2.5\times10^{-6}$ - $1.6\times10^{-5}$  & -\\    
    HD 69830 & n & $<7.8\times10^{-8}$ &  $<3.8\times10^{-8}$ - $2.2\times10^{-7}$  & -\\  
    HR 4796A & n & $<1.4\times10^{-5}$ &  $<1.3\times10^{-5}$ - $8.8\times10^{-5}$  & -\\    

   \bottomrule
  \end{tabular}
\end{table*}


We note that indeed the non-LTE (low collider density) masses derived are larger than previously derived in LTE. This is mainly because in the radiation dominated regime (i.e. far from LTE), it is common for $\sim 10$ K gas to not be as excited as in LTE, therefore needing a larger mass to reproduce the same flux upper limit. For the fluorescence calculation, we have taken SEDs from \citet{2017MNRAS.469..521K} and assumed that stellar radiation is absorbed at the radial locations\footnote{For the radius of HD 143675, we assumed a real to black body radius ratio of 2.87 \citep[the average of the mm-resolved sample from the R-L$_\star$ relationship,][]{2018ApJ...859...72M} and use the GAIA updated black body radius found in \citet{2017MNRAS.469..521K} of 16.4 au, leading to a true radius of 47 au. For the radius of HD 69830, we assumed 2 au, consistent with the range given in \citet{2009A&A...503..265S}, i.e. within 0.05-2.4 au.} derived in Table~\ref{tabbestfitcont}.  We show an example of the flux seen by a molecule at the radius of the dust belt in Fig.~\ref{spectre} for HD 117214 (the other SEDs are shown in the appendix \ref{figspectreappendix}), where the orange line is the ISRF, compared to the blue being the star at the belt radius. In addition, vertical grey lines delimit the CO photodissociation range. Red lines are electronic (UV) and rovibrational (IR) transitions that are accounted for in the non-LTE code used (which lead to fluorescence and can modify the rotational excitation). For the case of HD 117214, we find that the ISRF dominates in the CO photodissociation range but the stellar contribution is more important for the electronic (UV) and rovibrational (IR) excitation.

\begin{figure}
   \centering
   \includegraphics[width=9cm]{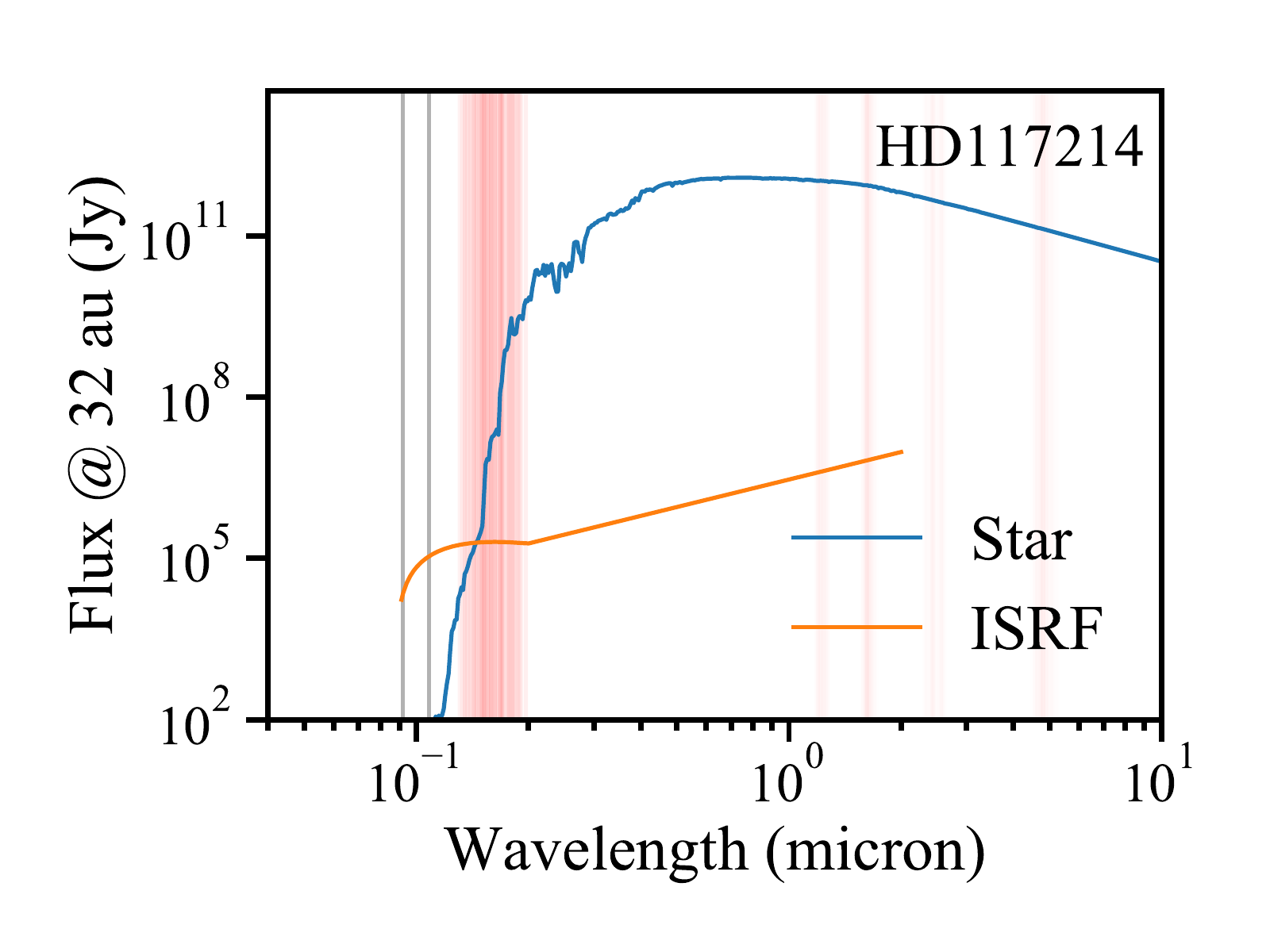}
   \caption{\label{spectre} Irradiation of the disc around HD 117214 at the belt radius (32 au). The orange line is the ISRF and the blue line is the star emission at the belt radius. Vertical grey lines delimit the CO photodissociation range. Red lines are electronic (UV) and rovibrational (IR) transitions that are accounted for in our non-LTE fluorescence calculation \citep{2015MNRAS.447.3936M,2018ApJ...853..147M}.}
\end{figure}

For the systems without clear detections, we assume the same geometrical parameters as the dust disc (see Table~\ref{tabbestfitcont}) as these would be unconstrained and fit for the temperature (assuming a uniform prior between 1 and 100 K) and $\Sigma_0$ to estimate an upper limit on the total mass of CO that could be hidden and not detected. We always find that the temperature is unconstrained, indicating that the CO mass that could be hidden for these non-detections may not be optically thick. We then derive an upper limit on the surface density $\Sigma_0$, which we list in Table~\ref{tabbestfitgas}. For all systems, we estimate the CO mass or upper limits from our MCMC simulations, which we list in Table~\ref{tabnlte}.

\begin{figure*}
   \centering
   \includegraphics[width=8cm]{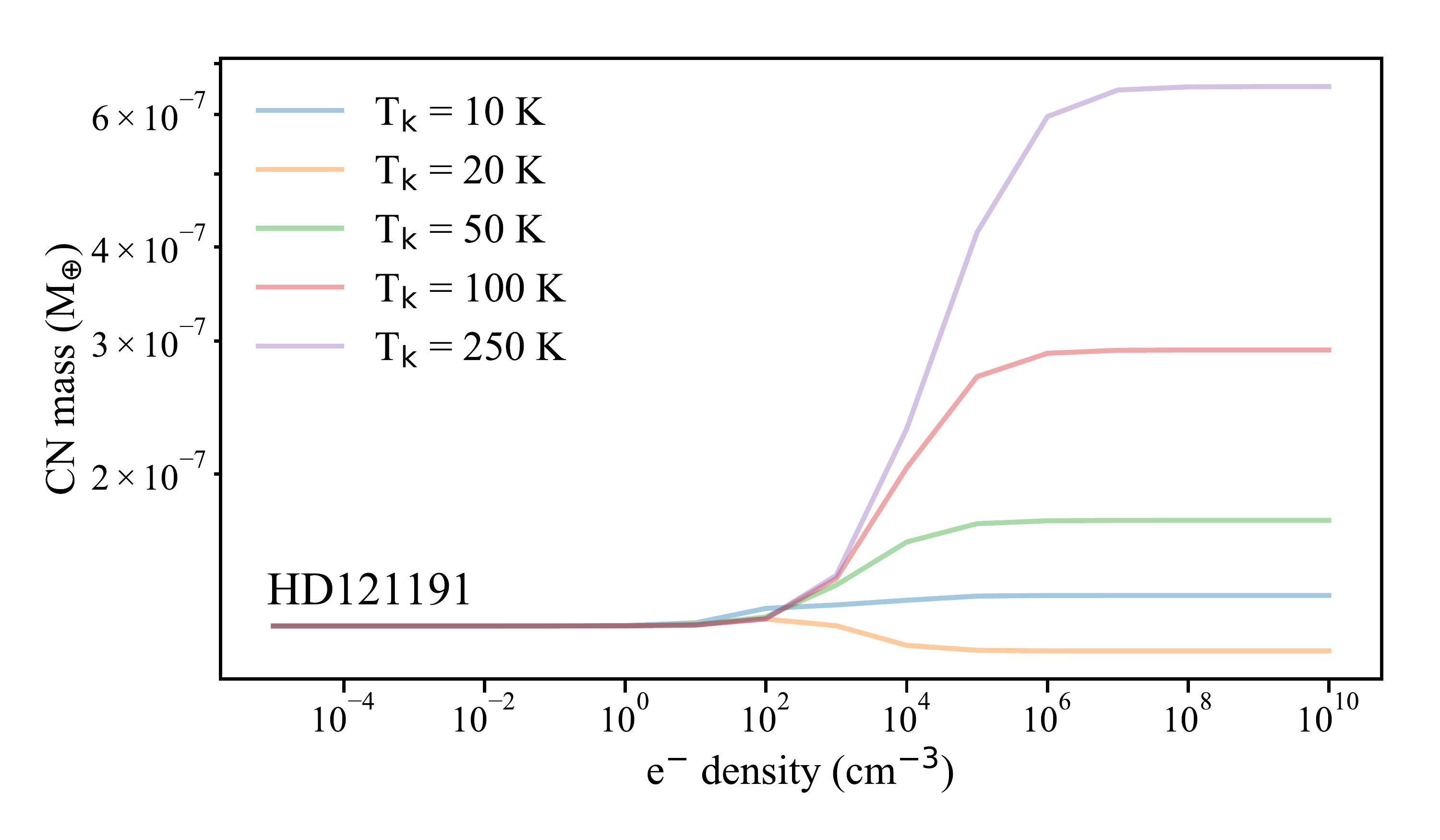}
      \includegraphics[width=8cm]{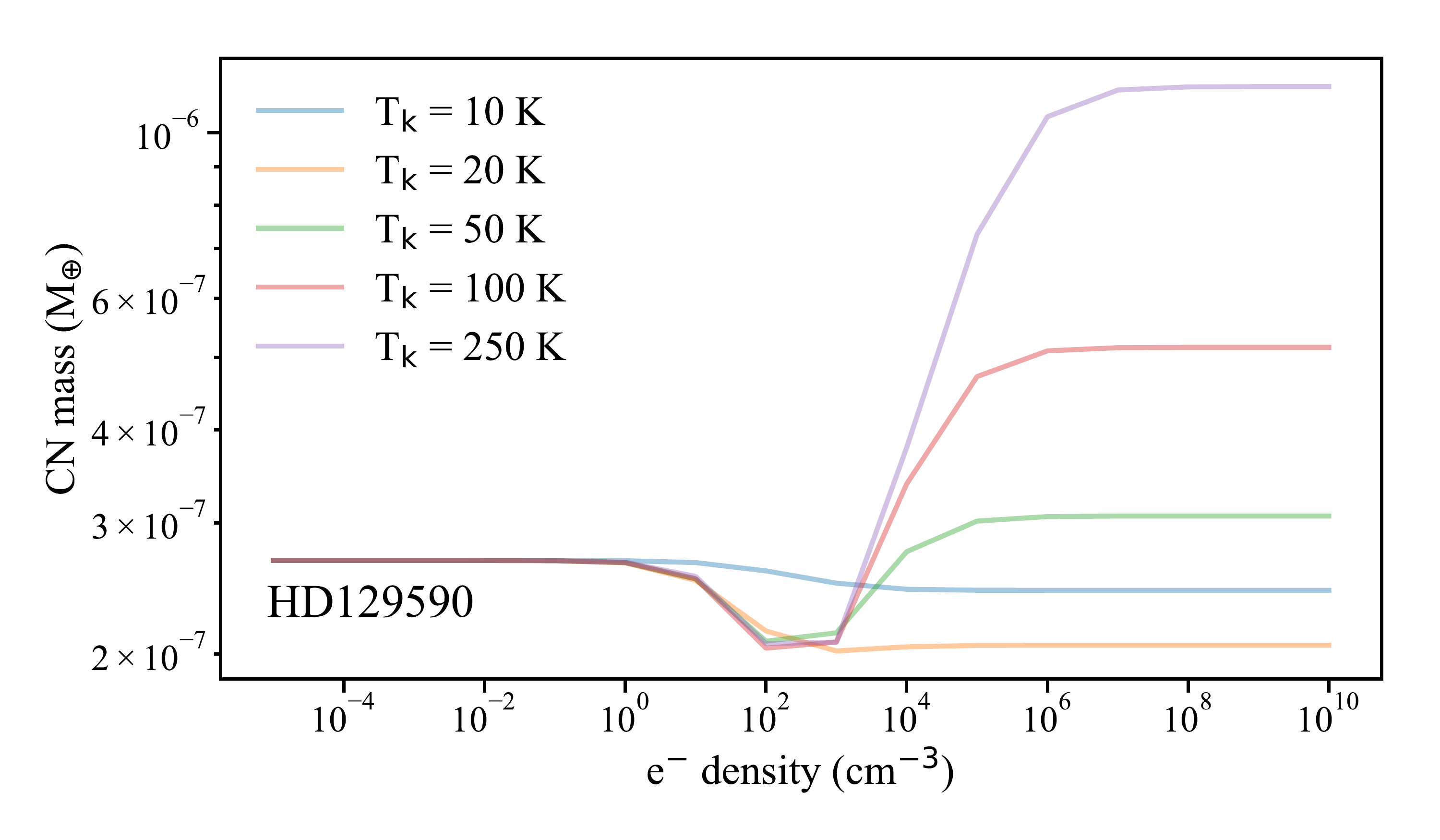}
   \caption{\label{figCNn} Upper limits on the CN ice mass fractions for HD 121191 and HD 129590 using the non-LTE fluorescence calculation from \citet{2015MNRAS.447.3936M,2018ApJ...853..147M}.}
\end{figure*}

We also compute CN masses in non-LTE assuming optically thin emission for the two discs with CO detected HD 121191 and HD 129590 in Fig.~\ref{figCNn}. We find very low upper limits in CN mass for both stars, between $1.2\times 10^{-7}$-$6.5\times 10^{-7}$ M$_\oplus$ for HD 121191 and $2.0\times 10^{-7}$-$1.2\times 10^{-6}$ M$_\oplus$ for HD 129590 (see Table~\ref{tabnlte}). We further discuss these values in Sec.~\ref{compo} as to what it means for the production mechanism of the gas in debris discs and how the upper limit we find can be translated to an upper limit in HCN to CO and compared to comets in our Solar System.

\section{Discussion}\label{discu}

\subsection{New continuum detections}\label{new}
Our main addition to the literature for the continuum section is given in Table~\ref{tabbestfitcont} as we derive fits of the geometry and mass of the dust discs surrounding the 10 stars of our sample. We now list some more detailed outcomes that emerge from our work.


\subsubsection{HD 106906}

Our observations of HD 106906 provide the first continuum detection in the mm for this target, which we also resolved. This system had already been targeted in \citet{2016ApJ...828...25L}, which led to a non-detection and an upper limit of 0.4 mJy (assuming it is extended over 3 beams) in agreement with our detection of $0.35 \pm 0.1$ mJy (our observations reached an rms $\sim$2.5 smaller than their observation). This new measurement can be used \citep[together with Herschel flux measurements,][]{2017arXiv170505693M} to put some constraints on the size distribution of grains which is linked to the slope of the modified black body $x$ (defined as $F_\nu \propto \lambda^{x}$) that can fit the Rayleigh-Jeans region of the SED \citep[e.g.][]{2012A&A...539L...6R}. We find a spectral index 160 $\mu$m-1.27 mm equals to $-2.64\pm0.15$ and assume it is the same as the modified black body slope (as the Rayleigh-Jeans regime starts to kick-in around 160 microns for HD 106906), which leads to a size distribution in $-3.36\pm0.13$ \citep[using the relation between slope and size distribution from][]{2006ApJ...636.1114D}. We find that this value is consistent with the size distribution expected from numerical simulations, though slightly shallower \citep[e.g.,][find values close to -3.6]{2012ApJ...754...74G,2013A&A...558A.121K}, which may indicate differences in composition and porosity leading to different collisional evolution.

Images in scattered light with GPI \citep{2015ApJ...814...32K} and SPHERE that show the location of $\mu$m-sized dust also exist \citep{2016A&A...586L...8L}. These images lead to a best fit with a clear inner hole inside of $50$ au and a ring peaking at about 73 au (after correcting for the new GAIA distance) detected to $>100$ au with a strong needle-like SE/NW asymmetry, with an inclination of $\sim 85^\circ$ and a PA of $\sim 104^\circ$. The geometrical values that fit the mm-disc from our work are in agreement with these results as we find a disc centered at $85\pm13$ au, an inclination greater than 59$^\circ$ and a PA of $112\pm7$ deg. In the mm-image, we do not see any obvious asymmetry but an image at higher resolution would be needed to assess whether an asymmetry is present. No asymmetry in the mm would seem to disfavour that the asymmetry observed in scattered light would be due to the otherwise (potentially eccentric) detected 11 $M_{\rm Jup}$ companion at a projected separation of 650 au \citep{2017ApJ...837L...6N}. 

\subsubsection{HD 114082}

SPHERE obtained the first resolved scattered light image of HD 114082 recently \citep{2016A&A...596L...4W}. They find that the $\mu$m-sized dust disc has a clear inner hole and is best fit as having an inner edge of $27.7^{+2.8}_{-3.5}$ au followed by a steep decreasing power law. This compares well to the value of $24^{+11}_{-9}$ au we find for the centre of this disc (although it is marginally resolved) and is the first mm-radius as it was not derived (see HIP 64184) in \citet{2016ApJ...828...25L}.

\subsubsection{HD 117214}

We have also improved the signal-to-noise of the previously observed system HD 117214 \citep[we have a 16$\sigma$ detection while it was detected at $\sim5\sigma$ in][]{2016ApJ...828...25L}. For HD 117214, because of the low S/N, \citet{2016ApJ...828...25L} did not fit a model to the data. We find that the mm-dust is located at $32\pm9$ au with a FWHM$<41$ au and a total dust mass of 0.1 M$_\oplus$. This disc was detected for the first time in scattered light recently with SPHERE \citep{2019arXiv191104793E} after several unsuccessful attempts \citep[e.g.][]{2019AJ....157...39G}. They find that the $\mu$m sized dust traced by optical wavelengths is located at around 45 au, which is consistent with our observations.

\subsubsection{HD 121191}

We have also improved the signal-to-noise of the previously observed system HD 121191 \citep[we have a 9$\sigma$ detection while it was detected at $\sim5\sigma$ in][]{2017ApJ...849..123M}. In \citet{2017ApJ...849..123M}, they only detect one side of the dust disc, which led to an underestimation of the total dust mass, which we find is close to 0.1 M$_\oplus$. Thanks to our deeper observation, we do not see any visual asymmetries in the continuum image of HD 121191 as was tentatively seen in \citet{2017ApJ...849..123M}. However, we find a displacement of the disc center from the expected stellar position, which could be either real or be due to ALMA calibration issues, and new observations would be needed to clarify this. According to our work, the mm-dust disc around HD 121191 is located at $52.1\pm11$ au. The mm-radius is smaller than the 195 au found by \citet{2016ApJ...833..263V} with marginally resolved Herschel observations. As they note, their large size leads to a large observed-to-black body disc radius, and the simplest explanation is that the disc radius is closer to the 52 au measured with ALMA but we note that Herschel emission could still be larger because it is sensitive to smaller grains in the halo, which are pushed on eccentric orbits by radiation pressure.

\subsubsection{HD 129590}

HD 129590 was recently detected in scattered light with SPHERE \citep{2017ApJ...843L..12M}. They find a disc with an inclination of $\sim 75^\circ$ and a PA of $\sim$121 degrees. The best fit of the dust distribution is for a single bright ring of radius 60-70 au and an inner clearing. In the mm, \citet{2016ApJ...828...25L} constrained the inclination of the disc around HD 129590 (HIP 72070) to be $>50^\circ$ (with a best-fit value of $70^\circ$) and a PA of $121_{-12}^{+17}$ degrees. They best fit the data with a disc having an inner edge $<40$ au and an outer edge at $110_{-30}^{+50}$ au. This is in agreement with our results as we find an inclination $>$65 deg and a PA of $117\pm5$ deg and the radius of the ring to be $74\pm6$ au.

\subsubsection{HD 143675}
The disc around HD 143675 has never been detected at wavelengths longer than 70 $\mu$m. Our new mm-observation put an upper limit on the total dust mass in the system equal to 0.08 M$_\oplus$. In addition, this non detection puts a constraint on the slope of the modified black body (and hence size distribution of the grains) that fits the SED, which should be steeper than -2.36 as otherwise the disc would have been detected. This leads to a size distribution steeper than -3.2. A recent observation in scattered light with GPI puts constraints on the $\mu$m sized-dust distance to its host star of $\sim50$ au \citep{2019arXiv191109667H}.

\subsubsection{HD 15745}

HD 15745 was detected in the sub-mm thanks to the JCMT \citep[at 850 $\mu$m, see][]{2017MNRAS.470.3606H}. The main improvement of our study is the resolution of the observation as the FWHM of the JCMT is around 13\arcsec \,at 850 $\mu$m to be compared to $\sim1.5$\arcsec \, in our ALMA survey. There were doubts concerning contamination by background galaxies in the JCMT images (as the mm-slope from far-IR to JCMT wavelength was close to -2, which is suspiciously flat), which we can now test. We can actually see two background galaxies within the JCMT field of view in our ALMA image. One galaxy (with a total integrated flux of 1.3 mJy) can be seen at the North-East of the image (see Fig.~\ref{figcont}) offset from HD 15745 by 9.4" in RA and 6" in Dec, and the main contaminating galaxy (with a total integrated flux of 2.9 mJy) is out of the image on the East, offset from HD 15745 by 15.2" in RA and -2.8" in Dec.
Extrapolating the observed JCMT flux for HD 15745 at 850 $\mu$m ($12\pm1.4$ mJy) to 1.27 mm gives 4.2 mJy (using a Rayleigh-Jeans slope of -2.6). With our new observations, we found a total flux of $1.2 \pm 0.32$ mJy (see Table~\ref{tabobscon}), which indeed seems inconsistent with the JCMT observations. We find that the slope of the modified black body would have to be close to -5.7 to agree with our new ALMA data. As this is too extreme, we conclude that the JCMT observations of HD 15745 were indeed contaminated by some extra flux from background galaxies.  Ignoring the JCMT flux and using our new ALMA result, we find that the mm-slope is of $-2.6\pm0.08$, which is typical for debris discs \citep{2016ApJ...823...79M}.

For HD 15745, we find an offset compared to the star of 0.59" in declination, which may potentially be real (as it is roughly aligned with the PA of the disc) and indicating that there would be a NE/SW asymmetry in this disc in the mm. Indeed, in scattered light, \citet{2014AJ....148...59S} find a clear NE/SW asymmetry, which we may see in the mm. Higher resolution data would be needed to confirm it. In scattered light the disc's inclination and PA are constrained to be $\sim67^\circ$ and $\sim23^\circ$, respectively, but it is not trivial to constrain in scattered light due to the fan shape of the disc \citep{2007ApJ...671L.161K,2014AJ....148...59S}. We find that our lower limit of 52 deg in inclination is consistent with these observations but the PA we derive of $39\pm5$ deg is slightly larger than in scattered light, which may reveal something yet to be understood about the underlying mechanism producing fan-like shaped debris discs. The $\mu$m-sized dust radial location is also constrained to be at $\sim68$ au (after correcting for the new GAIA distance) from the scattered light image \citep{2007ApJ...671L.161K}, which is close to the value found in our mm-study of $72\pm6$ au.

\subsubsection{HD 191089}
HD 191089 was also detected in the sub-mm thanks to the JCMT \citep[at 850 $\mu$m, see][]{2017MNRAS.470.3606H}. Proceeding in a similar way as for HD 15745, we find that for HD 191089, the JCMT flux is consistent with our new observation for a modified black body slope of -2.45, typical of values in other debris discs (i.e., the JCMT flux was probably not contaminated by background galaxies as for HD 15745). 

The first resolved image of HD 191089 was in the mid-IR \citep{2011MNRAS.410....2C}. They found a PA of $80\pm10$ deg and the emission was best-fitted with a belt from 26 to 84 au (using new GAIA distance), inclined at $55\pm5$ deg with an inner hole inside of 26 au. The first resolved scattered light observation was with the HST where they found inclination and PA consistent with mid-IR observations and that the disc should be within 70 au (using new GAIA distance). More recent work coupling HST (STIS and NICMOS), and GPI data, place the $\mu$m-sized dust traced at optical wavelengths at 46 au with a FWHM of 25 au, an inclination of $59^{+4}_{-2}$ deg and a PA of $70^{+4}_{-3}$ deg \citep{2019ApJ...882...64R}. We find that the sub-mm dust traced by ALMA is at a radius $\sim$43 au (with a FWHM$<45$ au), inclined by at least 52 deg, with a PA of $73\pm4$ deg consistent with previous findings. We do not see any obvious asymmetry (as in scattered light) in the mm-image and more work that goes beyond the scope of this paper would be needed to characterise that in further detail.

\subsubsection{HD 69830}
We do not detect mm-dust but rather the star's photosphere in this case as is explained in further detail in subsection \ref{star}.

\subsubsection{HR 4796 A}
As shown in Sec.~\ref{dustfitt}, we find that our findings are consistent with previous studies but we do not derive new updated constraints as our observations are at a lower resolution than previous work \citep[e.g.][]{2018MNRAS.475.4924K}.

\subsection{First gas detection around a G-type star: HD 129590}

A crucial result of this paper for the gas part is the new CO gas detection around the G1V (i.e. Sun-like) star HD 129590. HD 129590 is a young star member of the ScoCen association with an age of 10-16 Myr \citep{2011ApJ...738..122C}. The very first detections of CO were around A-type stars, which led people to believe it was an A-star phenomenon. The model by \citet{2017MNRAS.469..521K} showed that it was expected that gas would be released at a lower rate (on average) around later-type stars (see their Eqs. 1 and 2), which was later better quantified by \citet{2019AJ....157..117M}. Other than A-stars, we now know of two F stars\footnote{There is also CO marginally detected around the F-star $\eta$ Crv but very close-in and the gas origin is expected to be different and coming from exocomets being sent into the inner region of the system rather than being released in the debris belt itself \citep{2017MNRAS.465.2595M}.} surrounded by CO gas \citep[HD 146897 and HD 181327,][]{2016ApJ...828...25L,2016MNRAS.460.2933M}, one around a G star (this work) and one around an M star \citep{2019AJ....157..117M}. We have to integrate for longer to reach the sensitivity necessary to detect gas around these later spectral types but ALMA has the sensitivity to do so \citep{2017MNRAS.469..521K} and will likely lead to the first detection around a K-type star in the coming years.

We derive a mass range (in non-LTE assuming optically thin emission) of $2.1\times10^{-5}$ - $1.3\times10^{-4}$ M$_\oplus$. Higher S/N data and/or looking for CO isotopologues or neutral carbon would be needed to pinpoint the gas mass in this system and assess whether the line is optically thin and if this disc is shielded or not \citep{2018arXiv181108439K}. 

This new finding is important to understand the origin of our Solar System as it shows that this late gas can be present around G-type stars very similar to our Sun. Such a quantity of gas in the early time of our Solar System could have influenced the late stages of planet formation by changing the metallicity or C/O ratio of giant planets and providing some volatiles to the already formed terrestrial planets from early active planetesimal belts \citep{Kral19}.

\subsection{New constraints from the CO gas detection in HD 121191}\label{cons1211}

Our MCMC modelling finds that both the temperature $T_0$ and surface density $\Sigma_0$ at the gas disc centre is constrained (see Fig.~\ref{figcornergas}). Given that these parameters should be degenerate we tried to understand where these potential constraints come from, for example whether optical depth effects cause the line shape to vary in ways that are inconsistent with the data. At low-T, the gas becomes even more optically thick and most of the emission comes from large radii (where radial velocities are small) and the line becomes single-peaked, leading to a fit that is not as good as for the fiducial best-fit model. At high-T the constraint is harder to understand. We find that the CO disc becomes smaller because it is less optically thick, which does not allow for a good fit. Some more dedicated efforts would be needed such as exploring a larger parameter space and/or a different density profiles to validate these constraints but the cpu resources needed for that go beyond what is doable in a reasonable amount of time. New data at higher resolution and in a different band would help to further confirm our constraints on $\Sigma_0$ and $T_0$.

We now test whether the gas disc around HD 121191 is self-shielded. We find that $\Sigma_0 \sim 10^{-5}$ kg/m$^2$, i.e. the (vertical) column at $R_0 \sim 22$ au is $2 \times 10^{16}$cm$^{-2}$. Using \citet{2009A&A...503..323V}, we find that for such a CO column, the CO photodissociation timescale is longer than if unshielded by a factor $\sim$30. Neutral carbon could also provide some shielding, which would lower the predicted gas production rate and affect the predictions of the composition of exocomets in this system \citep{2018arXiv181108439K}. This shielding may also explain why the CO gas disc is observed in the inner region of the system (at $\sim$20 au) rather than at the parent belt location (at $\sim60$ au). Indeed, hydrodynamical simulations taking into account carbon and CO self-shielding \citep{2018arXiv181108439K,2019arXiv190809685M,Marino} show that the CO surface density can go up with decreasing distance to the host star when CO is shielded. Higher resolution images of this system along with new carbon observations with ALMA would be able to test the spreading model in great detail but it goes beyond the scope of this paper to model more precisely the low-resolution image we have at present.

Detecting CO gas closer to its parent belt is not unprecedented. The inner radius of the gas disk around the A3V star HD 21997 is located at $<26$ au \citep{2013ApJ...776...77K} and at $22\pm6$ au (similar to HD 121191) for the tentative gas detection around the F2V star $\eta$ Crv \citep{2017MNRAS.465.2595M}. Viscous evolution in a shielded disk might better explain the former while comets being sent inwards from an outer belt may better explain the latter. In HD 121191, shielding may be able to explain the results but we cannot rule out that comets are being sent inwards in this system that has an unusual mid-infrared emission feature and large quantities of warm dust \citep{2013ApJ...778...12M}. These comets would then release molecular species closer than the parent belt. Comets at 22 au around HD 121191 would have an equilibrium temperature close to  92 K, which is close to 80 K where desorption rates are maximum for CO$_2$ in isolation \citep{2004MNRAS.354.1133C}. CO$_2$ could then be released and create some CO, by dissociating into CO+O and/or because CO trapped in CO$_2$ ices would also be released. Desorption of water ice could also release trapped CO and/or CO$_2$ but it reaches a maximum at $\sim$140 K, which is at about 10 au in HD 121191. Higher resolution observations would allow us to distinguish between the CO$_2$ ice line and shielding models by looking at the radial distribution of the gas and its velocity at different locations in the disk.

\subsection{Detection of the star HD 69830}\label{star}
There is a disc of warm ($>200$K) dust \citep[][]{2005ApJ...626.1061B} detected at a few au from HD 69830 but does this system also possess a colder belt? Very little emission is expected in the mm from this warm component, and Herschel did not detect cold dust in the infrared \citep{2018MNRAS.475.3046S}. We detect some continuum emission at the star location but this system being very nearby (12.6 pc), it is possible that we actually detect continuum emission from the star itself. To test that hypothesis, we fit a PHOENIX stellar spectrum \citep{2013A&A...553A...6H} to the SED of that G8V star and allow for some extra emission coming from warm dust. We find that the emission at 1.27 mm is indeed dominated by the stellar flux, which reaches $0.0507\pm0.002$ mJy. This is clearly in agreement with the observed flux of $0.05 \pm 0.01$ mJy (see Table~\ref{tabobscon}). We thus confirm that there is no significant mm-wavelength emission coming from the warm dust beyond a few tens of microns and we do not detect any colder belt either. The total flux we detect is in agreement with expectations from emission from a G8V star and we do not detect any extra excess that could be attributed to chromospheric stellar emission such as in the $\alpha$ Cen system \citep{2015A&A...573L...4L}.


\subsection{Constraints on exocomet compositions}\label{compo}

\subsubsection{CO content}\label{COlab}

\begin{figure*}
   \centering
   \includegraphics[width=16cm]{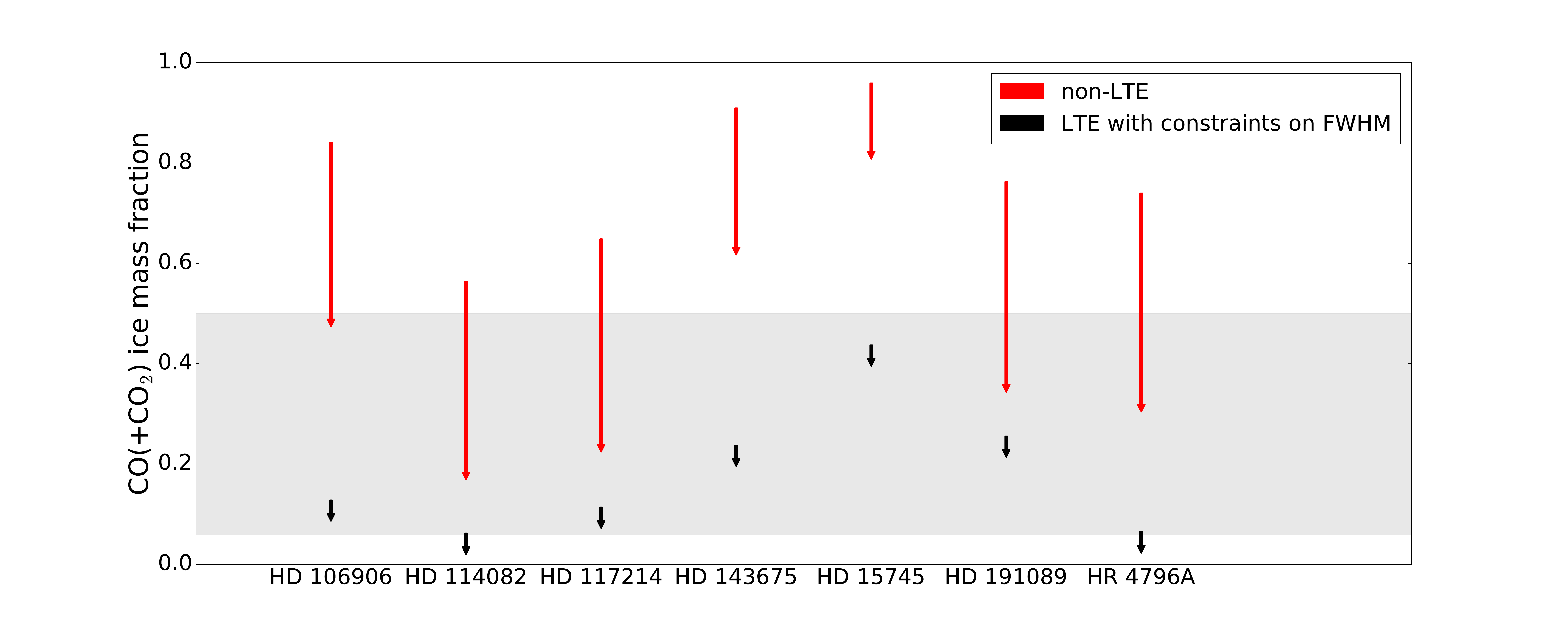}
   \caption{\label{figicemass} Upper limits on the CO(+CO$_2$) ice mass fractions for the 7 systems listed on the x-axis of the plot. In red, we show the extent of upper limits based on non-LTE calculations for a wide range of temperatures and electron densities (see Fig.~\ref{fignlte}). In black, we show the upper limit derived from our LTE MCMC calculations where we fixed the FWHM to observed values (see Sec.~\ref{COlab}). The grey shaded region is the range of CO(+CO$_2$) ice mass fractions in comets in the Solar System \citep{2011IAUS..280..261B}.}
\end{figure*}

We do not attempt to derive the CO content in the planetesimals of HD 121191 and HD 129590 as these discs are most probably self-shielded and may be shielded by carbon. However, we note that the total CO mass we found for HD 121191 agrees with our predictions from \citet{2017MNRAS.469..521K} (but there could be some extra shielding from carbon which is not accounted for in the 2017 model). For HD 129590, the gas model in \citet{2017MNRAS.469..521K} predicts masses of order $10^{-2}$ M$_\oplus$, which is much higher than found in our study. For these systems more observations are needed and more careful self-consistent modelling must be done to account for shielding to give a coherent CO content.
However, for non-detections (with much lower masses), we convert the upper limits in mass we found in Sec.~\ref{gasfit} (Table~\ref{tabnlte}) to an upper limit in CO(+CO$_2$) ice mass fraction on the planetesimals of these systems. To do so, we assume that the production rate of gas and dust are at steady state and compare them to get the CO mass fraction, which we calculate from Eq. 2 in \citet{2017ApJ...842....9M}, where we use \citet{2017MNRAS.469..521K} to compute CO photodissociation timescales. We note that these CO discs (if they exist at all) are most likely not self-shielded because the upper limits in surface density we find in Table~\ref{tabbestfitgas} are at least 2 orders of magnitude smaller than for HD 121191 and checking in \citet{2009A&A...503..323V}, these discs would have very little self-shielding\footnote{With no carbon observations at hand, we cannot say that carbon shielding is non-existent. If carbon is present in sufficient quantity to shield CO then we would predict even lower amount of CO ice trapped on the grains, hence our upper limits would still be valid.}. We show the results in Figure~\ref{figicemass}, where the black arrows are for upper limits from our LTE MCMC calculations and the red arrows account for the non-LTE range of masses we found. In Figure~\ref{figicemass}, the grey shaded region corresponds to the range of CO(+CO$_2$) mass fractions observed for comets in our Solar System, i.e. between 6 and 50\% \citep{2011IAUS..280..261B}. 

We conclude that the upper limits we obtained with ALMA are, so far, not at odds with the composition of Solar System comets. Assuming that the LTE upper limits are correct and using the width of belts obtained from other studies \citep{2018MNRAS.475.4924K,2016A&A...596L...4W,2016A&A...586L...8L,2014AJ....148...59S,2011MNRAS.410....2C,2019arXiv191104793E,2019arXiv191109667H}, we can put even stronger constraints on the CO content of planetesimals in these systems (see black arrows in Fig.~\ref{figicemass}). For instance, we find that the CO content in HD 114082 ($<$1.6\%) and HR 4796A ($<$1.9\%) is already below the range of CO content compared to the Solar System comets. This result for HR 4796A was already pointed out in \citet{2018MNRAS.475.4924K} where they used CO upper limits in band 7 to put a constraint of $<$1.8\%, which may mean that the composition of planetesimals in this system is different to what can be found in our Solar System or that CO is not released for some reasons yet to be understood \citep[e.g.][suggest that the belt in that system may have formed within the CO snowline and thus its planetesimals are CO depleted compared to comets]{Marino}. We also find that the CO content in the planetesimals of HD 106906 ($<$8\%) and of HD 117214 ($<7$\%) is already at the lower end of the composition range seen in Solar System comets.

\subsubsection{HCN content}\label{HCNa}

For the two systems with CO detections (HD 121191 and HD 129590), we estimate the maximum ratio of HCN/CO production rate. We use Eq.~3 in \citet{2018ApJ...853..147M} to compute this outgassing rate assuming no shielding and where we derived the level populations $x_{u, {\rm CN}}$ and $x_{u, {\rm CO}}$ using the same NLTE excitation code including fluorescence. Photodissociation timescales are mainly driven by the ISRF. As CN is not detected in both cases, we obtain upper limits (assuming that all CN comes from photodissociation of HCN, which is mostly true in our Solar System) on the HCN/[CO(+CO$_2$)] outgassing rate ratio from exocomets of $<2.4$\% for HD 129590 and $<0.46$\% for HD 121191. For HD 121191, this is about 5 times more constraining than the upper limit on the HCN/CO production rate derived in $\beta$ Pic \citep[of $\sim2.5$\%,][]{2018ApJ...853..147M}. It leads to much stronger constraints when comparing to measured outgassing rates in our Solar System. For instance, we find that these values for the HCN/CO production rate are much lower than observed for comets with short perihelia but is close to values observed for comets at greater distances \citep[$> 5$ au, see Fig.~5 in][]{2018ApJ...853..147M}. We note that these calculations assumed no shielding for either CO or CN (and that the emission is optically thin); CO could self-shield and carbon could shield CO even more, but CN should be less affected. For HD 121191, we found that CO could be shielded by at least a factor of 30 (see Sec.~\ref{cons1211}), which means that less CO per unit time is produced than assumed in our first guess estimate and the HCN/CO production rate will go up by at least a factor 30 (unless CN is strongly shielded). For HD 129590, this effect may be less important but still has to be quantified better from deeper observations of CO, CN and targeting carbon with ALMA.

These results could be explained by different scenarios. Either HCN is truly depleted compared to CO in these exocomets \citep[similar to the low HCN/CO ratio found for the interstellar comet 2I/Borisov,][]{2020NatAs.tmp...84C}, or CO is being preferentially released (which would favour some release mechanisms over others) or this ratio is actually higher due to CO shielding that makes CO more easily observable. The latter is certainly true for HD 121191 and should be checked further for HD 129590.

\begin{figure*}
   \centering
   \includegraphics[width=8cm]{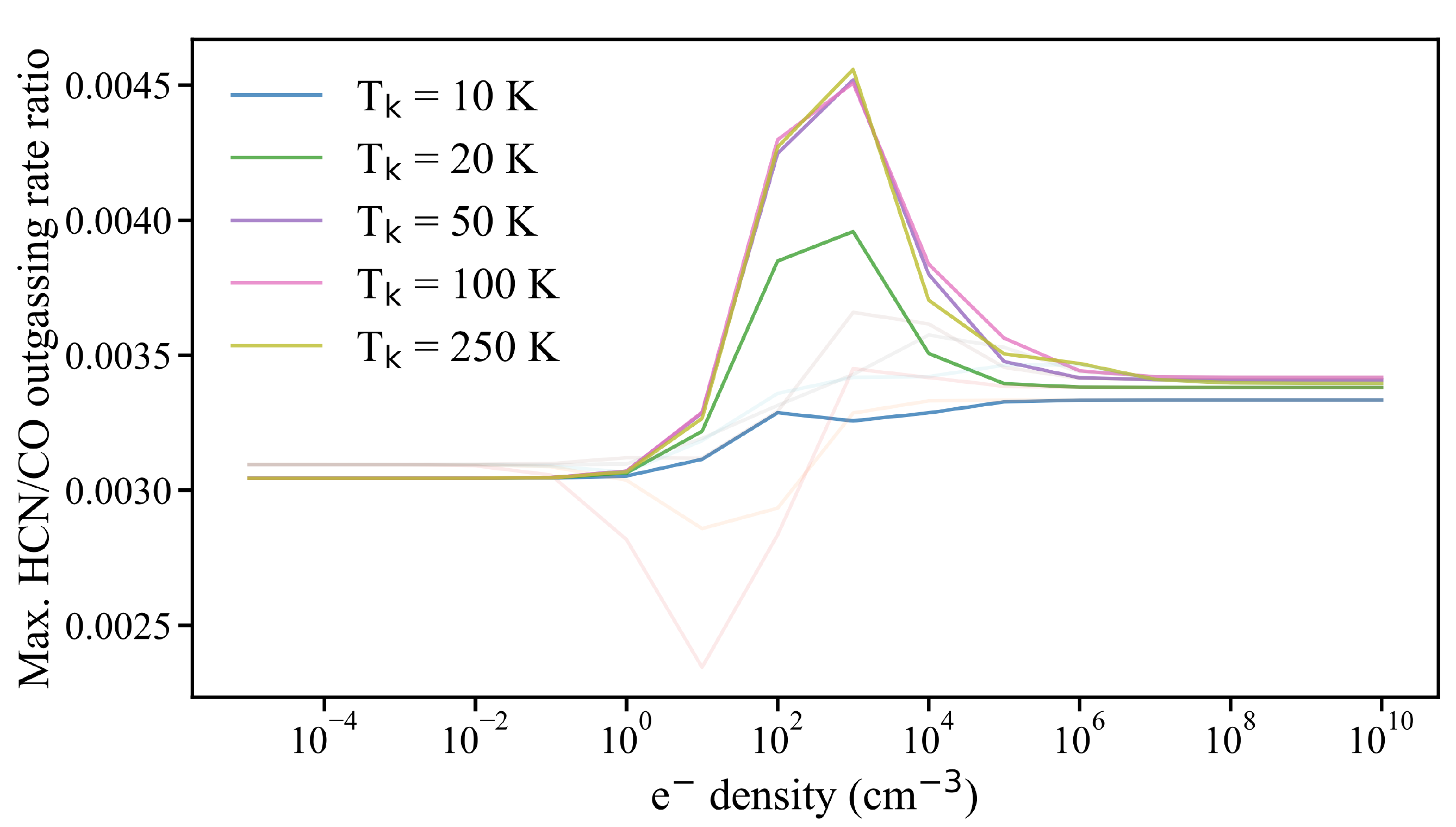}
      \includegraphics[width=8cm]{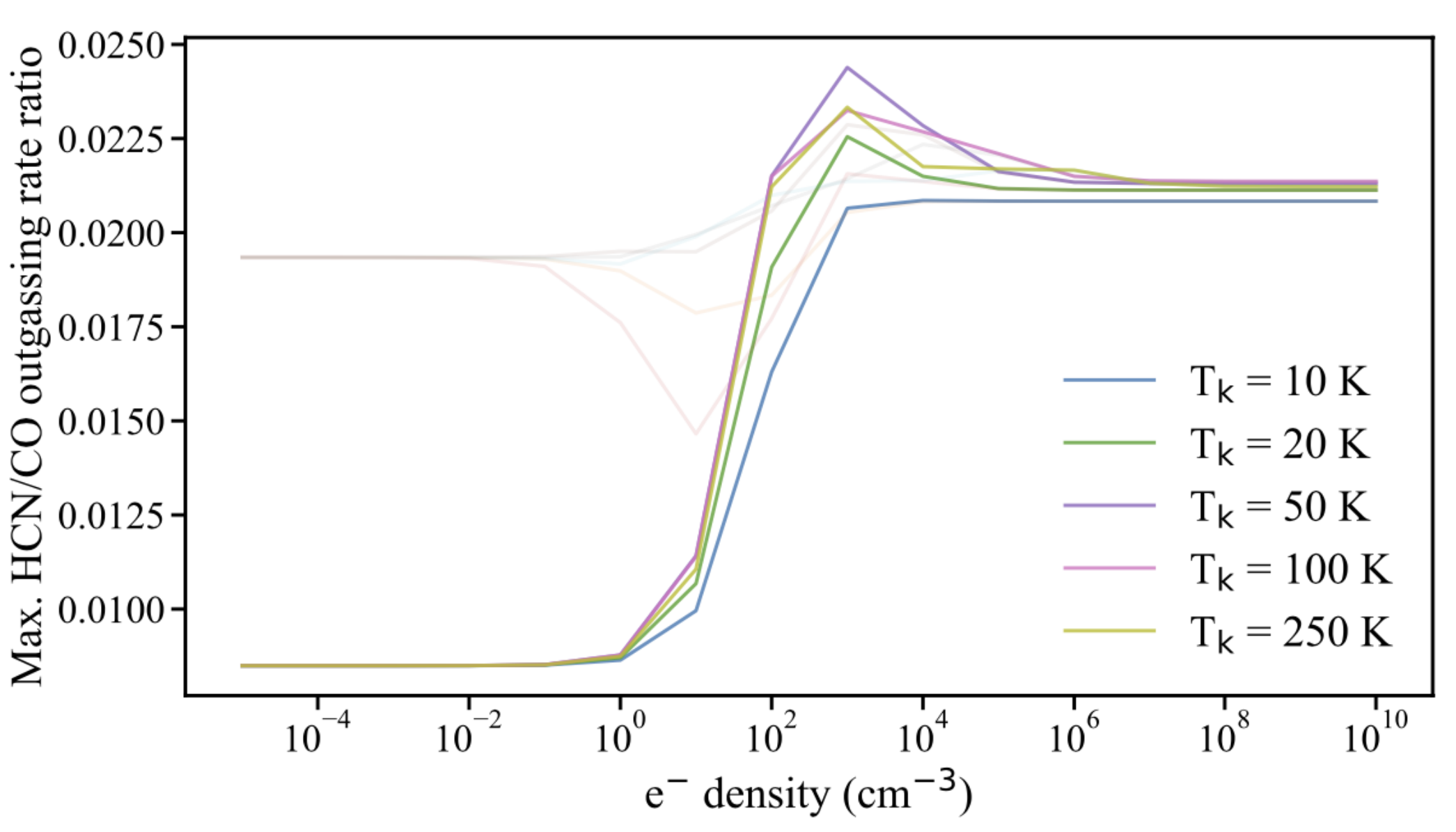}
   \caption{\label{figCN} Upper limits on the HCN to CO outgassing rate for HD 121191 and HD 129590.}
\end{figure*}

    
\subsection{Unexplained CO gas detection close to HD 106906}\label{1069}
We also detect some significant CO emission on the SE side of HD 106906 at about 3-4" when following the continuum disc PA (see Fig.~\ref{fig106}). The coordinates of the brightest pixel of the CO emission are RA 12:17:53.425 and Dec -55:58:35.776 (ICRS, observed on the third June 2018). However, the PA of the CO putative detection seems to be almost perpendicular to the PA of the continuum, so the CO may be due to a cloud along the line of sight. But we note that the cloud is roughly at the stellar velocity and the spectrum shows a double peaked profile, which is usually due to circumstellar discs. We just report this unexplained CO detection without finding convincing explanations owing to its origin. If it is a cloud then its intra-cloud motions could be complex enough to create a double peaked profile and if it is circumstellar then the PA of the gas disc is definitely not well oriented compared to the PA of the dust disc. HD 106906 is, however, a complex system with a planet at hundreds of au on the NW side of the star and with, supposedly, recent fly-bys, which may have influenced its evolution and created asymmetries, which we see in the dust component \citep{2019AJ....157..125D} that may have impacted any gas disc present. Further observations at higher resolution would be needed to confirm whether or not this CO is bound to the star and start investigating different possible scenarios.

\section{Summary and Conclusions}\label{ccl}
In this study, we carried out a survey with ALMA (band 6) of the 10 following stars: HD 106906, HD 114082, HD 117214, HD 121191, HD 129590, HD 143675, HD 15745, HD 191089, HD 69830 and HR 4796. We looked for dust, as well as gas emission (CO and CN).
We detect continuum emission in 9/10 systems, sometimes for the first time, and are able to derive the geometry (position, extent, inclination, PA) and masses of these dust discs. For instance, we provide the first detection in the mm for HD 106906, which we also resolve (see Fig.~\ref{figcont}). We also provide the first mm-radius for HD 114082, 117214, HD 15745, HD 191089 and the first radius at all for HD 121191 (see Table~\ref{tabbestfitcont}). We detect emission from the star HD 69830 (located at 12.6 pc) with ALMA, which seems consistent with model expectations of emission of a G8V star with no excess that could be attributed to chromospheric stellar emission.

We also detect circumstellar gas around HD 121191 and HD 129590 (see Fig.~\ref{figcodetec}) as well as some CO gas at a few arcsec from HD 106906 (which does not seem to orbit the star itself, see Fig.~\ref{fig106}) and some CO along the line-of-sight of HD 114082 (not at the stellar velocity, see Fig.~\ref{figcocloud}). This is the first detection of gas around HD 129590 and around any G-type star. The gas around HD 129590 may be colocated with its planetesimal belt ($\sim$ 70 au) and its total mass is likely between $2-10 \times 10^{-5}$ M$_\oplus$. This detection around a star similar to our Sun raises questions as to whether gas could also have been released early in the Solar System lifetime and contributed to feeding the atmospheres of its 8 (giant and terrestrial) planets early-on with gas released by its early planetesimal belts \citep{Kral19}.
From our gas detection around HD 121191 (at higher S/N than previous study), we are able to derive the position of the belt, which we find is very close in ($\sim 20$ au, see Table~\ref{tabbestfitgas}) compared to the planetesimal belt that is located at $\sim 50$ au. We speculate that this discrepancy may be explained from CO self-shielding and shielding from carbon that would prolong the CO lifetime and allow CO to viscously spread in the inner region \citep[see][for more detail]{2018arXiv181108439K}.

For the systems with no CO detections, we derive upper limits in CO mass as well as upper limits on the CO content in planetesimals of these belts. We find systems (assuming LTE) where the CO ice mass fraction is already low compared to Solar System comets (HD 106906, HD 114082, HD 117214, HR 4796 A), which may indicate that the CO content in these planetesimals is very different than in our Solar System or that CO is not released at all (or much less efficiently) in these systems for reasons that are yet to be understood. Finally, we derive an HCN/CO outgassing rate for planetesimals orbiting HD 121191 and HD 129590, which we find similar to Solar System comets orbiting at large distances ($>$5 au) from the Sun but should be refined with further observations to quantify shielding that may happen in these systems and could affect the HCN/CO predictions.

\section*{Acknowledgments}
This paper is dedicated to L\'ea. We thank the referee for their very thoughtful report that improved the paper.
LM acknowledges support from the Smithsonian Institution as a Submillimeter Array (SMA) Fellow. GMK is supported by the Royal Society as a Royal Society University Research Fellow.
This paper makes use of the
following ALMA data: ADS/JAO.ALMA\#2017.1.00704.S. 
ALMA is a partnership of ESO
(representing its member states), NSF (USA) and NINS (Japan),
together with NRC (Canada) and NSC and ASIAA (Taiwan)
and KASI (Republic of Korea), in cooperation with the Republic of Chile. The Joint ALMA Observatory is operated by ESO,
AUI/NRAO and NAOJ. 
This work has made use of data from the European Space Agency (ESA) mission
{\it Gaia} (\url{https://www.cosmos.esa.int/gaia}), processed by the {\it Gaia}
Data Processing and Analysis Consortium (DPAC,
\url{https://www.cosmos.esa.int/web/gaia/dpac/consortium}). Funding for the DPAC
has been provided by national institutions, in particular the institutions
participating in the {\it Gaia} Multilateral Agreement.

\section*{Data availability}
The data underlying this article are available in the ALMA archive: ADS/JAO.ALMA\#2017.1.00704.S.

\appendix





\section{Irradiation of the 10 discs in our sample}

\begin{figure*}
   \centering
\begin{tabular}{cc}
\includegraphics[width=6cm]{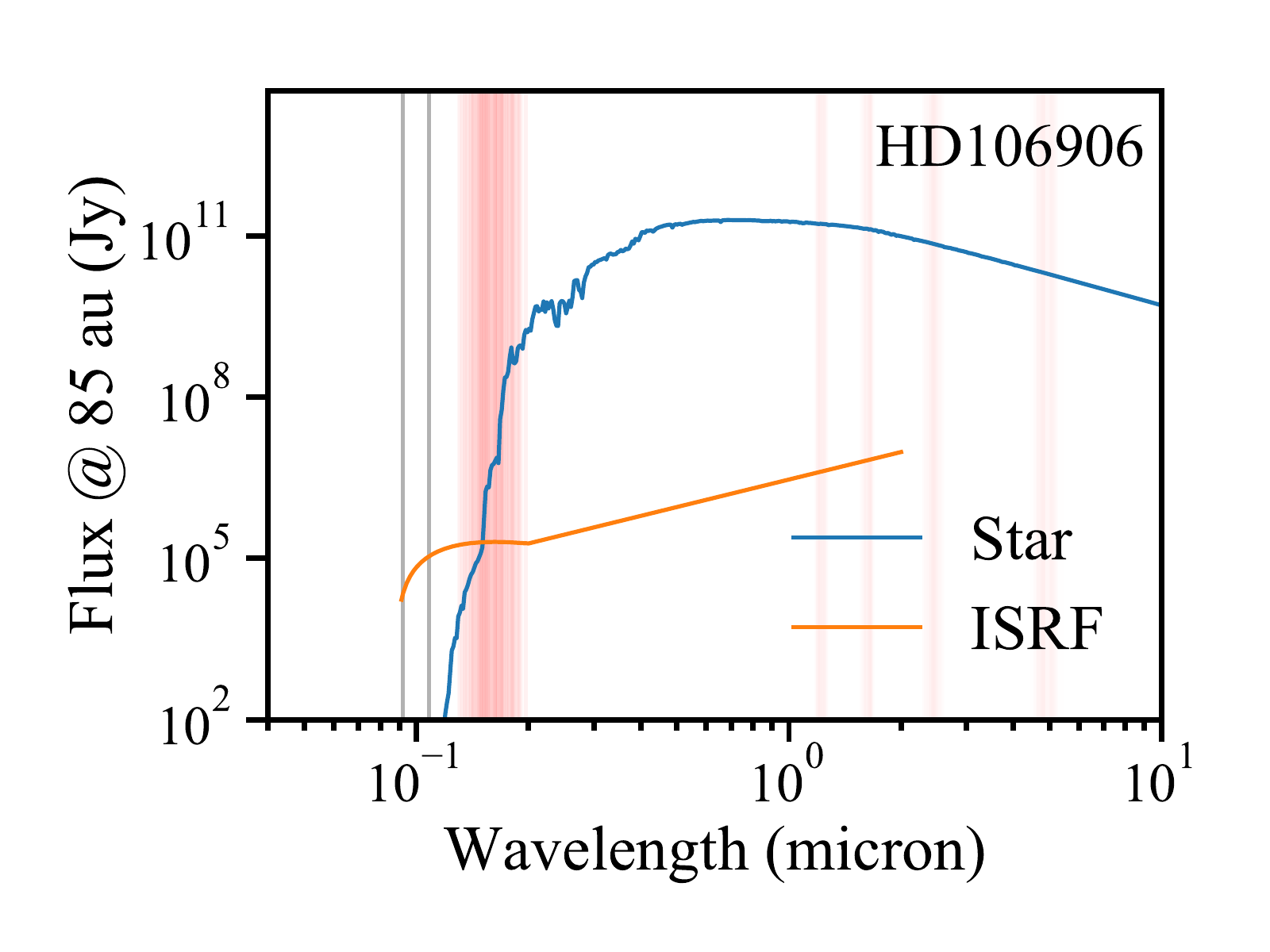}&
\includegraphics[width=6cm]{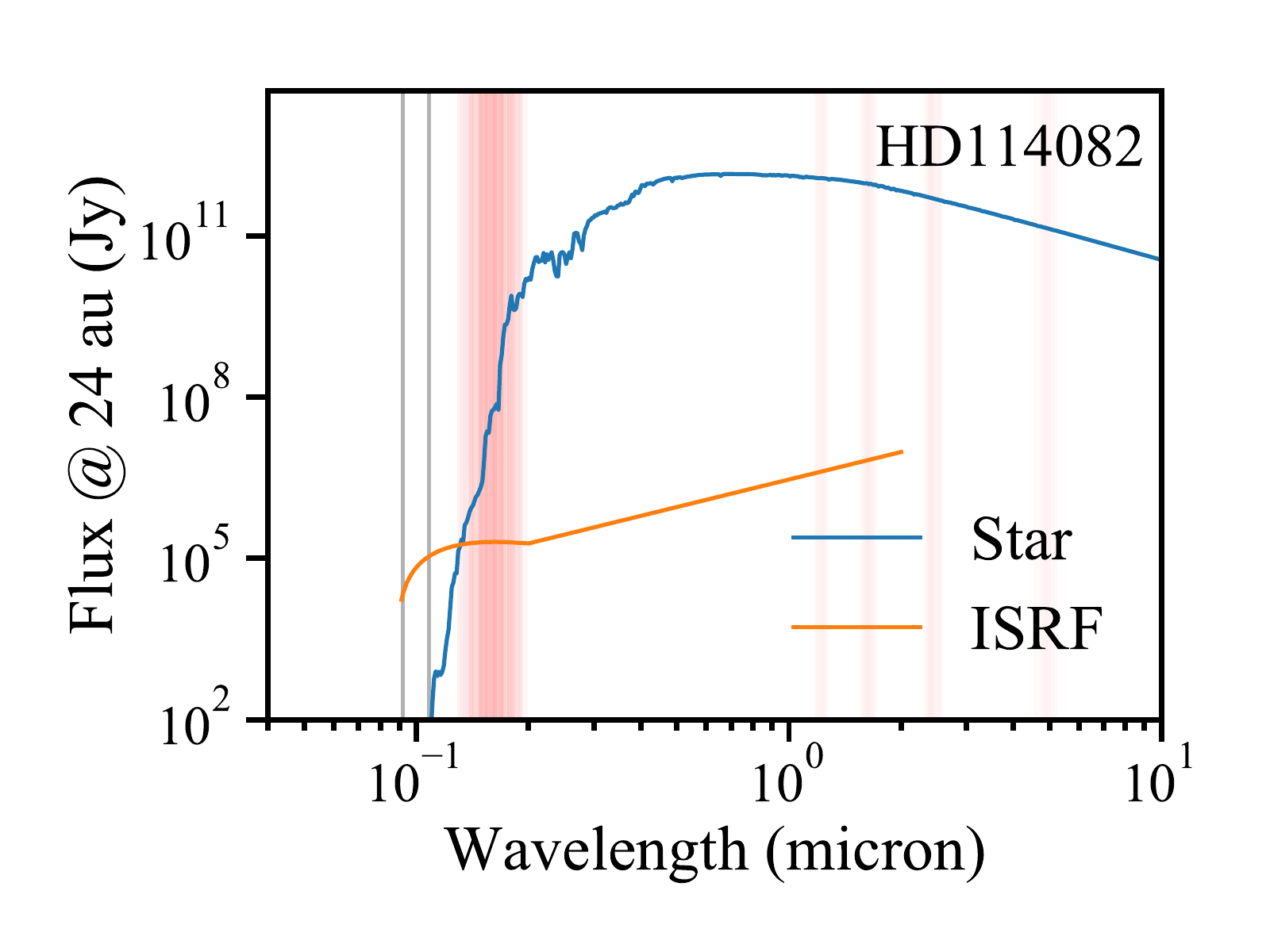}\\

\includegraphics[width=6cm]{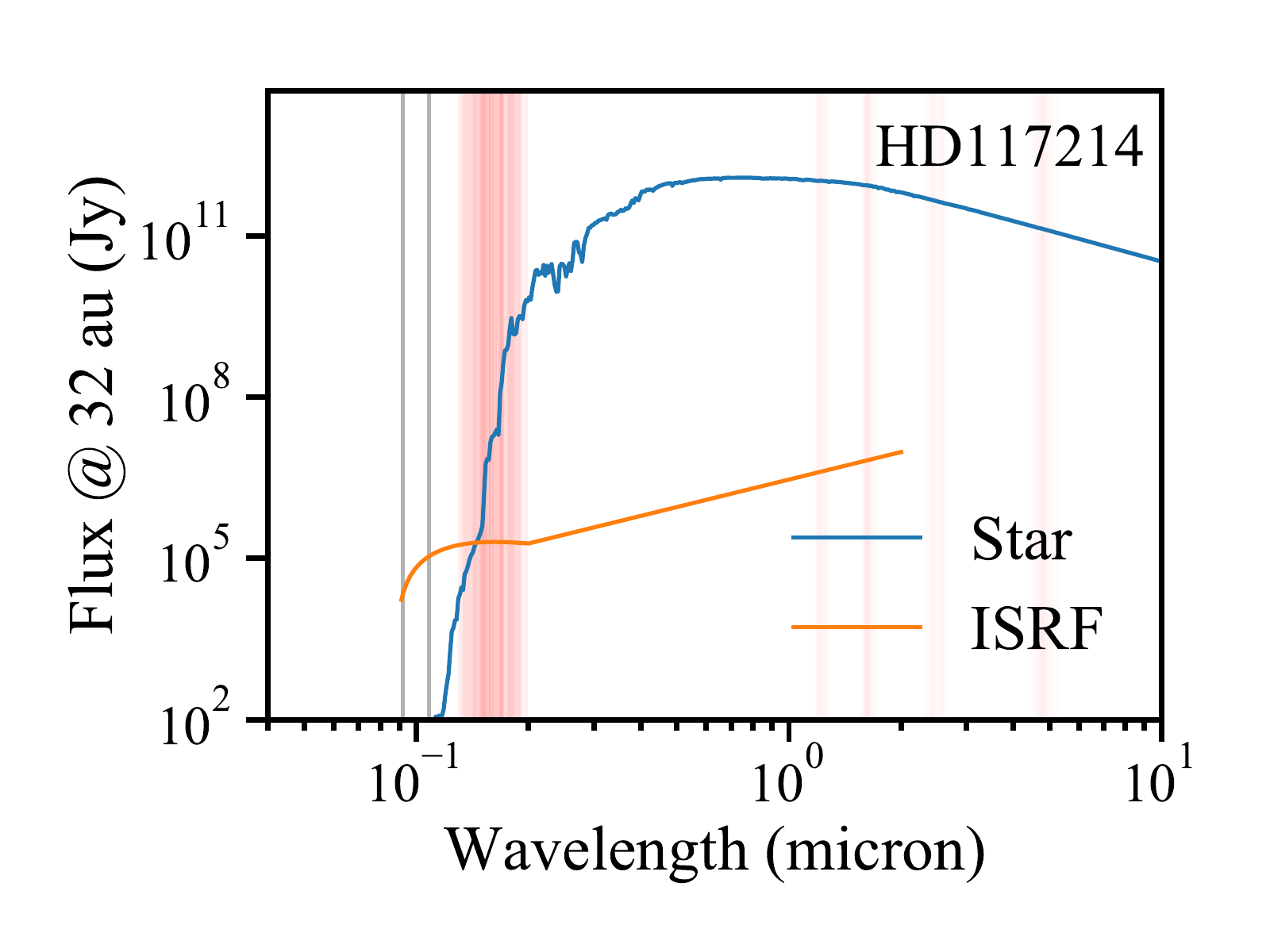}&
\includegraphics[width=6cm]{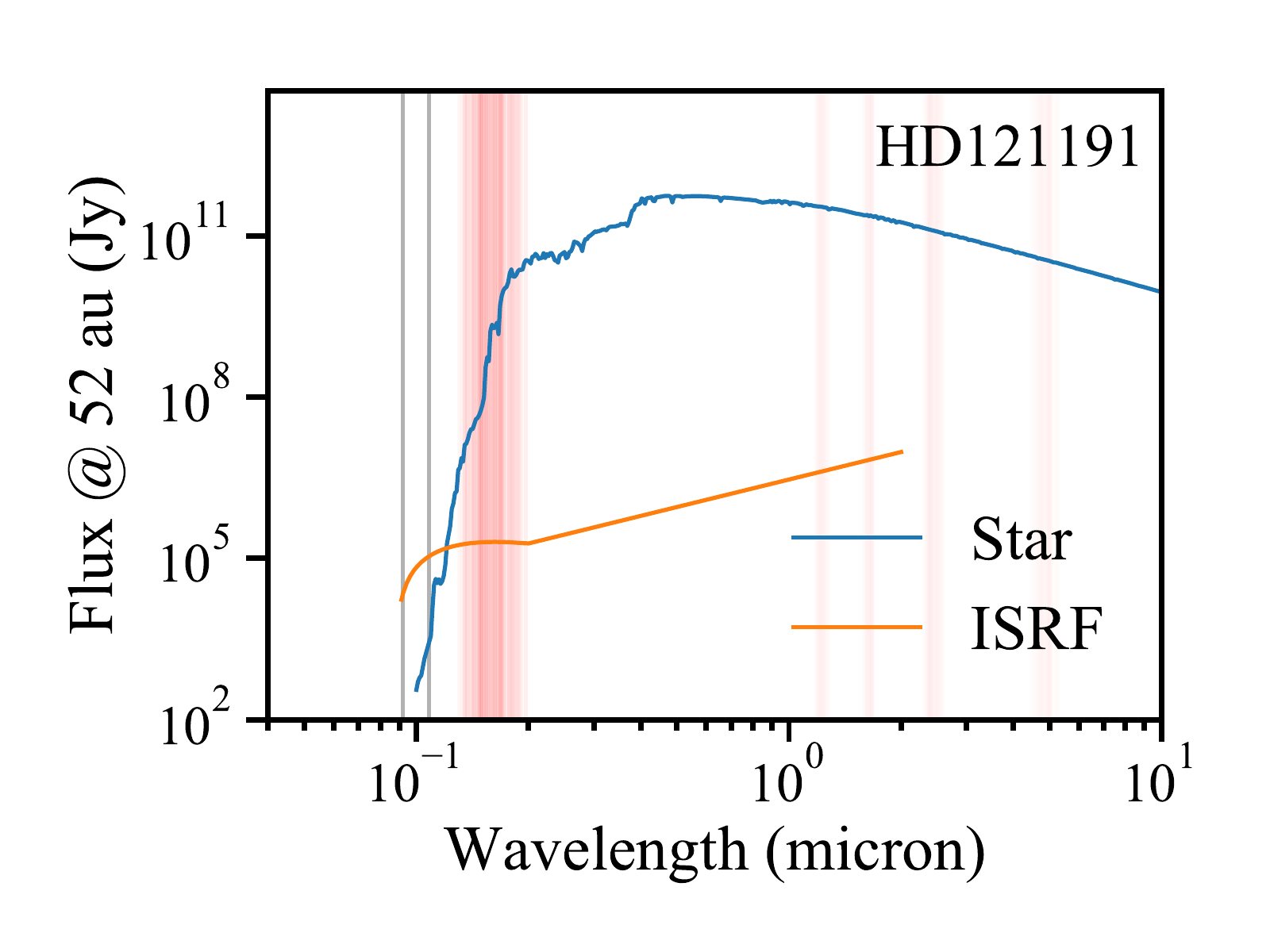}\\

\includegraphics[width=6cm]{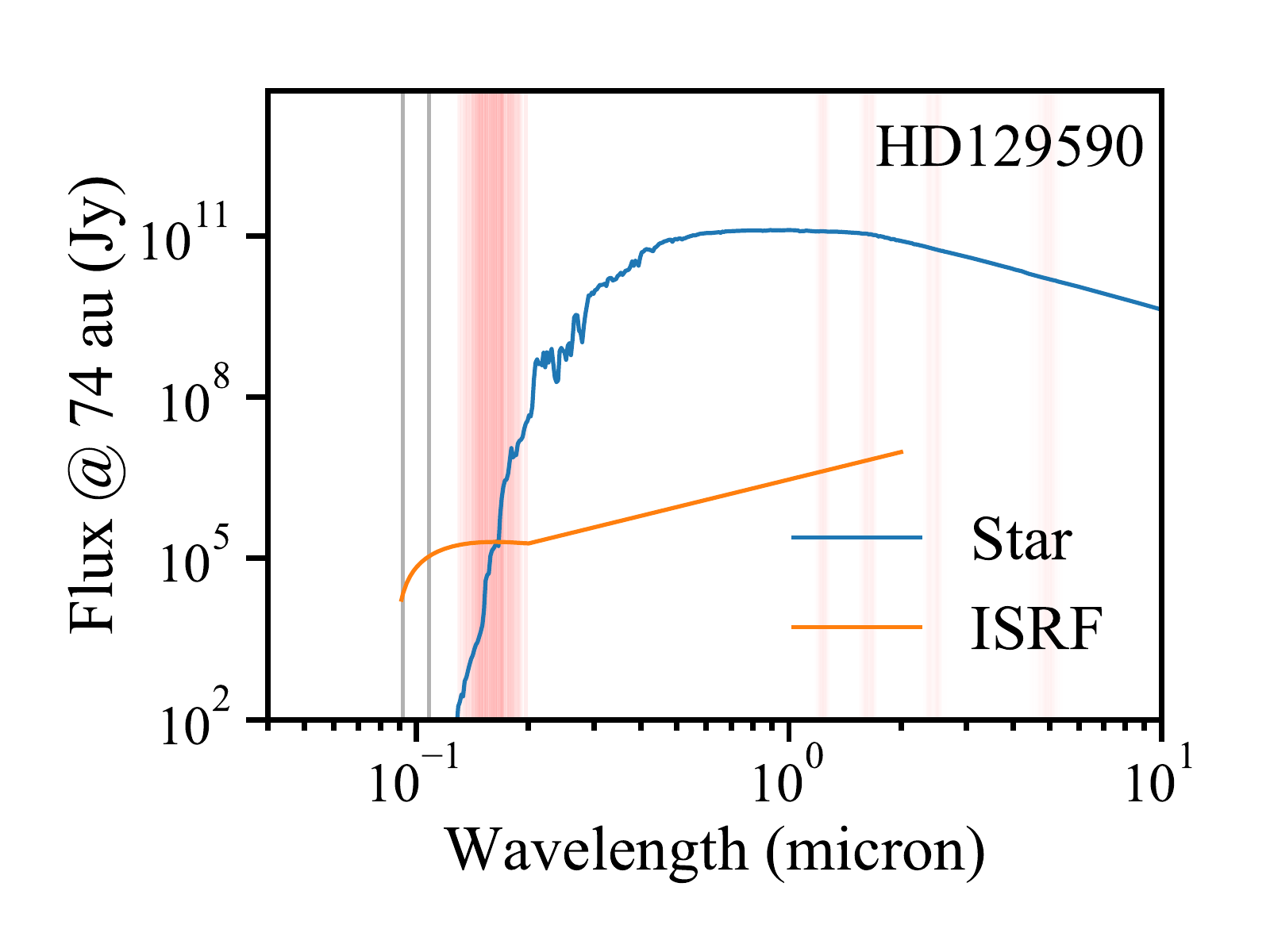}&
\includegraphics[width=6cm]{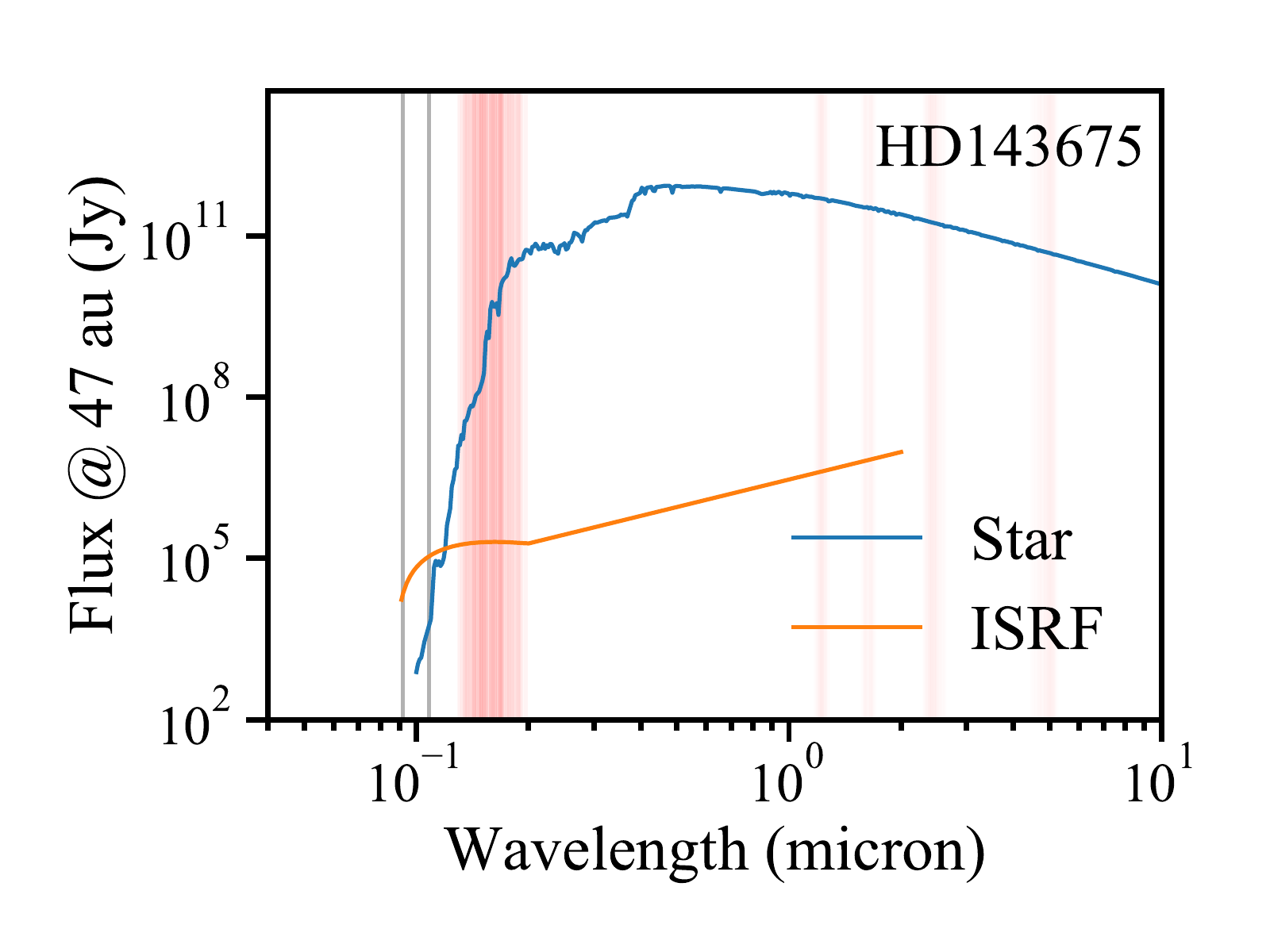}\\

\includegraphics[width=6cm]{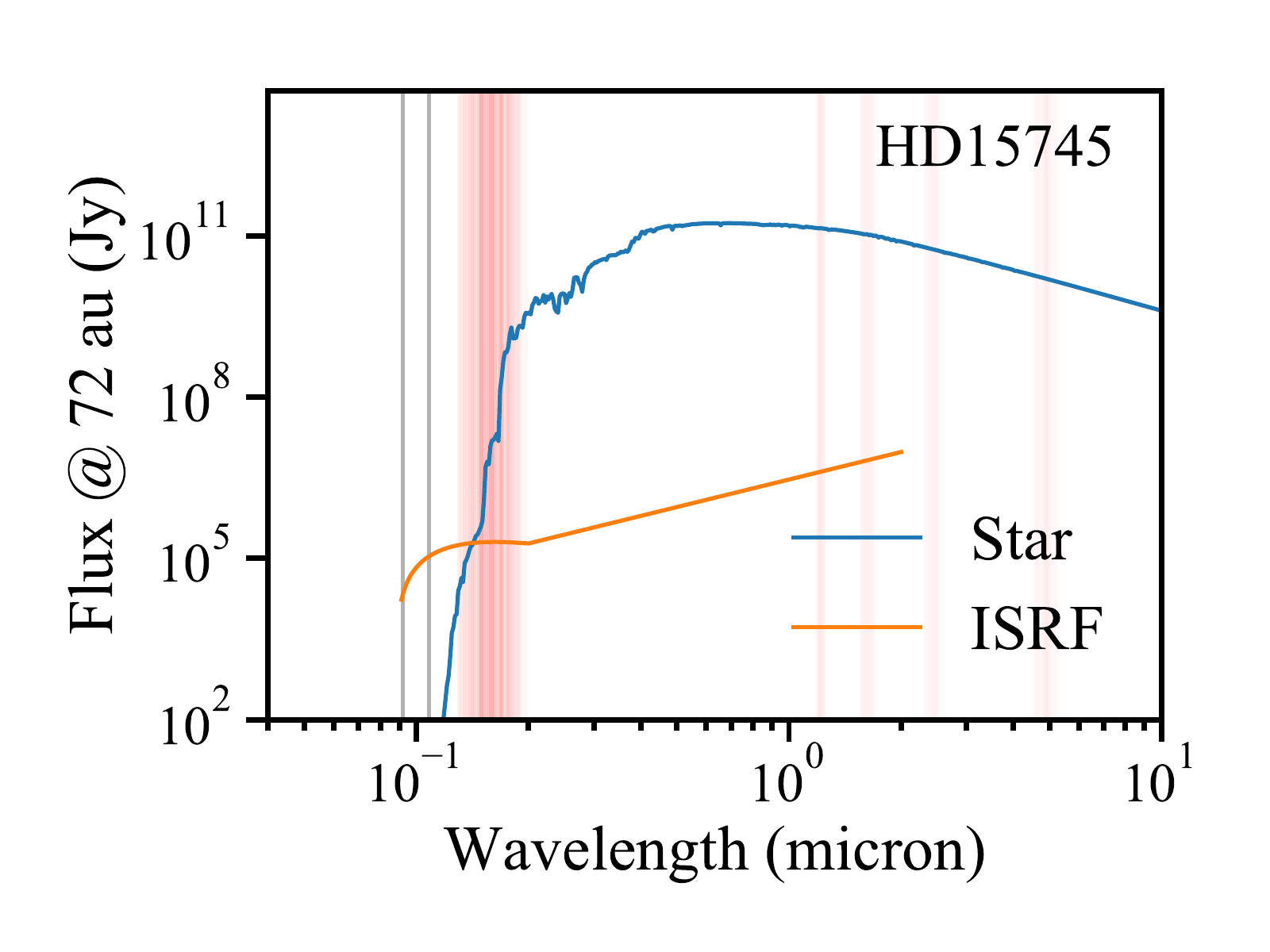}&
\includegraphics[width=6cm]{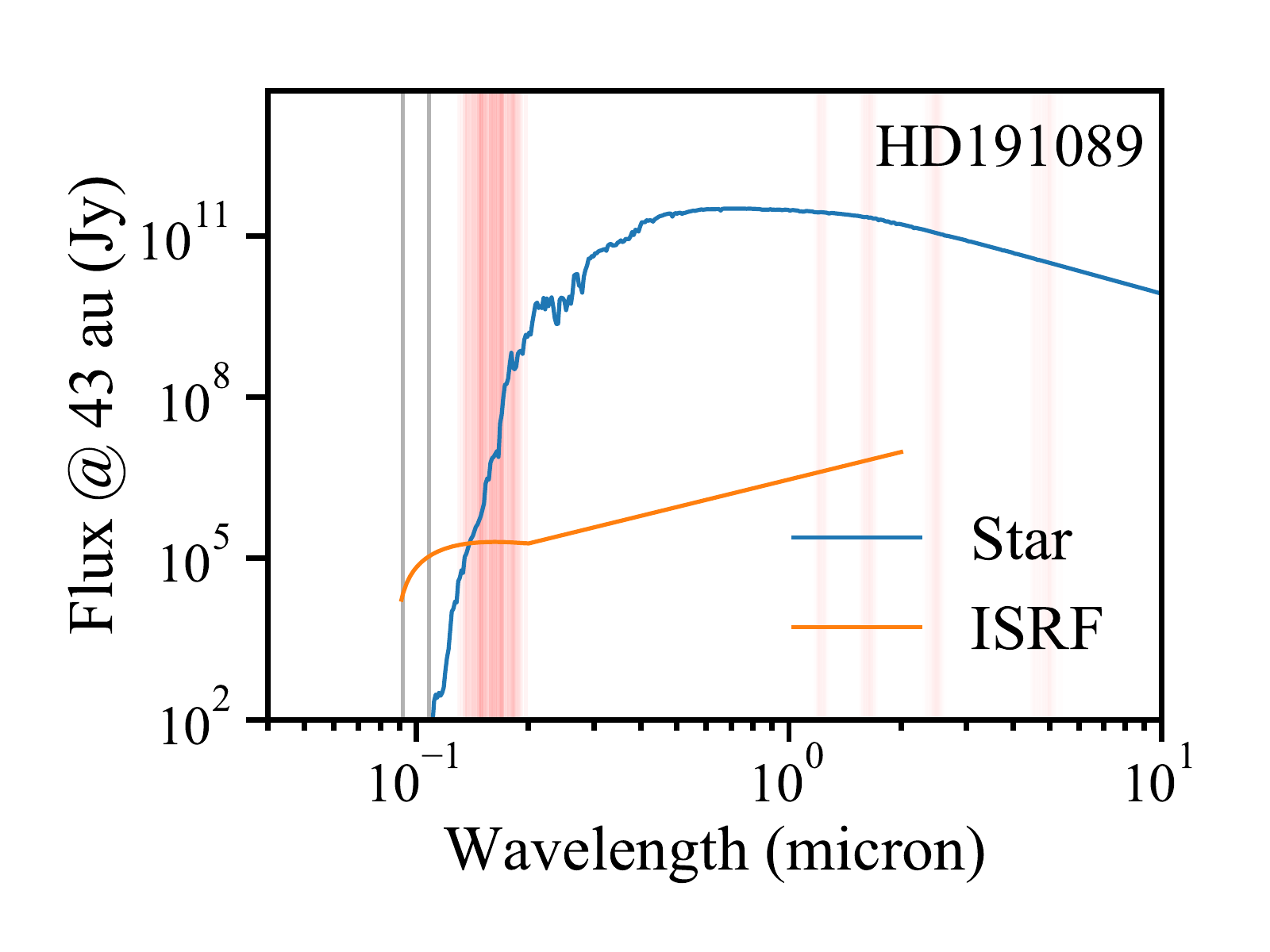}\\

\includegraphics[width=6cm]{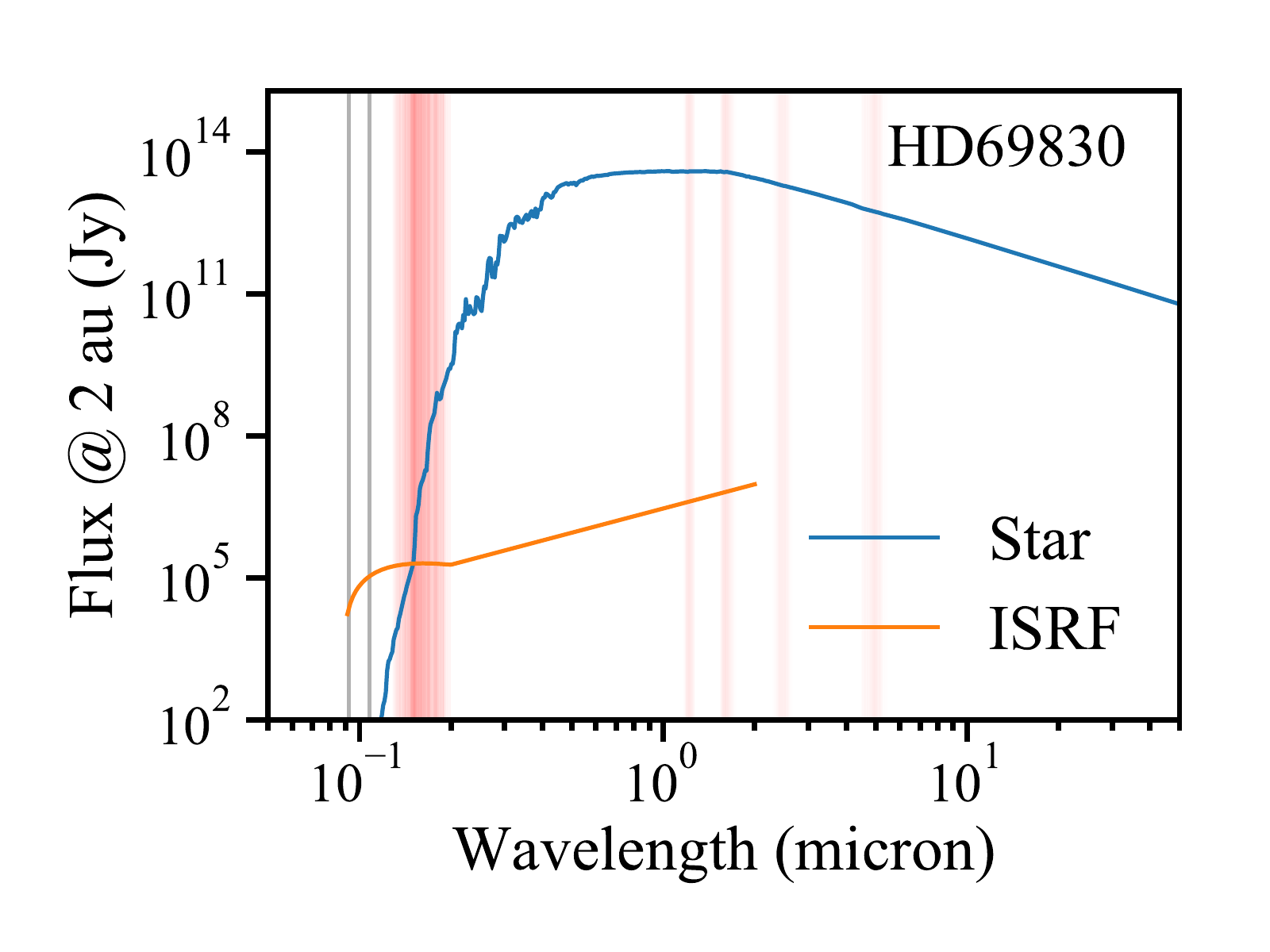}&
\includegraphics[width=6cm]{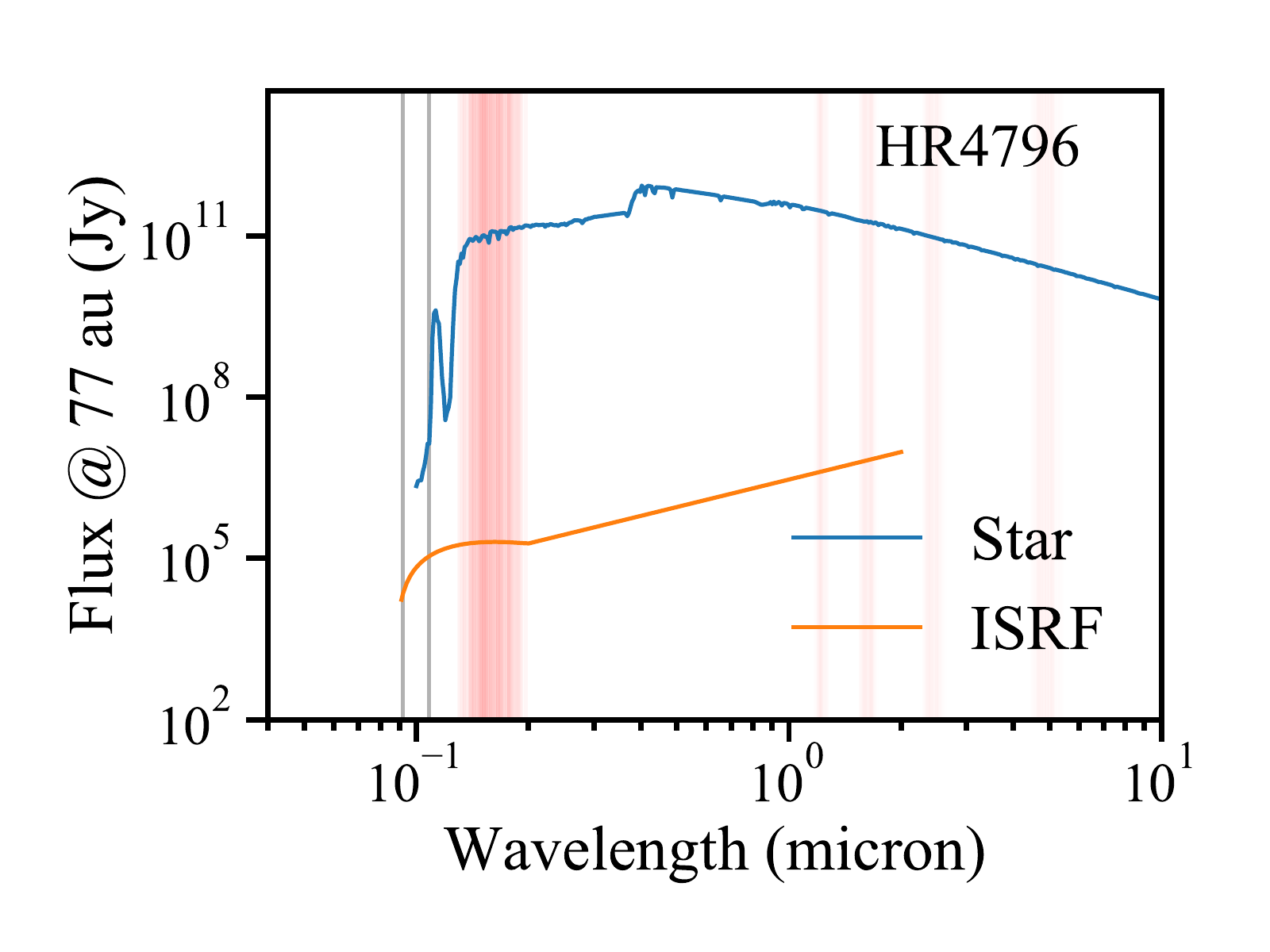}

\end{tabular}

    \caption{Irradiation of the discs for our sample of 10 stars. The orange line is the ISRF and the blue line is the star emission at the belt radius. Vertical grey lines delimit the CO photodissociation range. Red lines are electronic (UV) and rovibrational (IR) transitions that are accounted for in the non-LTE code we use \citep{2015MNRAS.447.3936M}.}
    \label{figspectreappendix} 
\end{figure*}

\label{lastpage}

\end{document}